\documentclass[physics,dissertation,unsrt,coversheets,nochapterblankpages,nononchapterblankpages]{puthesis}
\usepackage{subeqn}
\usepackage{subfigure}

\usepackage{hyperref}

\usepackage{multicol}

\title{Long-Range Multiplicity Correlations in Relativistic Heavy Ion Collisions as a Signal for Dense Partonic Matter}

\author{Terence J. Tarnowsky}{Tarnowsky, Terence J.}

\pudegree{Doctor of Philosophy}{Ph.D.}{May}{2008}

\majorprof{Rolf P. Scharenberg}

\campus{West Lafayette}

\newcommand{\be}{\begin{equation}}
\newcommand{\ee}{\end{equation}}

\begin{document}

\setlength{\pdfpageheight}{\paperheight}
\setlength{\pdfpagewidth}{\paperwidth}

\volume

%
%
%
%
%


\begin{acknowledgments}
I would like to thank my advisor Rolf Scharenberg for providing an engaging research experience, supporting me during my studies, and providing a clear view of physics that sometimes gets lost in the big picture. I would also like to thank Brijesh Srivastava, who was an extremely valuable and encouraging colleague while carrying out this research. Both individuals have helped shape the work presented here. I would also like to thank my committee members: John Finley, Andrew Hirsch, Denes Molnar, Fuqiang Wang, Scharenberg, and Srivastava. My fondest regards go out to the past and present members of the high-energy nuclear physics group at Purdue including Blair Stringfellow and Norbert Porile, along with all the post-docs and fellow graduate students I have worked with over the years. I also thank the leaders of the FTPC effort in STAR: Peter Seyboth, Janet Seyboth, and Alexei Lebedev for being willing to dedicate much of their time to instructing me on FTPC calibration and operation. My thanks go out to the colleagues and friends I have made through working in the STAR collaboration. Finally, I would like to thank my family and friends for being supportive throughout my academic career and humoring me when I try to describe my research to them.
 
This work was supported by the DOE grant: DE-FG02-88ER40412 and the Purdue Research Foundation grant: PRF-690 1396-3955.
\end{acknowledgments}

%
%
%

\tableofcontents

\listoftables

\listoffigures





\begin{abstract}

A dense form of matter is formed in relativistic heavy ion collisions. The constituent degrees of freedom in this dense matter are currently unknown. Long-range, forward-backward multiplicity correlations (LRC) are expected to arise due to multiple partonic interactions. Model independent and dependent arguments suggest that such correlations are due to multiple partonic interactions. These correlations are predicted in the context of the Dual Parton Model (DPM). The DPM describes soft partonic processes and hadronization. 
This model indicates that the underlying mechanism creating these long-range multiplicity correlations in the bulk matter is due to multiple partonic interactions. 

In this thesis, long-range multiplicity correlations have been studied in heavy ion (Au+Au and Cu+Cu) and hadron-hadron ({\it pp}) collisions. The behavior has been studied as a function of pseudorapidity gap ($\Delta\eta$) about $\eta$ = 0, the centrality, atomic number, and incident energy dependence of the colliding particles. Strong, long-range correlations ($\Delta\eta >$ 1.0) as a function of $\Delta\eta$ are found for central collisions of 
heavy ions at an energy of $\sqrt{s_{NN}}$ = 200 GeV. 
This indicates substantial amounts of dense partonic matter are formed in central heavy ion collisions at an energy of $\sqrt{s_{NN}}$ = 200 GeV.

\end{abstract}

%
%
%

\chapter{Introduction}
%
%

\subsection{Quark Searches}
Searches for quarks in high-energy accelerator and cosmic ray experiments, and even experiments similar to the Millikan oil-drop experiment, have not observed free quarks. However, in deep inelastic scattering events, where a high-energy electron scatters off a proton or neutron, the 
cross section of wide-angle scattering was inconsistent with an electron interacting with 
a point-like proton. In the 1960s, high-energy electron experiments became available to probe the fundamental structure of nucleons via deep inelastic scattering. At larger momentum transfers (Q$^{2}$), it is possible to probe short-range features. The physical existence of quarks and gluons inside protons and neutrons was discovered in these high-energy inelastic electron scattering experiments. 

\subsection{Experimental Basis for Quantum Chromodynamics}
The basic tenets of the strong interaction and its characterization by quantum chromodynamics (QCD) have been based on experimental results. The existence of hadrons consisting of three quarks in the same spin state suggested that quarks must possess an additional distinguishing characteristic. One example, the $\Delta^{++}$ (or $\Delta^{-}$) resonance consists of three {\it u} ({\it d}) quarks in the same spin state ($\frac{1}{2}$). Since {\it u} and {\it d} quarks are fermions, they will obey the Pauli Exclusion principle. As such, the ground state wave function must be antisymmetric with respect to the interchange of any two quarks in coordinate space. However, based on the known properties of the $\Delta^{++}$ (or $\Delta^{-}$) it appeared that the wavefunction was symmetric. The problem was solved 
by the addition of a new quantum number, which forms a fully antisymmetric wave function. The concept was proposed by Oscar Greenberg, who discussed quarks in terms of a new quantum number. 
It was Murray Gell-Mann who named the new quantity ``color''.  In the case of $\Delta^{++}$ (or $\Delta^{-}$), the three quarks have different colors and the overall wave function of the resonance is antisymmetric with respect to interchange of any quark. The quarks combine in such a fashion to make all baryons and mesons color neutral objects (Table \ref{HadronTable}).
\\
\begin{table}
\begin{center}
\begin{tabular}{|*{4}{c|}}
\hline
      Name &     Symbol & Mass ($GeV/c^{2}$) & Quark Content \\
\hline
      Pion & $\pi^{+}$ &      0.140 & $u\overline{d}$ \\
\hline
      Kaon &   $K^{+}$ &      0.494 & $u\overline{s}$ \\
\hline
      Phi &   $\Phi$ &      1.019 & $s\overline{s}$ \\
\hline
      J/$\Psi$  & J/$\Psi$ &  3.097 &  $c\overline{c}$ \\
\hline
      Proton &       $p$ &    0.938 &     $uud$ \\
\hline
      Neutron &       $n$ &    0.940 &     $udd$ \\
\hline
      Delta &       $\Delta^{++}$ &    1.232 &     $uuu$ \\
\hline
      Lambda  & $\Lambda^{0}$ &  1.116 &  $uds$ \\
\hline
      Xi  & $\Xi^{0}$ &  1.315 &  $uss$ \\
\hline
      Omega  & $\Omega^{-}$ &  1.672 &  $sss$ \\
\hline
\end{tabular}
\caption[Table of representative hadrons.]{A listing of some hadrons, their masses, and quark content \protect\cite{pdg}.} 
\label{HadronTable}
\end{center}
\end{table}

In the highest energy scattering events quarks were never observed to exist outside of their parent hadrons. This indicated that the force between two quarks must increase as the distance between them increases. The presence of a massless particle to mediate the force between two quarks was introduced. This particle was referred to as a gluon. The gluon is the analogue to the photon, a massless boson that mediates the electromagnetic force between charged objects in quantum electrodynamics (QED). Unlike the photon, which carries no electric charge, the gluon cannot be neutral with respect to the color charge. If this were the case, the potential energy for a quark-quark or quark-antiquark state would yield an average potential that for mesons would be negative (attractive), while that for baryons would be positive (repulsive). If gluons are colored, the potential for both mesons and baryons is attractive. Since they are colored, gluons not only interact with quarks, but other gluons as well. Their colored nature leads to gluon self-interaction. 
\\

Since color charge is a conserved quantity, and the emission or absorption of a gluon leads to a color change, gluons can be considered as consisting of a $color-\overline{color}$ (anti-color) charge. There are nine possible combinations of $color-\overline{color}$, which leads to the prediction that there should exist nine gluons. However, the ninth combination is actually a linear combination of all $color-\overline{color}$, and therefore colorless. If this gluon existed, colorless baryons would be able to emit these particular gluons and interact with each other via a short-range gluon force, which is not the case experimentally.
%


\begin{table}
\begin{center}
\begin{tabular}{|*{4}{c|}}
\hline
      Flavor &  Symbol & Current Mass ($GeV/c^{2}$) & Electric Charge \\
\hline
      up    &   $u$   &  $\sim$ 0.003     &      $\frac{2}{3}$ \\
\hline
      down  &   $d$   &  $\sim$ 0.006     &      $-\frac{1}{3}$ \\
\hline
\hline
      charm &   $c$   &     1.25          &      $\frac{2}{3}$ \\
\hline
      strange & $s$   &  $\sim$ 0.10      &      $-\frac{1}{3}$ \\
\hline
\hline
      top   &   $t$   &  $\sim$ 174     &      $\frac{2}{3}$ \\
\hline
      bottom &   $b$   &  4.20     &      $-\frac{1}{3}$ \\
\hline
\end{tabular}
\caption[Table of quarks.]{A listing of quarks, their current masses, and electric charge \protect\cite{pdg}.} 
\label{QuarkTable}
\end{center}
\end{table}


The potential between two quarks as a function of distance can be parameterized by,

\begin{equation}\label{QCDpotential} V(r) = - \frac{C\alpha_{s}}{r} + kr \end{equation}

\noindent where $\alpha_{s}$ is the strong coupling constant, $r$ the separation between the quarks, and $k$ a constant of about 1 GeV/fm. The form of this potential is such that at small separation distances ($r$ $\Rightarrow$ 0), the linear term vanishes and one is left with the form of an attractive Coulomb potential. The strong coupling constant $\alpha_{s}$ has been observed to decrease with increasing momentum transfer. This property is referred to as ``asymptotic freedom''. For large values of $r$, on the order of 10$^{-16}$ m (0.1 fm), the linear term dominates. As separation distance increases, an infinite amount of energy becomes required to separate the two quarks. However, as the energy in the color field between the two quarks increases with the separation distance, it will eventually exceed some threshold where it becomes more energetically favorable to create a $q-\overline{q}$ pair from the vacuum than to continue separating the quarks. This process forms two mesons, i.e., colorless objects. This property of the strong interaction makes the observation of isolated colored objects impossible. A schematic drawing of the chromoelectric flux between a quark-antiquark pair is shown in Figure \ref{QuarkFlux}. The six known quark flavors are shown in Table \ref{QuarkTable} along with their electric charge and estimates of their current mass (quark mass when deconfined) \cite{pdg}. The constituent mass is the effective mass of a quark in a binding potential and is larger than the current mass.

\begin{figure}
\centering
\includegraphics[width=3in]{qq_flux.eps}
\caption[Flux lines between a quark-antiquark pair.]{Flux lines between a quark-antiquark pair, which collapse into a tube between the two particles as their separation distance increases \protect\cite{Hands}.}
\label{QuarkFlux}
\end{figure}

\section{Phase Diagram of Strongly Interacting Bulk Matter}

\begin{figure}
\centering
\includegraphics[width=5.5in]{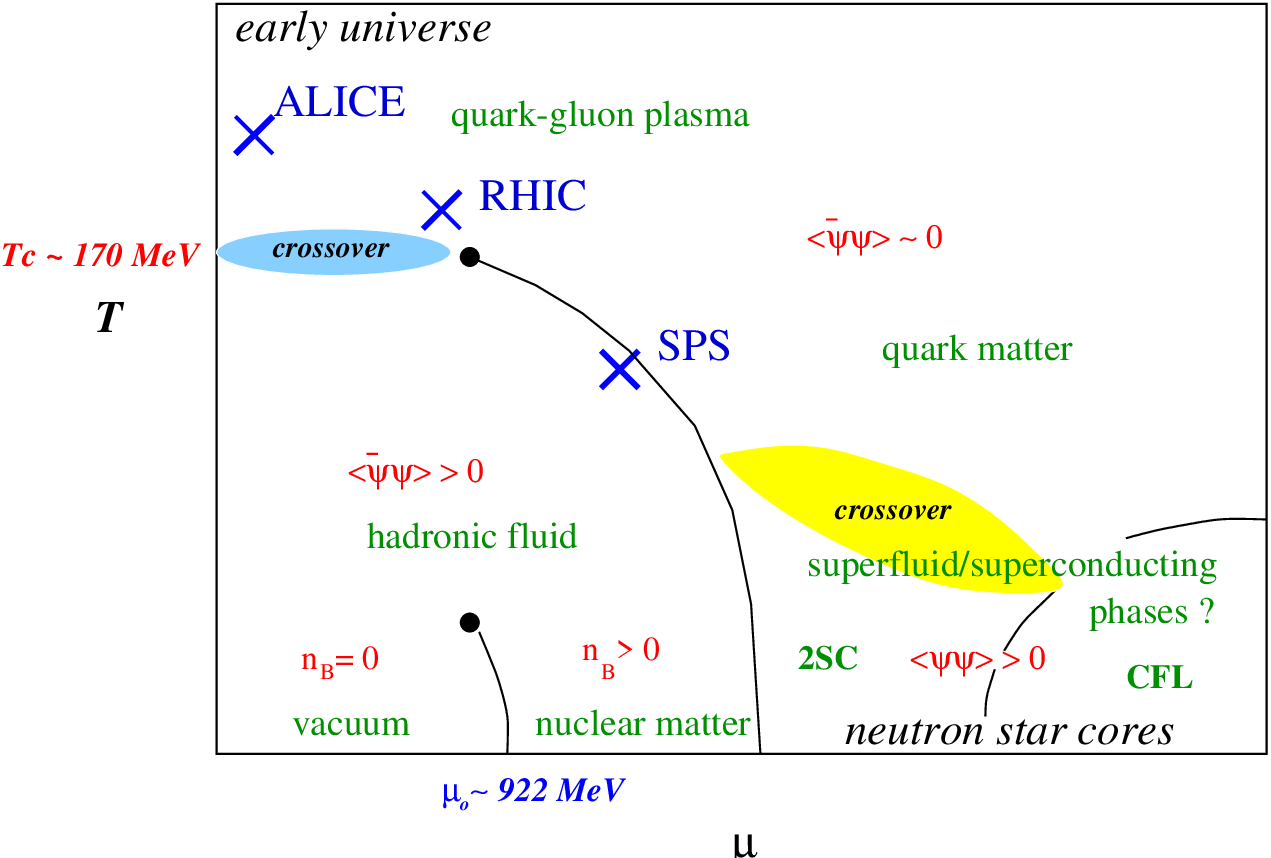}
\caption[A proposed phase diagram for QCD.]{A proposed phase diagram for QCD showing temperature as a function of baryon chemical potential ($\mu_{B}$) \protect\cite{Alton}. High-energy heavy ion collisions probe the regime of low $\mu_{B}$ and high temperature. The small curve denotes the nuclear liquid-gas phase transition \protect\cite{LiquidGas1, LiquidGas2, LiquidGas3}.}
\label{QCDPhaseDiagram}
\end{figure}

The quantum chromodynamic theory of strongly interacting matter predicts a phase transition between hadronic matter and quark-gluon degrees of freedom as the temperature or density of hadronic matter is increased. A schematic representation of the proposed QCD phase diagram is shown in Figure \ref{QCDPhaseDiagram} \cite{Alton}.

Normal nuclear matter exists at low temperature and a baryon chemical potential of approximately 922 MeV, corresponding to an energy density of $\approx$ 0.15 GeV/fm$^{3}$. 
At non-zero temperatures in this region the dominant contribution to the vacuum pressure is from pions, the lightest mesons. In a pion gas, there are only three hadronic degrees of freedom. Neglecting the pion mass, the pressure of an ideal pion gas can be expressed as \cite{WongBook},

\begin{equation}\label{Pressurepiongas} P_{pion \ gas} = \frac{3\pi^{2}}{90}T{^4} \end{equation}

\noindent which is the pressure of a Stefan-Boltzmann blackbody with three degrees of freedom. 
Here it is assumed that the relation between pressure and energy density is $P = \frac{1}{3}\epsilon$. The number of degrees of freedom in a state of matter consisting of deconfined quarks and gluons is much greater than that of a pion gas. In a system of deconfined light quarks (u, d, and s) and gluons there are a total of N $\approx$ 48 partonic degrees of freedom, with 16 bosonic (N$_{boson}$) degrees of freedom and 36 fermionic (N$_{fermion}$) degrees of freedom. The bosonic degrees of freedom arise from the gluons that have 2 spin states and 8 color combinations ($2*8$). The fermionic degrees of freedom are contributed by both the quarks and antiquarks (2) with 2 spin states, 3 color states, and n$_{f}$ flavor states each ($2*2*3*n_{f}$). The factor of $\frac{7}{8}$ in Equation \ref{NdofQGP} arises from differences in Bose-Einstein (gluons of spin 1) and Fermi-Dirac (quarks of spin $\pm\frac{1}{2}$) statistics in the Boltzmann factor. $n_{f} = 3 $ if only {\it u}, {\it d}, and {\it s} quarks are considered. In an ideal quark-gluon plasma (QGP) with massless quarks, the system can be modeled as a massless, relativistic gas with energy density $\epsilon = 3P$.
\begin{subequations}
\begin{equation}\label{NdofQGP} N = N_{boson} + \frac{7}{8}N_{fermion} \end{equation}
\begin{equation}\label{NdofQGP2} N = 16 + \frac{21}{2}n_{f} \end{equation}
\begin{equation}\label{PQGP} P_{QGP} = \frac{N\pi^{2}}{90}T^{4} \end{equation}
\end{subequations}

\noindent When the pressures of the pion gas and the QGP are equal, both phases should exist in equilibrium. Calculations have been carried out using lattice QCD methods to determine the deconfinement temperature, T$_{c}$. Lattice QCD models quarks and gluons on a discrete lattice. From these results, T$_{c}$ is predicted to lie at $\approx$ 170 $\pm$ 10 MeV, corresponding to temperatures on the order of 10$^{12}$ Kelvin.
 
\begin{figure}
\centering
\includegraphics[width=5in]{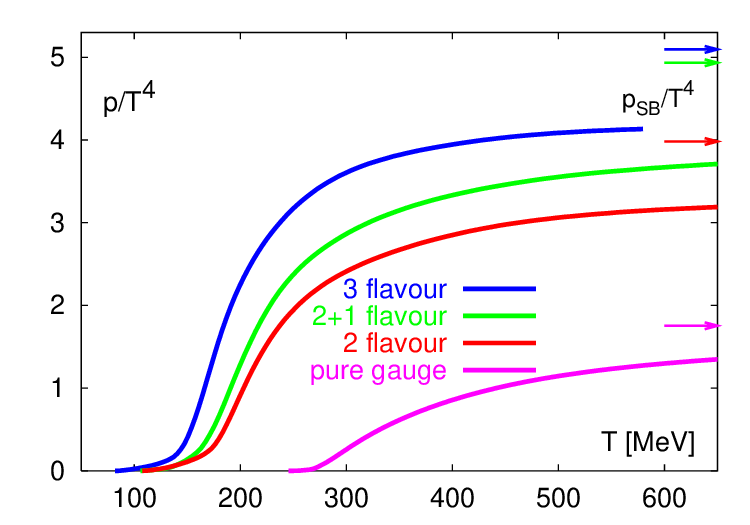} 
\caption[Lattice QCD calculations for pressure as a function of temperature.]{Lattice QCD calculations for pressure as a function of temperature. At $\frac{T}{T_{c}} = 1$ ($\approx$ 170 MeV, the critical temperature) there is an increase in pressure, potentially indicating a phase transition \protect\cite{Karsch}.}
\label{LQCD}
\end{figure}

\begin{figure}
\centering
\includegraphics[width=5in]{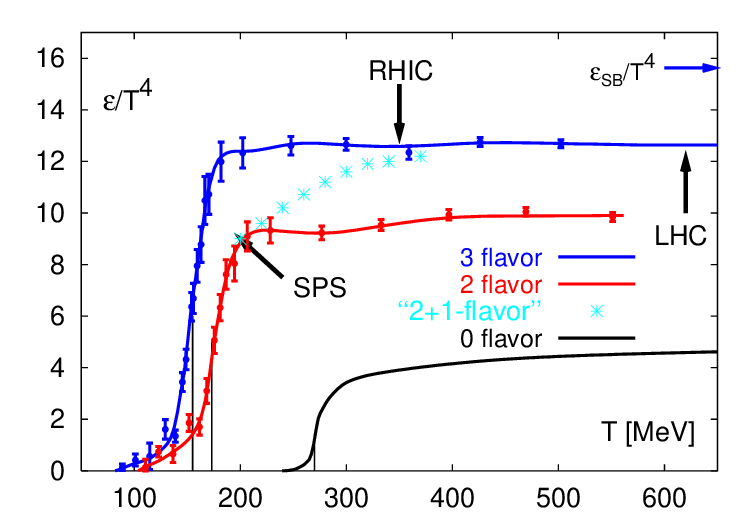} 
\caption[Lattice QCD calculations for energy density as a function of temperature.]{Lattice QCD calculations for energy density as a function of temperature. At $\frac{T}{T_{c}} = 1$ ($\approx$ 170 MeV) there is a rapid change in energy density, indicating a rapid increase in the number of degrees of freedom and a potential phase transition \protect\cite{Karsch}. The arrows indicate the estimated temperatures in collisions at SPS, RHIC, and LHC.}
\label{LQCD2}
\end{figure}

Figures \ref{LQCD} and \ref{LQCD2} show lattice QCD calculations for the pressure and energy density normalized by T$^{4}$ as a function of temperature \cite{Karsch}. The calculations are for three light quark flavors, two light and one heavy quark flavor, 2 light quark flavors, and the pure gauge case, i.e., with infinite quark masses. The factor $e^{-\frac{T}{m_{s}}}$, taking into account the mass of the strange quark $m_{s}$, accounts for the difference between the two light flavor and the ``2+1'' case.
 
At approximately $T = T_{c}$ (170 MeV) there is a rapid change in the energy density of the system. For $T > 1.5T_{c}$, the energy density (related to the number of degrees of freedom) appears to be nearly constant, but below the Stefan-Boltzmann limit for an ideal gas of quarks and gluons. 
Since the pressure rises more slowly than the energy density it has been suggested the QGP that may be produced at RHIC (up to $\approx$ 2 T$_{c}$), may still consist of 
clusters of strongly interacting quarks and gluons \cite{Shuryak}. Thus, at temperatures greater than 2T$_{c}$ the coupling may continue to decrease, eventually (T$>>$T$_{c}$) yielding a weakly interacting plasma. 
\\

Lattice QCD calculations with realistic quark masses indicate the presence of a critical point (region) \cite{Fodor}. The exact location of the critical point is dependent on the quark masses used in the calculation. The critical point is shown in Figure \ref{QCDPhaseDiagram} as the point lying on the phase boundary between hadronic matter and a QGP, at $\mu_{B}$ (baryon chemical potential) equal to normal nuclear matter density. It is expected that for a transition to the left of the critical point, there exists a smooth crossover region, whereas to the right of the critical point a first order phase transition is expected. In the crossover regime a second order phase transition is expected, where there is no discontinuity in the entropy, i.e., no latent heat. For the quark-gluon to hadronic phase transition, it is possible that chiral symmetry is restored in the partonic phase. 
If chiral symmetry is exact, the pion would be massless. The mass of the pion (140 MeV) is non-zero because chiral symmetry is not exact and leads to non-zero quark masses ($m_{u,d} \neq 0$).

\section{Signatures of the Quark Gluon Plasma}\label{Signatures_QGP}

There are several suggested signatures of quark gluon plasma formation: $J/\Psi$ suppression \cite{JPsiSatz}, strangeness enhancement \cite{StrangenessRafelski}, direct photons, Hanbury Brown-Twiss (HBT) determined source radii, etc. \cite{QGPSigBass, QGPSigWong}. Two other observables that can indicate QGP formation are related to the bulk, collective nature of the system and the presence of energy densities high enough to support a deconfined quark-gluon system. These two measurements involve transverse 
elliptic flow of the bulk particles and the suppression of high $p_{T}$ particles in the medium. However, none of these signatures are conclusive evidence for the formation of quark-gluon matter.


%
\subsection{Strangeness Enhancement}

In {\it pp} collisions the production of strange particles is suppressed compared to particles containing light ({\it u} and {\it d}) quarks. Equal yields of {\it u}, {\it d}, and {\it s} quarks are not measured. One explanation for the suppression is the higher production threshold due to the mass of the {\it s} and $\overline{s}$ quark pair, if thermal production goes as $e^{-\frac{m_{q}}{T}}$, where $m_{q}$ is the mass of the produced quark. Strangeness can be enhanced due to the higher temperature of a quark-gluon medium, which leads to a lower effective quark mass (related to chiral symmetry restoration) \cite{QGPSigWong}. A deconfined system of quarks and gluons can also lead to the production of {\it s} and $\overline{s}$ quark pairs by gluon fusion. Strangeness enhancement has been seen experimentally at AGS, SPS, and RHIC, across a wide range of energies in both nucleus-nucleus and proton-nucleus collisions. This makes the signature of strangeness enhancement an ambiguous signal of the transition to quark-gluon matter.

\subsection{High-$p_{T}$ Particle Suppression}

The production of a dense, colored medium in heavy ion collisions was predicted to lead to the attenuation of high-$p_{T}$ partons that traverse the medium. If a high density, colored medium is produced, a colored parton that passes through the medium will interact with the medium constituents via gluon Bremsstrahlung \cite{TannenbaumSummary}. The final consequence of this is a suppression of high $p_{T}$ ($>$ 2 GeV) particles. This effect has been seen in Au+Au collisions at an energy of $\sqrt{s_{NN}}$ = 200 GeV, but is absent in d+Au at the same energy \cite{STARQuenchingPRL}. 
The energy loss of a parton via gluon radiation through a particular medium is dependent on the transport coefficient $\hat{q}$ \cite{Baier}. The calculation of $\hat{q}$ yields the same energy loss for an ideal pion gas and an ideal quark-gluon plasma, at equal energy density \cite{Baier, EntropyMuller}. Independently, jet quenching cannot distinguish a deconfined system of quarks and gluons from a hadronic system with similar gluon content \cite{EntropyMuller}.
\\

\begin{figure}
\centering
\includegraphics[width=4in]{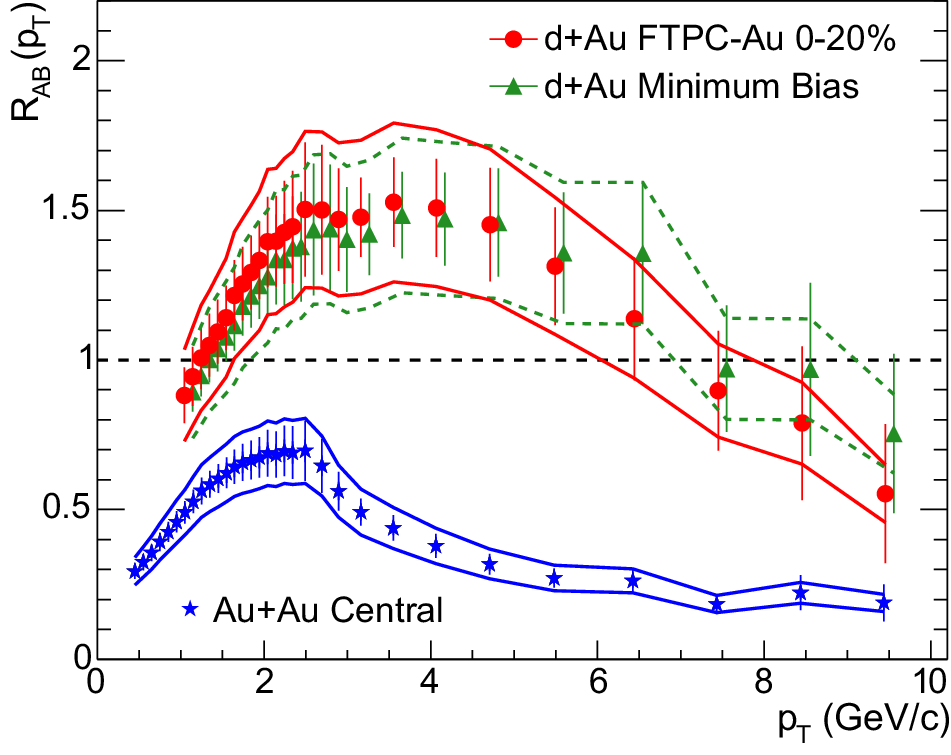} 
\caption[Nuclear modification factor ($R_{AB}$).]{Nuclear modification factor ($R_{AB}$) for central and minbias d+Au and central Au+Au collisions at an energy of $\sqrt{s_{NN}}$ = 200 GeV, showing the strong suppression of high-$p_{T}$ particle production in central Au+Au collisions \protect\cite{STARQuenchingPRL}.}
\label{STAR_RAA}
\end{figure}

In the study of nucleus-nucleus collisions there may be initial state nuclear effects that influence hadron production. This effect can be quantified using the nuclear modification factor. The nuclear modification factor is generally defined as \cite{STARwhitepaper},

\begin{equation}\label{RAA}
R_{AB}(p_{T}) = \frac{dN{_AB}/d\eta d^{2}p_{T}}{T_{AB} d\sigma_{NN}/d\eta d^{2}p_{T}}
\end{equation}

where $T_{AB} = \left<N_{bin}\right>/\sigma_{inelas}^{pp}$ accounts for the collision centrality. This is the ratio of inclusive charged hadron yields from A+B collisions to {\it pp}, normalized by the number of binary nucleon-nucleon collisions. In the absence of nuclear effects, if particle production scales with the number of binary collisions ($N_{bin}$) then $R_{AB}$ = 1. Figure \ref{STAR_RAA} shows $R_{AB}$ for central and minimum bias d+Au and central Au+Au data at an energy of $\sqrt{s_{NN}}$ = 200 GeV. However, there is a large suppression in particle production at high-$p_{T}$ in central Au+Au data.

\subsection{Open Questions}

While the signatures discussed in Section \ref{Signatures_QGP} provide evidence that dense matter is formed in heavy ion collisions, there remain a number of fundamental questions that must be addressed. The signals of deconfinement and the bulk properties of the system (thermalization, number of degrees of freedom, etc.) remain important open questions. Models including partonic or hadronic systems have been used to fit the data. The analysis of forward-backward (FB) multiplicity correlations has the potential to distinguish between hadron-hadron and parton-parton interactions. Both model independent and model dependent arguments suggest that long-range forward-backward correlations are due to multiple partonic interactions. This measurement can provide a clear signal that partonic degrees of freedom are involved. 

\section{Forward-Backward Multiplicity Correlations}

Correlations that are produced across a wide range in rapidity are thought to reflect the earliest stages of a heavy ion collision, free from final state effects \cite{LRCCGC}.
\\

The study of forward-backward (FB) multiplicity correlations has a long history. Studies were carried out in $e^{+}-e^{-}$, $\mu^{+}-p$, $p-p$, $p-\overline{p}$, and other experiments \cite{LRCee, LRCee2, LRCee3, LRCmup, LRCpp, LRCpp2, LRCpp3, LRCpp4}. 
Correlations that extend over a long range in pseudorapidity ($\eta$) have the potential to probe the early stages of heavy ion collisions. Forward-backward correlations have been characterized by the forward-backward correlation strength, {\it b}, the slope extracted from a linear relationship between the average multiplicity measured in the backward rapidity hemisphere ($<N_{b}>$) and the multiplicity in the forward rapidity hemisphere, $N_{f}$. This relationship was predicted theoretically, seen in hadron-hadron experiments, and expressed as \cite{LRC, bLinear},

\begin{equation}\label{linear} 
<N_{b}(N_{f})> = a + bN_{f} 
\end{equation}

In this definition, the correlation strength {\it b} can be positive or negative with a range of $|b| < 1$. This maximum (minimum) represents total correlation (anti-correlation) of the produced particles separated in rapidity. $b = 0$ is the limiting case of entirely uncorrelated particle production. Experimentally, the slope of {\it b} in hadron-hadron experiments is found to be positive. The intercept of Equation \ref{linear} {\it a} is related to the number of uncorrelated particles. 
\\

The correlation strength can be expressed as the ratio of the covariance of the forward-backward multiplicity and the variance of the forward multiplicity. This is done by performing a linear regression of Equation \ref{linear} and minimizing R$^{2}$,

\begin{equation}\label{regression}
\displaystyle R^{2} = \sum_{i}^{n}\left[N_{b_i} - \left(a + bN_{f_i}\right)\right]^{2}
\end{equation}

The minimization condition implies $\frac{\delta R^{2}}{\delta a} = \frac{\delta R^{2}}{\delta b} = 0$. This produces a set of equations that can be expressed in matrix form,


\begin{equation}\label{bmatrix}
\left[
\begin {array}{c}
a\\
\noalign{\medskip}
b
\end {array}
\displaystyle \right] = \left(n\sum_{i=1}^{n}N_{f_i}^{2}-\left(\sum_{i=1}^{n}N_{f_i}\right)^{2}\right)^{-1} \left[
\begin {array}{c}
\displaystyle \sum_{i=1}^{n}N_{b_i}\sum_{i=1}^{n}N_{f_i}^{2}-\sum_{i=1}^{n}N_{f_i}\sum_{i=1}^{n}N_{f_i}N_{b_i}\\
\noalign{\medskip}
\displaystyle n\sum_{i=1}^{n}N_{f_i}N_{b_i}-\sum_{i=1}^{n}N_{f_i}\sum_{i=1}^{n}N_{b_i}
\end {array}
\right]
\end{equation}

When expanded to solve for {\it b}, the matrix in Equation \ref{bmatrix} yields,

\begin{equation}\label{bexpanded}
b = \frac{\displaystyle n\sum_{i=1}^{n}N_{f_i}N_{b_i}-\sum_{i=1}^{n}N_{f_i}\sum_{i=1}^{n}N_{b_i}}{\displaystyle n\sum_{i=1}^{n}N_{f_i}^{2}-\left(\sum_{i=1}^{n}N_{f_i}\right)^{2}}
\end{equation}

the covariance of $N_{b}$ and $N_{f}$ normalized by the variance of $N_{f}$. Equation \ref{bexpanded} can be expressed in terms of the following calculable average values,

\begin{equation}\label{expectationvalue}
b = \frac{<N_{f}N_{b}>-<N_{f}><N_{b}>}{<N_{f}^{2}>-<N_{f}>^{2}}
\end{equation}

with \cite{LRCbasis},

%
%
%
\begin{subequations}
\begin{equation}\label{IntegralEqnCov} <N_{f}N_{b}>-<N_{f}><N_{b}> = \int_{y<0}dy \int_{y'>0}dy'\left[<N(y)N(y')>-<N(y)><N(y')>\right]
\end{equation}
\begin{equation}\label{IntegralEqnVar} <N_{f}^{2}>-<N_{f}>^{2} = \int_{y>0}dy \int_{y'>0}dy'\left[<N(y)N(y')>-<N(y)><N(y')>\right]
\end{equation}
\end{subequations}

%
%
%
%

\subsection{Short-Range Correlations}\label{S-RC}

Short-range correlations are correlations that extend over a small range of pseudorapidity ($|\eta| <$ 1.0). Short-range correlations are due to various short-range order effects \cite{LRCbasis}. These effects can include particles produced from cluster decay, resonance decay, or jet correlations. The particles produced in a single inelastic collision are known to only exhibit short-range correlations \cite{SRC}. 
%
%

%
\subsection{Long-Range Correlations}

Long-range correlations are correlations that extend over a wide range in pseudorapidity, beyond $|\eta| > 1.0$. The presence of long-range correlations is a violation of short-range order. Short-range order is expected to hold as long as ``unitarity constraints are neglected'' \cite{LRCbasis}. In the approximation of short-range order, only single scattering is considered. Therefore, quantum mechanical probability is not conserved, since it is possible to have multiple scattering terms.
\\

The presence of short-range order in each inelastic collision produces short-range correlations, discussed in Section \ref{S-RC}. The consideration of unitarity leads to the existence of multiple inelastic scattering, in addition to the single scattering that determines the short-range correlations. Multiple inelastic elementary scatterings are the source of the long-range correlation \cite{LRC}. Due to the possibility of 
multiparton scatterings in hadron-nucleus and nucleus-nucleus collisions compared to {\it pp}, if a long-range correlation exists at a given energy it should be enhanced in collisions involving nuclei \cite{LRC}. Long-range correlations have been seen in high-energy ($\sqrt{s_{NN}} >$ 1 TeV) {\it p$\overline{p}$} collisions where multiparton excitations are seen in high multiplicity events \cite{Walker2004, LRCpbarp}.
%
%
%

\chapter{Experiment}

\section{RHIC}

\begin{figure}
\centering
\includegraphics[width=6in]{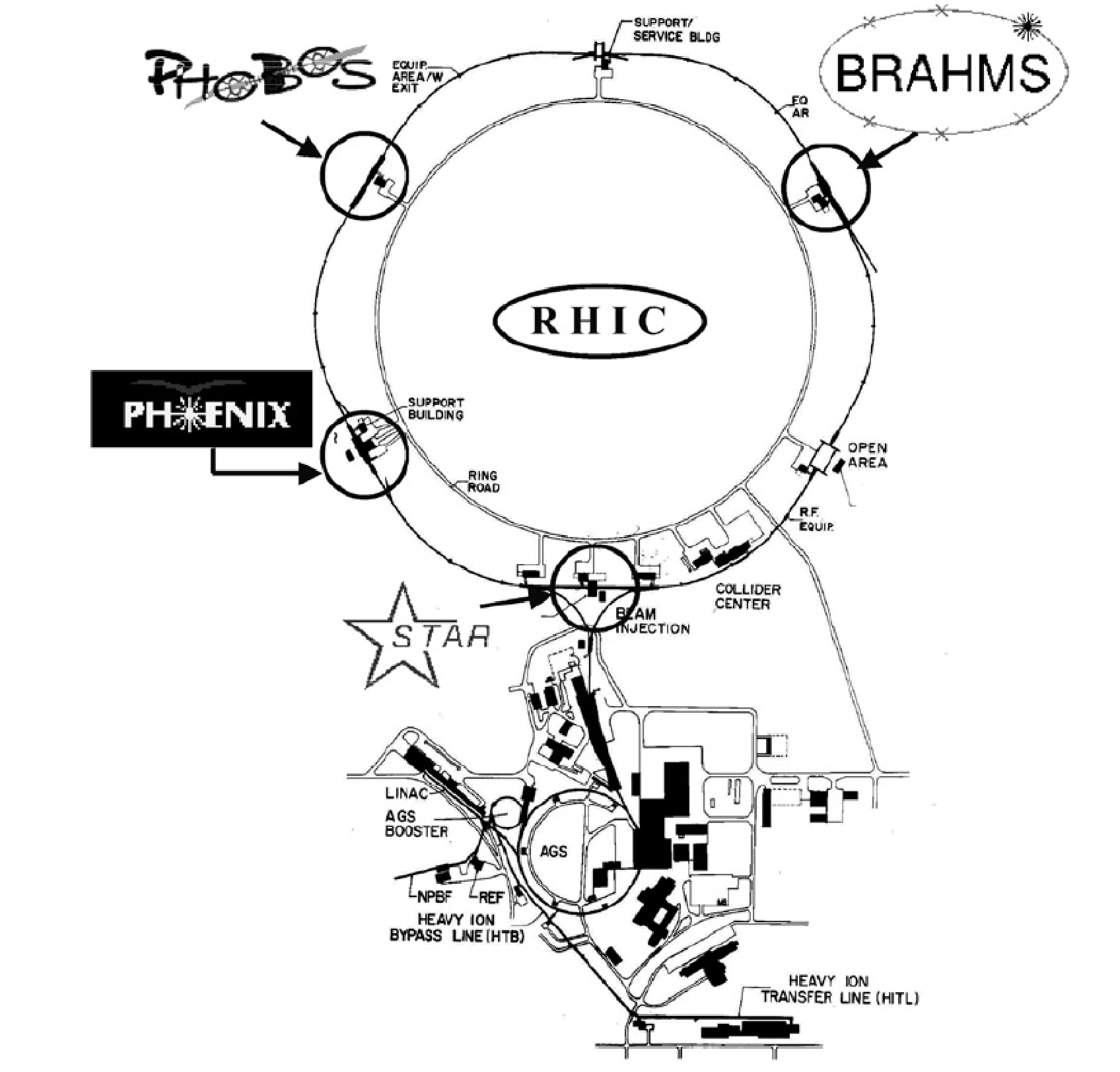} 
\caption[The Relativistic Heavy Ion Collider (RHIC) complex.]{The Relativistic Heavy Ion Collider (RHIC) complex \protect\cite{STAR}.}
\label{RHIC}
\end{figure}

The Relativistic Heavy Ion Collider (RHIC) complex (Figure \ref{RHIC}) at Brookhaven National Laboratory (BNL) consists of a succession of coupled accelerators that provide RHIC with fully ionized heavy ions, or polarized protons. RHIC is a superconducting hadron collider with two concentric rings (colloquially referred to as ``Blue'' (clockwise ion revolution) and ``Yellow'' (counter-clockwise ion revolution), 3.8 km in circumference. As Figure \ref{RHIC} shows, the rings are not circular, but consist of several arc sections mated to straight, insertion sections. The center of the insertion sections contains the intersection points. There are six intersection points around the RHIC ring. Four are (or were) occupied by experiments, BRAHMS at 2 o'clock, STAR at 6 o'clock, PHENIX at 8 o'clock, and PHOBOS at 10 o'clock. As of 2007, only PHENIX and STAR are actively taking data. The superconducting magnets in each ring operate at a magnetic rigidity, $B\rho = 8400 \ Tesla \cdot m$, which yields a kinetic energy of 100 $GeV/A$ for Au ions, with an $\frac{A}{Z}$ ratio of 2.5. The magnetic field strength is approximately 3.5 T. At this rigidity, lighter ion species can reach 125 $GeV/A$, while protons, with $\frac{A}{Z}$ = 1, can be accelerated to 250 GeV. Due to the presence of two independent rings, RHIC has the capability of colliding particles ranging from protons ($A=1$) to gold (Au) ($A=197$) in both symmetric and asymmetric collisions (e.g., deuteron+Au). Collisions involving yet heavier particle species up to uranium (U) will be possible with the upgrade of an Electron Beam Ion Source (EBIS). 

The injection of heavy nuclei into RHIC requires several successive steps. For Au, negatively charged ions are created from a pulsed sputter ion source, partially stripped of electrons with a foil at the high voltage terminal of the Tandem Van de Graaff, and accelerated by the second stage of the tandem to an energy of 1 MeV/A. After leaving the Tandem and passing through yet another stripping foil, the ions have a charge of +32. They are then transferred to the AGS Booster synchrotron, where their energy is increased to 95 MeV/A. Upon exit from the Booster the ions are further stripped to a charge of +77, leaving only two electrons of the initial 79 in the gold atom. These are injected into the Alternating Gradient Synchrotron (AGS), the final boosting stage prior to injection into RHIC and accelerated to an energy of 8.86 GeV/A. They then leave the AGS and are stripped of their final electrons as they travel through the transfer line to RHIC, and are then fully ionized to a charge of +79. Once in RHIC, the ions are accelerated to their final colliding energy and stored for several hours. 
\\
%
%


\section{STAR}

\begin{figure}
\centering
\includegraphics[width=5in]{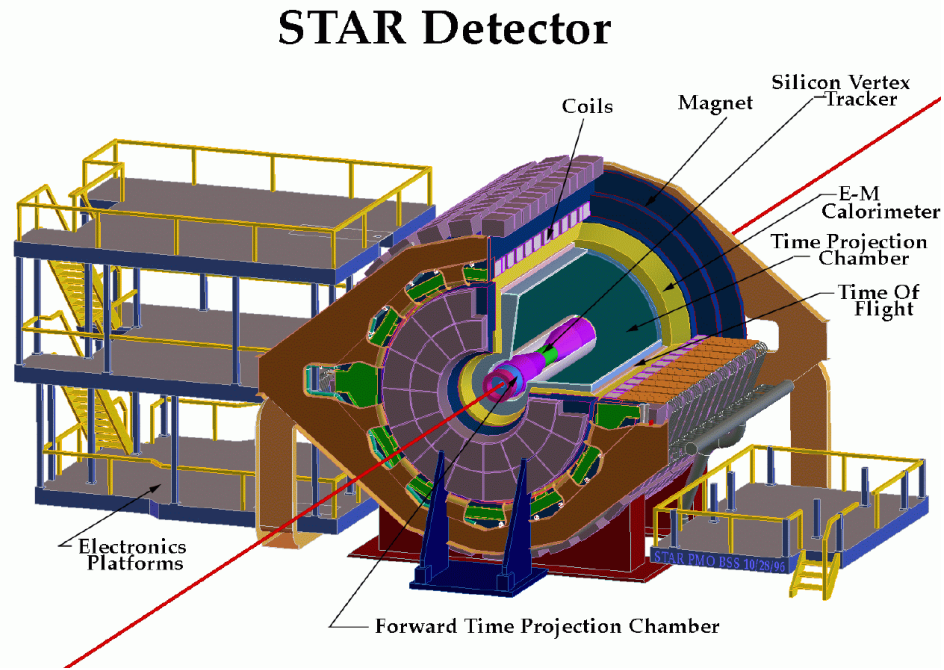} 
\caption[The STAR detector.]{The STAR detector.}
\label{STAR}
\end{figure}

The Solenoidal Tracker at RHIC (STAR), is one of the two large experiments at RHIC (Figure \ref{STAR} \cite{STAR}). STAR is a large acceptance detector, covering a pseudorapidity range of $|\eta| < 1.8$, with additional coverage at forward pseudorapidity $|\eta| \approx 4.0$. The main tracking detectors in STAR have full azimuthal symmetry, covering $\Delta\phi = 2\pi$. The large $\eta$ and $\phi$ acceptance, symmetry about $\eta$ = 0, and accurate tracking makes STAR ideally suited for characterizing the bulk particles ($p_{T} <$ 2.0 GeV) and for performing detailed correlation studies.

\subsection{Time Projection Chamber}

\begin{figure}
\centering
\includegraphics[width=5in]{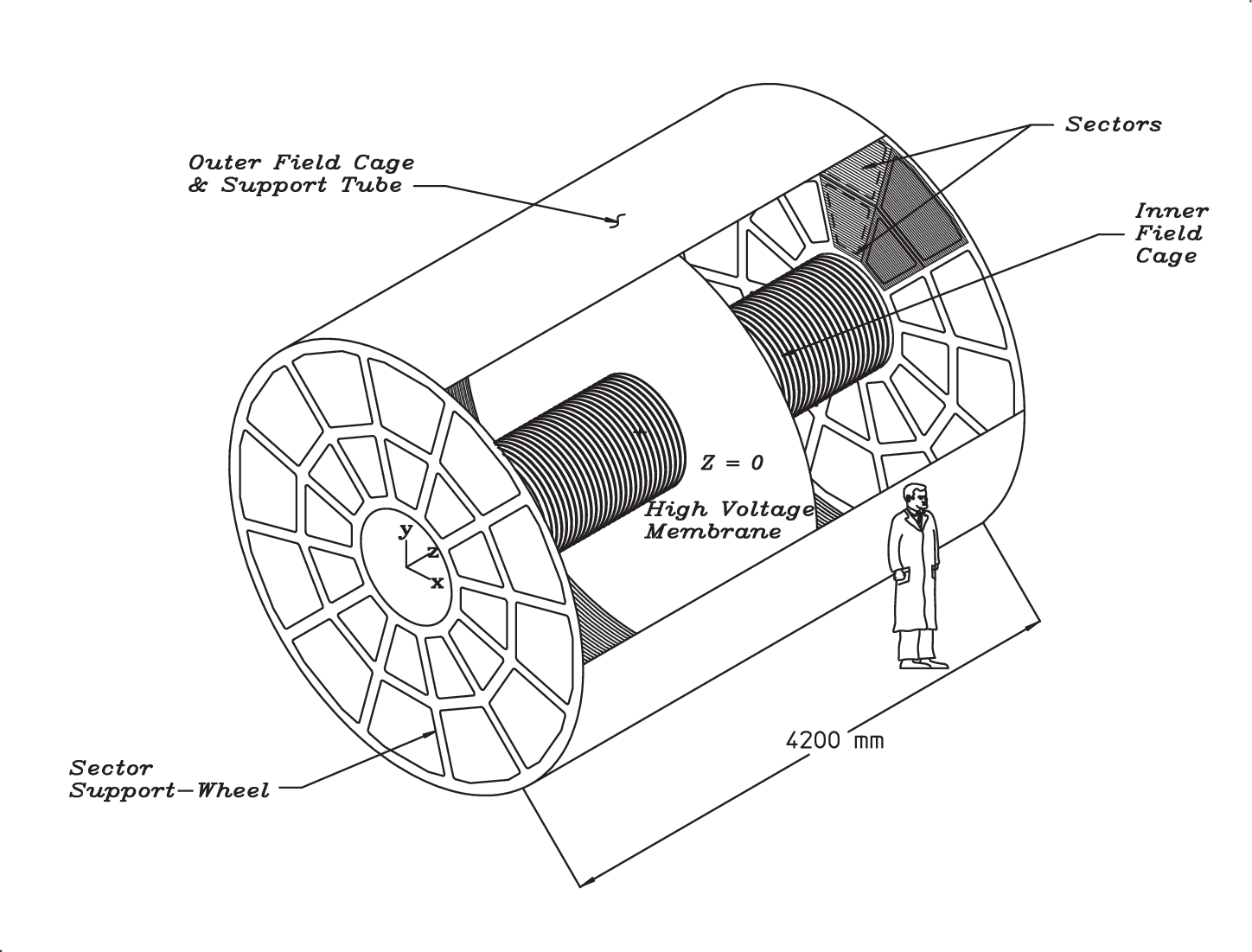} 
\caption[The STAR Time Projection Chamber].{A schematic diagram of the STAR Time Projection Chamber. \protect\cite{STARTPC}}
\label{TPCSchematic}
\end{figure}

The main detector at STAR is the Time Projection Chamber (TPC), seen in Figure \ref{TPCSchematic}. The TPC is a large cylinder, 4.2 meters in length by 2 meters radius. The inner radius of the TPC is 50 cm from the beam axis. Using a TPC, 3-dimensional reconstruction of charged particle tracks from a collision is possible. The TPC allows charged particle tracking, momentum determination, and identification from energy loss (dE/dx) over a range of $\pm$ 1.8 units in pseudorapidity ($\eta$, see Appendix A for definitions of kinematic variables). The momentum range of particles identifiable by the TPC extends from 0.1-1 GeV/c, while the total range of measured momentum extends from 0.1-30 GeV/c. The TPC gas volume is filled with P10, a mixture of 90\% argon (Ar) and 10\% methane ($CH_{4}$). A central cathode membrane at a potential of 28 kV, with the end caps at ground potential, supplies the drift electric field in the TPC.  Electrons liberated from gas atoms by the passage of charged particles through the TPC drift longitudinally toward the end caps, where the tracking information is read out. The drift time is approximately 40 $\mu$sec, corresponding to a drift velocity of $\approx$ 5.5 cm/$\mu$sec. The minimum two-track resolution of the TPC is 2 cm. The TPC (along with all STAR tracking detectors) lies inside the field generated by the STAR magnet. At its maximum, this uniform, longitudinal magnetic field has a magnitude of 0.5 T. 

\subsection{Forward Time Projection Chambers}

Two additional TPCs lie at forward rapidities covering the phase space $2.5 < |\eta| < 4.0$ over the full azimuth. These Forward Time Projection Chambers (FTPCs) are unlike the main TPC in that they utilize a radial drift electric field, compared to the longitudinal drift for the main TPC. This was a necessary result of the small extent of the FTPCs, which are only $\approx$ 1 m in length. The radial drift improves two-track resolution to $\approx$ 2 mm, an order of magnitude better than the main TPC. The drift electric field is created between a cylindrical inner cathode and the outer wall at ground potential. The electrons ionized by the passage of charged particles through the active gas volume drift to the outer radius of the FTPCs where their signals are read out by 9600 pads. The maximum radial drift distance is 22.32 cm. The FTPC gas volume consists of a 50-50\% mixture of Ar and carbon dioxide ($CO_2$). Further information regarding the work done for the FTPCs is explained in Appendix \ref{AppendixB}.

\begin{figure}
\centering
\includegraphics[width=5.5in]{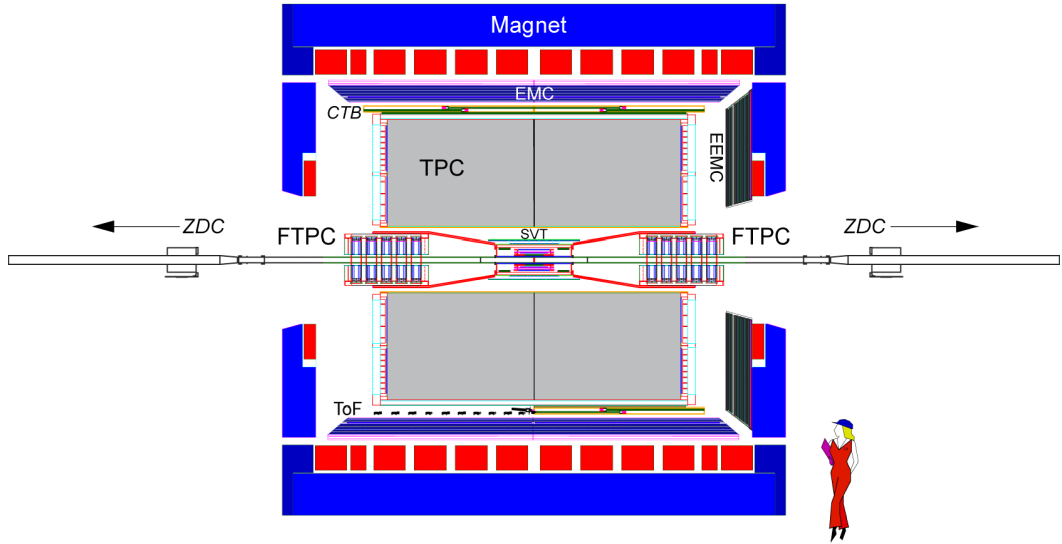} 
\caption{Cross section of the STAR detector and subsystems.}
\label{STAR_CrossSection}
\end{figure}

\subsection{Trigger and Data Acquisition}

The trigger detectors at STAR consist of the central trigger barrel (CTB), the zero-degree calorimeters (ZDCs), and the Beam-Beam Counters (BBCs). The CTB measures charged particles within $\eta = \pm$ 1 with full azimuthal coverage. The ZDCs are located up and downstream of the STAR detector, at the position of the RHIC DX magnets, the point where the incoming (outgoing) ion beams are brought together (split). The ZDCs measure spectator neutrons from the collision that propagate with the beam momentum.
The location of the primary collision vertex can be determined from time of flight differences in a coincidence signal in the two ZDCs. The BBCs provide an additional coincidence measure for triggering and are also used to monitor beam luminosity. The ZDCs and BBCs provide complementary vertex determination and are used in the majority of minimum bias triggers.
\\

The data acquisition system (DAQ) acquires data at rates up to 100 Hz. Due to the large multiplicities in heavy-ion collisions, raw data file sizes can be as large as 200 MB per event. At typical trigger rates, this is far too large for DAQ to handle. To reduce the file size, zero-suppression is applied for each detector. Events are written to tape storage at the RHIC Computing Facility (RCF) at a rate of 30-50 MB/s.
%
%
%

%
%

\chapter{Theoretical String Model}

\section{Color Strings/Chains}

Multiparticle production at high energies can be characterized by a string model (from a chromoelectric flux tube) via Schwinger pair production, which was first derived for $e^{+}e^{-}$ collisions \cite{Fluxtube1979}. It was later generalized to the case of hadron collisions, including heavy ion collisions \cite{Colorrope1984, Colortube1986}. Between a produced $q-\overline{q}$ pair is a constant, chromoelectric field that manifests as a tube (Figure \ref{QuarkFlux}). In this field additional $q-\overline{q}$ pairs can be generated. The creation of new $q-\overline{q}$ pairs screens the original field. This leads to pairing of new quarks with the original ones. The process can be repeated many times, leading to multiple hadron production of mesons and baryons.
\\

In heavy ion ($A+A$) collisions, it is expected the number density of these strings is much higher than in $e^{+}e^{-}$ or {\it pp} collisions. In {\it pp} or $p+A$ collisions, there may only be one string formed. In $A+A$ collisions there can be many strings. Immediately after the collision of two heavy ions a chromoelectric field is formed between the receding nuclei. These fields are confined in a small transverse area, forming color flux tubes (also referred to as ``strings'' or ``chains''). In central (full overlap) collisions, some strings may overlap with each other. 
%
%
%
\begin{figure}
\centering
\includegraphics[width=4in]{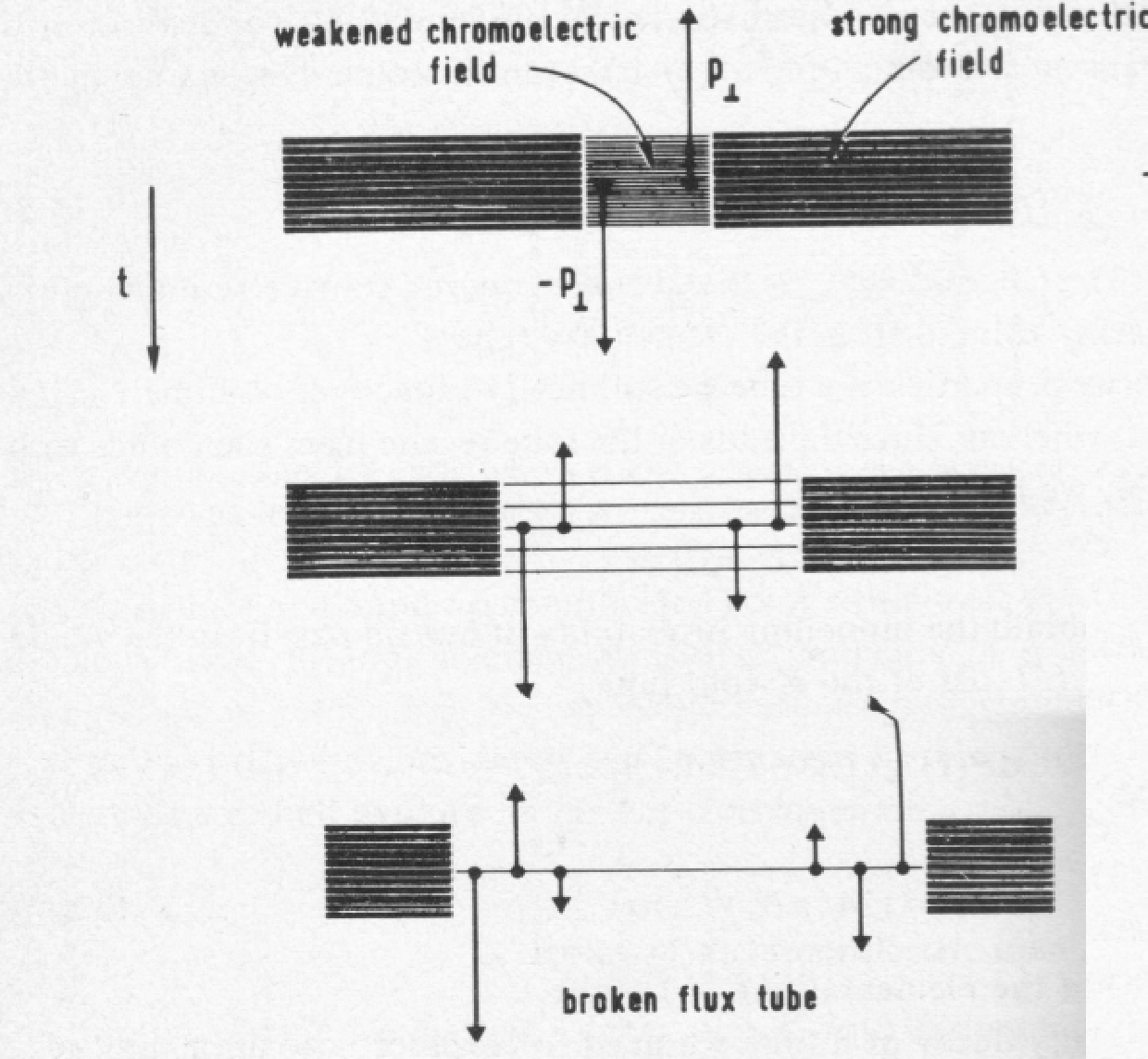} 
\caption[Particle emission from string decay.]{Particle emission from string decay. Each $q-\overline{q}$ pair weakens the color field \protect \cite{Colortube1986}.}
\label{StringDecay}
\end{figure}
%
%
%
%
%

\section{Dual Parton Model}

The Dual Parton Model (DPM) \cite{DPM} (and the similar Quark Gluon String Model \cite{QGSM}) is intended to describe soft hadronic physics in nucleus-nucleus, hadron-nucleus, and hadron-hadron collisions. Unlike hard processes, which can be treated perturbatively, soft physics occurs in the QCD regime of strong coupling. The DPM has been used to predict several observables of high-energy collisions including: multiplicity, pseudorapidity ($\eta$), and p$_{T}$ distributions; particle ratios; charge distributions; KNO scaling; heavy particle production; nuclear stopping; J/$\Psi$ suppression; strangeness enhancement; and long and short-range multiplicity correlations in pseudorapidity ($\eta$).
\\

\begin{figure}
\centering
\includegraphics[width=5in]{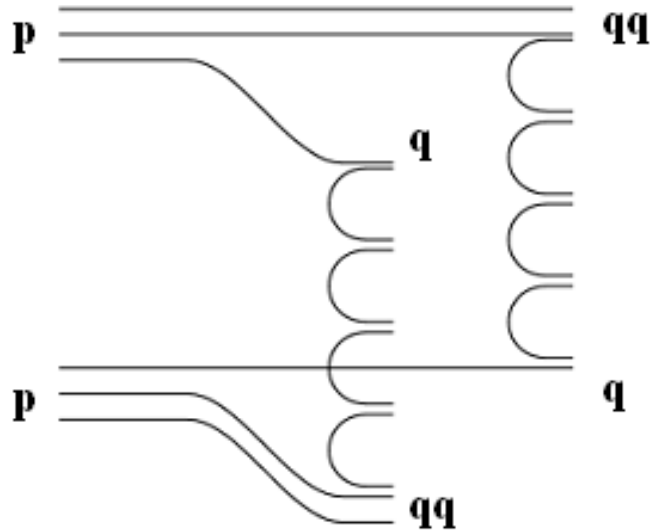} 
\caption[Dominant two string particle production mechanism in the DPM.]{Dominant two string particle production mechanism in the DPM for {\it pp} interactions at high energy \protect \cite{DPM2}.}
\label{2String}
\end{figure}

\begin{figure}
\centering
\includegraphics[width=5in]{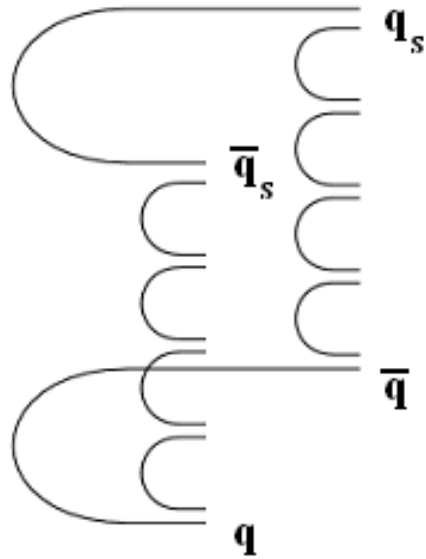} 
\caption[Four string particle production mechanism in the DPM.]{Four string particle production mechanism in the DPM for {\it pp} interactions\protect \cite{DPM2}. This includes the two string process (Figure \protect\ref{2String}) with the addition of two strings of the type $q-\overline{q}$ as shown.}
\label{4String}
\end{figure}

In {\it pp} collisions, the dominant production mechanism is a two string process. Figure \ref{2String} shows this process. The two strings are comprised of the valence quarks and therefore create strings with a quark and diquark at the ends. The strings are formed by soft gluon exchange between partons. A one string process dominates in cases where a q or $\overline{q}$ in the projectile can annihilate with its antiparticle in the target, such as in {\it p$\overline{p}$} interactions. Higher order contributions also exist. The second order contribution is from a four string process, shown in Figure \ref{4String}. The additional strings are formed between sea quarks and antiquarks. 

\subsection{Hadronization in the DPM}

Particle production in the DPM is due to the hadronization of chains (or strings) stretched between partons. Similar to the Schwinger particle production mechanism, (where at some threshold separation distance between a $q-\overline{q}$ pair, a new $q-\overline{q}$ pair is created from the vacuum), the chains in the DPM hadronize at some threshold to form $q-\overline{q}$ pairs. In high-energy particle collisions, the number of chains produced is equal to twice the number of inelastic nucleon-nucleon collisions. The main assumption in the particle production process put forth by the DPM is that the chains hadronize independently. Though the superposition of chains is possible, this independent hadronization scheme is the basis for all DPM predictions. In {\it pp}, the single-particle inclusive cross section is described by the superposition of two chains \cite{DPM}. For any one cut Pomeron, particles are produced with only short-range rapidity correlations. It is the superposition of several cut Pomerons (appropriately weighted) that contribute to the fluctuation in the number of strings and produces long-range rapidity correlations. The Pomeron was introduced as the particle exchanged to explain the increasing scattering cross-section in high energy experiments at energies greater than $\approx$ 20 GeV \cite{Pomeron_Gribov,Pomeron_Chew,Pomeron_GPT}. The exchange of other particles, such as mesons, leads to a predicted decrease in the cross-section as a function of energy, whereas Pomeron exchange predicts an increasing cross-section with energy. The Pomeron is predicted to consist of quarks and gluons combined in such a way that the particle carries no color charge. 

\subsection{Correlations in the DPM}

From the Dual Parton Model, rapidity correlations are described by the average multiplicity in a backward rapidity interval, $<N_{b}>$, as a function of total multiplicity in a forward rapidity interval, N$_{f}$. A linear expression relating these two was demonstrated in several high-energy experiments (\ref{linear}),

\begin{equation}\label{linear2} 
<N_{b}(N_{f})> = a + bN_{f} 
\end{equation}

where the slope, $b$, is the correlation strength and the intercept, $a$, is a measure of the uncorrelated particles. Both $b$ and $a$ are functions of the energy and atomic number of the colliding species. In terms of N$_{b}$ and N$_{f}$, $a$ and $b$ can be expressed as,

\begin{subequations}
\begin{equation}\label{a} a = \frac{<N_{b}><N_{f}^{2}>-<N_{f}N_{b}><N_{f}>}{<N_{f}^{2}>-<N_{f}>^{2}} \end{equation}
\begin{equation}\label{b} b = \frac{<N_{f}N_{b}>-<N_{f}><N_{b}>}{<N_{f}^{2}>-<N_{f}>^{2}} \end{equation}
\end{subequations}
Multiplicity fluctuations in high-energy collisions may have three sources: the fluctuation in the number of particles in each chain (with fixed ends), fluctuations in the position of the chain ends (and the subsequent invariant mass of the chain), or the fluctuation in the number of chains \cite{DPM}. 
\\

To emphasize long-range correlations, a gap can be introduced between the forward and backward rapidity windows, eliminating the midrapidity region from consideration. This has the effect of removing most of the short-range correlations, such that predominantly long-range correlations are present. Under the assumption that short-range correlations are confined to individual chains, these long-range correlations are due to the superposition of a fluctuating number of chains. This leads to the prediction that the short-range forward-backward correlation strength will decrease rapidly with increasing rapidity interval. 
\\

Though the DPM reproduces the results of a great deal of experimental data, the assumption of independent chains leads to the realization that even with a large density of non-interacting, overlapping strings, they would not provide a means to produce a thermalized system. It is predicted that even a small interaction between strings will lead to effects that precede the possible formation of a quark-gluon matter \cite{DPM}.

\subsection{Long-Range Correlations in the DPM}\label{LRCinDPM}

Long-range correlations span a pseudorapidity gap of $|\eta| > 1.0$. The physical mechanism that produces long-range correlations is the fluctuation in the number of elementary inelastic collisions, controlled by unitarity \cite{LRCbasis}. The requirement of unitarity conserves quantum mechanical probability by allowing multiple scattering to occur in the model.
%
%
\\

The forward-backward correlation in {\it pp} collisions can be expressed as \cite{CapellaLHC},

\begin{eqnarray}\label{SRCLRC}
<N_{f}N_{b}>-<N_{f}><N_{b}>  \:= \: <n>\left(<N_{0f}N_{0b}>-<N_{0f}><N_{0b}>\right) \nonumber\\
+ \left[\left(<n^{2}>-<n>^{2}\right)\right]<N_{0f}><N_{0b}>
\end{eqnarray}

The first term in Equation \ref{SRCLRC} corresponds to the correlation between particles from a single inelastic collision. These correlations have a short-range in rapidity. The second term details the contribution to the long-range correlation. The quantity $<n^{2}>-<n>^{2}$ is the fluctuation in the number of elementary inelastic collisions. Introducing a gap about midrapidity ($|\eta| > 1.0$) will substantially reduce the magnitude of the short-range term, which will eventually become negligible with increasing $\eta$ gap. The average charged particle multiplicity ($<N>$) and average number of inelastic collisions ($<n>$) is given by,

\begin{subequations}
\begin{equation}\label{meanmult} <N> = <n><N_{0}>
\end{equation}
\begin{equation}\label{meaninelas} <n> = \frac{\displaystyle \sum_{n=1}^{\infty}n\sigma_{n}}{\displaystyle \sum_{n=1}^{\infty}\sigma_{n}}
\end{equation}
\end{subequations}

where $N_{0}$ is the charged particle multiplicity per inelastic collision and $\sigma_{n}$ is the probability of {\it n} inelastic collisions. In more complicated systems (e.g., collisions of heavy nuclei), there are contributions from strings of two varieties, diquark-quark (qq-q) (from the first inelastic collision) and $q-\overline{q}$ (from subsequent inelastic collisions). Then, Equations \ref{meanmult} and \ref{SRCLRC} becomes (for large $\Delta\eta$) \cite{CapellaLHC},

\begin{equation}\label{meanmultgeneral}
<N> = \frac{1}{\displaystyle \sum_{n=1}^{\infty}\sigma_{n}}\left[\displaystyle \sum_{n=1}^{\infty}\sigma_{n}\left(2<N^{qq-q}>+(2n-2)<N^{q-\overline{q}}>\right)\right]
\end{equation}

\begin{equation}\label{LRCgeneral}
<N_{f}N_{b}>-<N_{f}><N_{b}>  \:= \: 4\left[\left(<n^{2}>-<n>^{2}\right)\right]<N^{q-\overline{q}}>_{f}<N^{q-\overline{q}}>_{b}>
\end{equation}

\section{Parton String Model}\label{PSM}

The introduction of string interaction into the DPM has been accomplished in the Parton String Model (PSM) \cite{PSM_AAPS, SFM_ABP, SFM_ABP2}. It is possible that at large string densities the assumption of string independence in the DPM may be too strong. If two strings overlap, in transverse area, their color fields will overlap, and they may fuse. This forms a single string with higher color at the ends. When the fused string hadronizes, one will see a reduction in the produced particle multiplicity due to fewer overall strings, and an increase in mean p$_{T}$, a result of energy-momentum conservation. This result has been seen experimentally in central high-energy heavy ion collisions. The produced particle yield is less than that expected from  N$_{bin}$ scaling of multiple independent pp collisions. $N_{bin}$ is the number of binary nucleon-nucleon collisions that occur when two nuclei collide. These results are also seen in parton saturation phenomena models, such as the color-glass condensate (CGC) \cite{CGC}. The PSM has been shown to be in good agreement with {\it pp} and heavy ion data at various energies and experiments including SPS at CERN and RHIC \cite{SFM_SPS_FPS, SFM_RHIC_APS}.
\\

There are two contributions to the multiplicity in heavy ion collisions. The first is proportional to N$_{part}$ (the number of nucleons participating in the collision), the second to N$_{bin}$ \cite{Ferreiro}. The experimental results from Au+Au collisions at RHIC are below the prediction of N$_{bin}$ scaling from {\it pp} to Au+Au. Fusion of strings can account for this multiplicity reduction, which is on the order of $\approx$ 30\% in central collisions. The PSM also predicts the average transverse momentum ($<p_{T}>$) enhancement seen in Au+Au collisions at RHIC. This is the basic result of overlapping strings conserving energy and momentum (fewer particles, but with higher $<p_{T}>$ per particle). There is also good agreement with the inclusive p$_{T}$ spectra and the pseudorapidity density.
\\

A predicted consequence of string fusion is the presence of long and short-range rapidity correlations. Equation \ref{b} for $b$ can be expressed in terms of the forward-backward dispersion squared normalized by the forward (or backward) dispersion squared,

\begin{equation}\label{dispersion} b = \frac{D_{bf}^{2}}{D_{ff}^{2}} 
\end{equation}

It is predicted that if one studies the long-range correlations by introducing a rapidity gap, at large values of N$_{f}$, such as those in central heavy ion collisions, $b$ will deviate from the expected linear increase \cite{DPM}. If there is soft (p$_{T} <$ 2 GeV) string fusion, the slope value is expected to decrease even further. Since the number of strings increases with energy, atomic number, and centrality, one expects the effects of fusion to grow in these cases. This Monte Carlo model (PSM) only considers the fusion of two strings close enough in transverse space, and does not fuse hard (high-$p_{T}$) strings at all. The model also incorporates a basic rescattering of produced particles and decay of resonances.
%

\section{Color Glass Condensate/Glasma}

\begin{figure}
\centering
\subfigure[Two Lorentz contracted, CGC nuclei incident on one another. The initial chromoelectric and magnetic fields are oriented transverse to the direction of travel.]{
\includegraphics[width=4.5in]{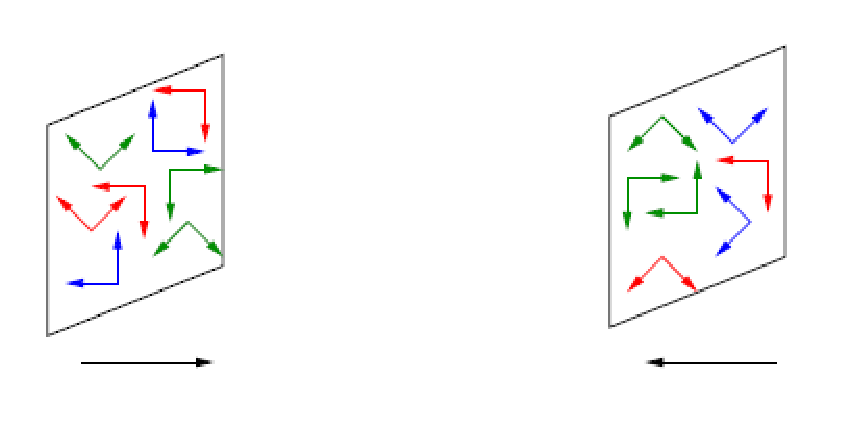} 
\label{Glasma1}
}
\subfigure[Following the collision, both nuclei recede from one another. Additional field sources are induced in each nucleus, creating longitudinal color electric and magnetic fields between the nuclei. These fields are similar to the strings in the DPM.]{
\includegraphics[width=4.5in]{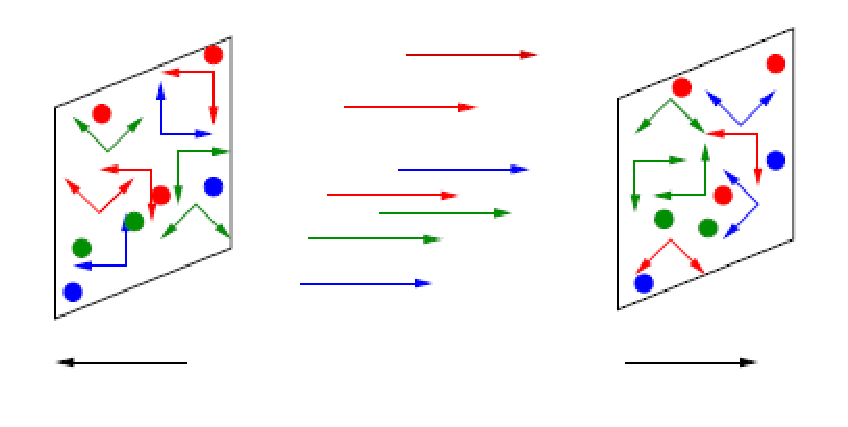} 
\label{Glasma2}
}
\caption[Schematic of the CGC/Glasma.]{Schematic of the CGC/Glasma \protect \cite{LRCCGC}.}
\label{Glasma}
\end{figure}

The Color Glass Condensate (CGC) model is a description of high density gluonic matter at small Bjorken x (see Appendix \ref{AppendixA}) \cite{CGC2}. In the CGC model, the incoming, Lorentz contracted nuclei, are envisioned as two sheets of colliding colored glass. In the initial stage of the collision, the color electric and magnetic fields are oriented transverse to the beam direction (and each other). After the collision there are additional sources formed that produce longitudinal color electric and magnetic fields. This state is referred to as the Glasma, an intermediate phase between the CGC and Quark Gluon Plasma (QGP) \cite{CGCGlasma}. This is shown schematically in Figure \ref{Glasma} \cite{LRCCGC}. The longitudinal fields provide the origin of the long-range correlation and are similar to the strings as formulated in the DPM.

%

\begin{figure}
\centering
\includegraphics[width=6in]{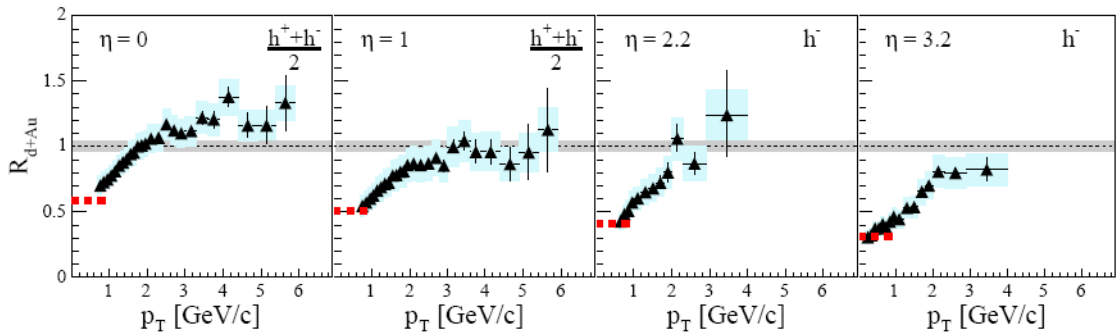} 
\caption[$R_{d+Au}$ measured by the BRAHMS experiment.]{The nuclear modification factor ($R_{d+Au}$) measured by the BRAHMS experiment as a function of pseudorapidity ($\eta$) in 10\% most central $d+Au$ collisions at an energy of $\sqrt{s_{NN}}$ = 200 GeV. The high-$p_{T}$ suppression at forward pseudorapidity has been suggested as a signal for the Color Glass Condensate (CGC) \protect\cite{Brahmswhitepaper}.}
\label{BrahmsCGC}
\end{figure}

Experimental evidence from the BRAHMS collaboration suggests that the CGC is a valid description of the initial state of colliding nuclei in $d+Au$ collisions \cite{Brahmswhitepaper}. BRAHMS observes a suppression of the nuclear modification factor $R_{d+Au}$ (Equation \ref{RAA}) at high $p_{T}$ and forward rapidity, in 0-10\% most central $d+Au$ events at an energy of $\sqrt{s_{NN}}$ = 200 GeV. This is shown in Figure \ref{BrahmsCGC}. 
\\

At midrapidity, for $d+Au$ collisions at an energy of $\sqrt{s_{NN}}$ = 200 GeV, there is no suppression of $R_{d+Au}$, whereas a large suppression is seen in Au+Au at the same energy both for mid- and forward rapidities. This indicates that the suppression in Au+Au is not due to an initial state effect. The suppression of $R_{d+Au}$ at forward rapidity should be related to the initial conditions of the colliding nuclei, as there is no contamination of the final state from a produced quark-gluon medium. 

\chapter{Experimental Analysis}\label{ExpAnalysis}

\section{Data Analyzed}

\begin{table}
\begin{center}
\begin{tabular}{|*{2}{c|}}
\hline
$\sqrt{s_{NN}}$ (GeV) & System    \\
\hline
       400 &         pp \\
\hline
       200 & Au+Au, Cu+Cu, pp \\
\hline
      62.4 & Au+Au, Cu+Cu, pp \\
\hline
      22.4 &      Cu+Cu \\
\hline
\end{tabular}
\caption[Summary of colliding systems and energies at RHIC.]{A summary of the available data for the various colliding systems and energies at RHIC.} 
\label{DatasetTable}
\end{center}
\end{table}

\begin{figure}
\centering
\includegraphics[width=5in]{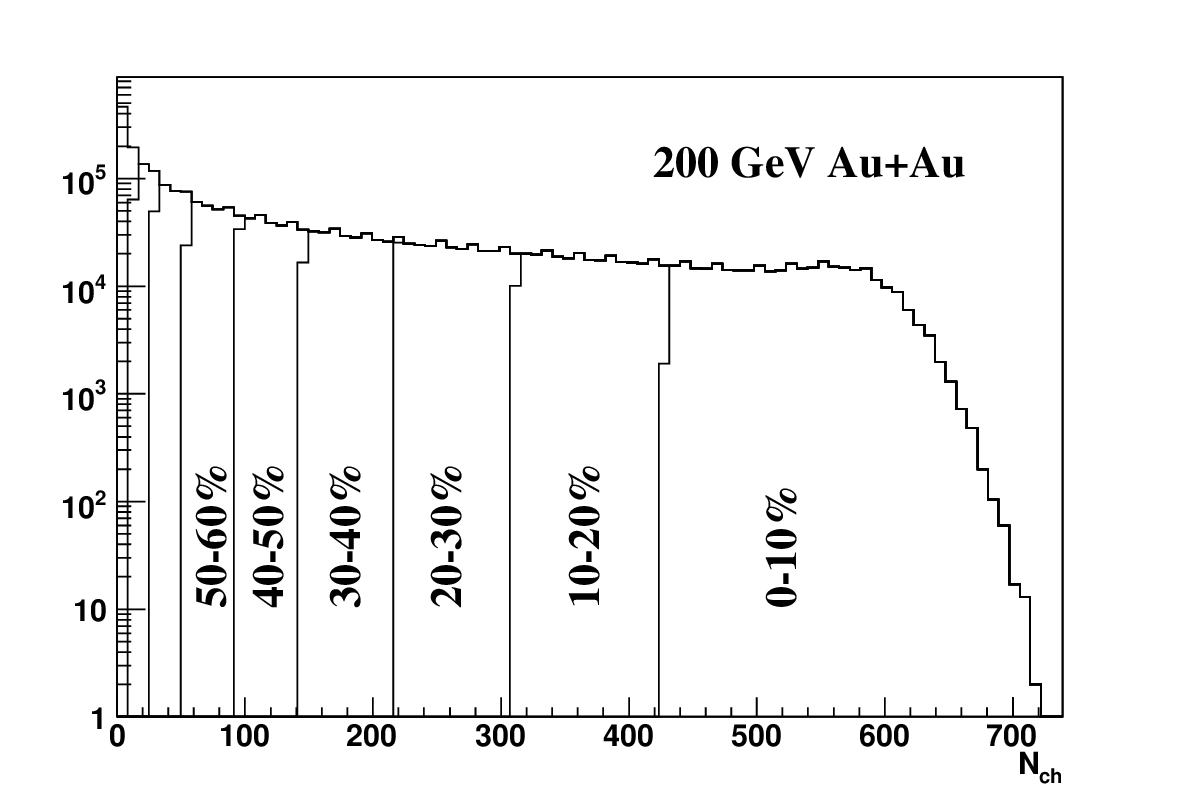} 
\caption[Raw charged particle multiplicity in $\sqrt{s_{NN}}$ = 200 GeV Au+Au.]{Raw charged particle multiplicity ($N_{ch}$) in $\sqrt{s_{NN}}$ = 200 GeV Au+Au showing divisions by centrality.}
\label{refmult}
\end{figure}

Events from the data samples in Table \ref{DatasetTable} were analyzed and the requisite quantities extracted from the files. This included event-by-event total charged particle multiplicities (for centrality determination), charged particle multiplicities in each forward and backward measurement interval, and z-vertex position. Every event analyzed was from the minimum bias triggered data sample. For heavy nucleus collisions these were subdivided into centrality bins. Centrality is a characterization of heavy ion events into categories based on the measured charged particle multiplicity (reference multiplicity). The criteria for determining the reference multiplicity is as follows: charged tracks from the primary vertex, fit points in the TPC $\geq$ 10, within the pseudorapidity region $-0.5 < \eta < 0.5$, and with a distance of closest approach (dca) of less than 3 cm to the primary vertex. Each centrality bin spans 10\% of the total multiplicity; from most central to most peripheral: 0-10, 10-20, 20-30, 30-40, 40-50, 50-60, 60-70, and 70-80\% of the total hadronic cross section. These are the centralities used for Au+Au data, while up to 50-60\% was considered for Cu+Cu collisions. The STAR reference multiplicity distribution at 200 GeV/A in Au+Au can be seen in Figure \ref{refmult}.

Several quality cuts were implemented during the data processing phase. These include some that are also used in the reference multiplicity determination, as mentioned above. The cuts include the number of fit points in the TPC (minimum of 10) and dca $<$ 3 cm. 

\begin{table}
\begin{center}
\begin{tabular}{|*{4}{c|}}
\hline
Centrality &   $N_{ch}$ &      $N_{part}$ &       $N_{bin}$ \\
\hline
    0-10\% & $N_{ch}\geq 431$           &  325.9 +5.4 -5.3 &  940.0 +66.9 -69.5 \\
\hline
   10-20\% & $312\leq N_{ch} \leq430$ &  234.6 +8.3 -9.3 &  591.3 +51.9 -59.9 \\
\hline
   20-30\% & $217\leq N_{ch} \leq311$ &  166.7 +9.0 -10.6 &  368.6 +41.1 -50.6 \\
\hline
   30-40\% & $146\leq N_{ch} \leq216$ & 115.5 +8.7 -11.2 &  220.2 +30.0 -38.3 \\
\hline
   40-50\% & $94\leq N_{ch} \leq145$ &  76.6 +8.5 -10.4 &  123.4 +22.7 -27.3 \\
\hline
   50-60\% &  $56\leq N_{ch} \leq93$ &  47.8 +7.6 -9.5 &  63.9 +14.1 -18.9 \\
\hline
   60-70\% &  $30\leq N_{ch} \leq55$ &  27.4 +5.5 -7.5 &  29.5 +8.2 -11.3 \\
\hline
   70-80\% &  $14\leq N_{ch} \leq29$ &  14.1 +3.6 -5.0 &  12.3 +4.4 -5.2 \\
\hline
\end{tabular}
\caption[Centrality definition for Au+Au at $\sqrt{s_{NN}}$ = 200 GeV.]{Centrality definition in terms of the number of charged particles for Au+Au at $\sqrt{s_{NN}}$ = 200 GeV. The number of nucleons participating in the collision ($N_{part}$) and the number of binary nucleon-nucleon collisions ($N_{bin}$) estimated from Monte Carlo Glauber simulations \protect\cite{STARGlauber} are also shown.}
\label{2001AuCentrality}
\end{center}
\end{table}

\section{Calculating the Forward-Backward Correlation Strength}

The forward-backward (FB) correlation strength {\it b} is calculated for all charged particles within the STAR TPC acceptance of $|\eta| <$ 1 and with transverse momentum $p_{T} >$ 0.15 GeV/c. The pseudorapidity range is subdivided into forward and backward measurement intervals of width 0.2 $\eta$. The FB correlation strength {\it b} was measured in forward and backward intervals that are symmetric about midrapidity ($\eta$ = 0). The separation between the intervals was measured from the center of each bin and included $\Delta\eta$ = 0.2, 0.4, 0.6, 0.8, 1.0, 1.2, 1.4, 1.6, and 1.8 units in pseudorapidity. The bins at $\Delta\eta$ = 0.2 are contiguous in $\eta$ space at $\eta$ = 0. This is the only bin for which there is no physical gap in $\eta$ space. 

\begin{figure}
\centering
\includegraphics[width=5.5in]{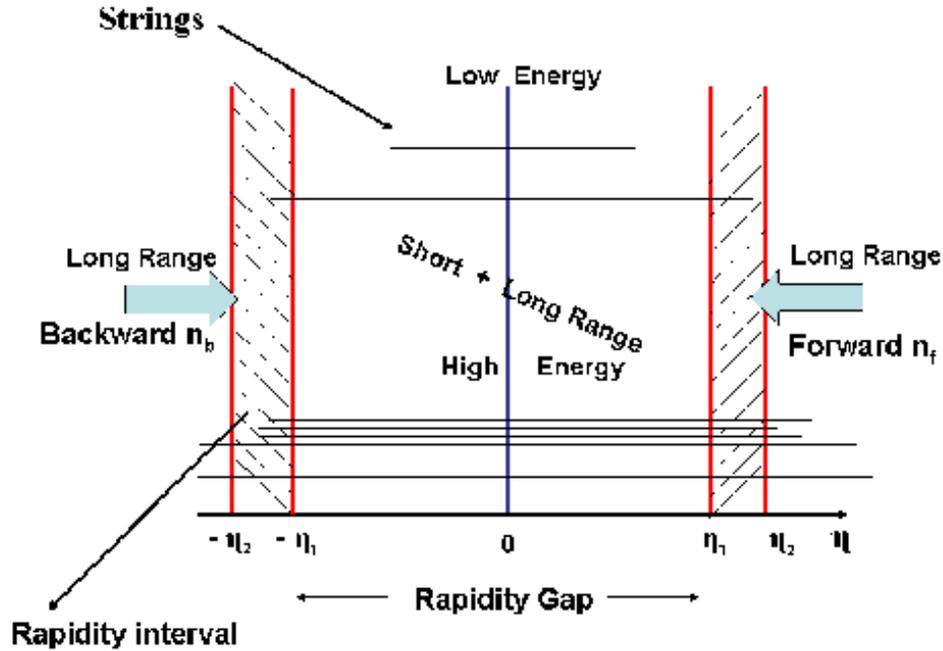}
\caption[Schematic diagram of the measurement of a forward-backward correlation.]{\footnotesize{Schematic diagram of the measurement of a forward-backward correlation.}}
\label{FB_Schematic}
\end{figure}

The coordinate system defined above differs from that of the standard two particle pseudorapidity correlation \cite{LRCpp3}. The FB correlation strength is measured in a coordinate system where $\eta$ = 0 is always physically located at midrapidity (Figure \ref{FB_Schematic}). The collision vertex (z = 0) defines this point. Therefore, all $\Delta\eta$ values are measured in an absolute (fixed) coordinate system about $\eta$ = 0. In the usual two particle correlation measurement, the relative $\eta$ difference between the particles is considered. This has particular implications for the definitions of the measurement interval and how the centrality determination is made.

\begin{figure}
\centering
\subfigure[]{
\includegraphics[width=2.85in]{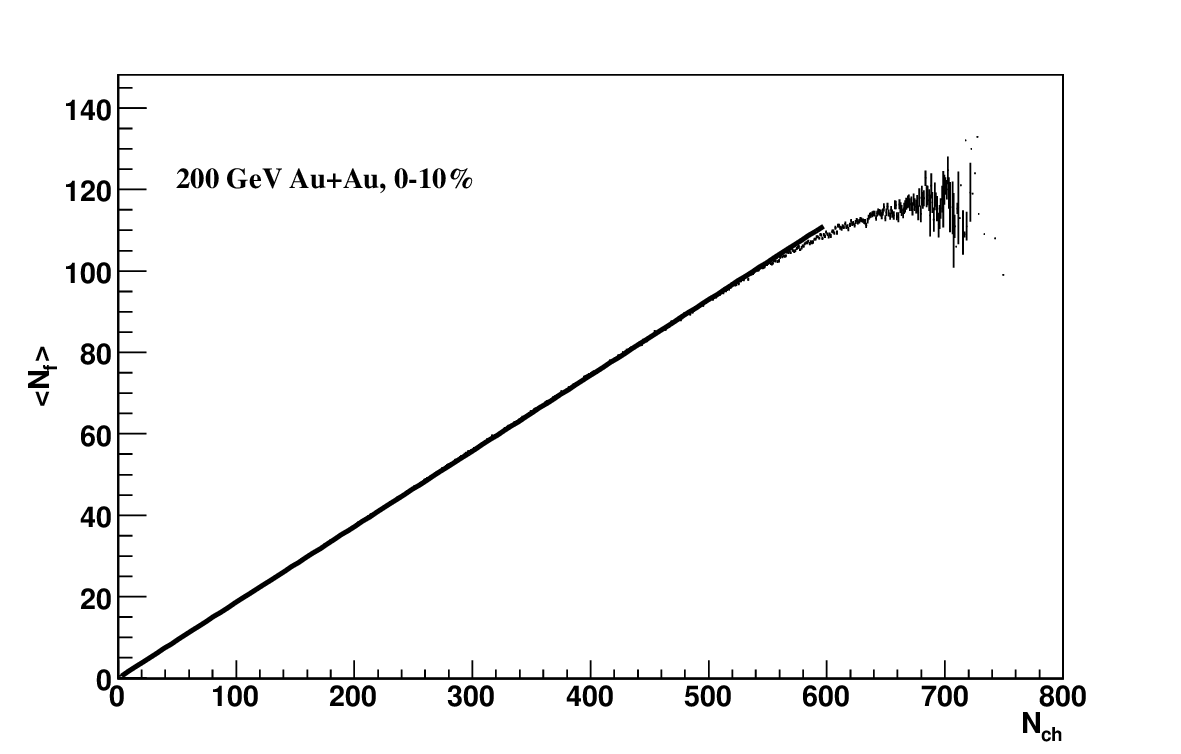} 
\label{Nf}
}
\subfigure[]{
\includegraphics[width=2.85in]{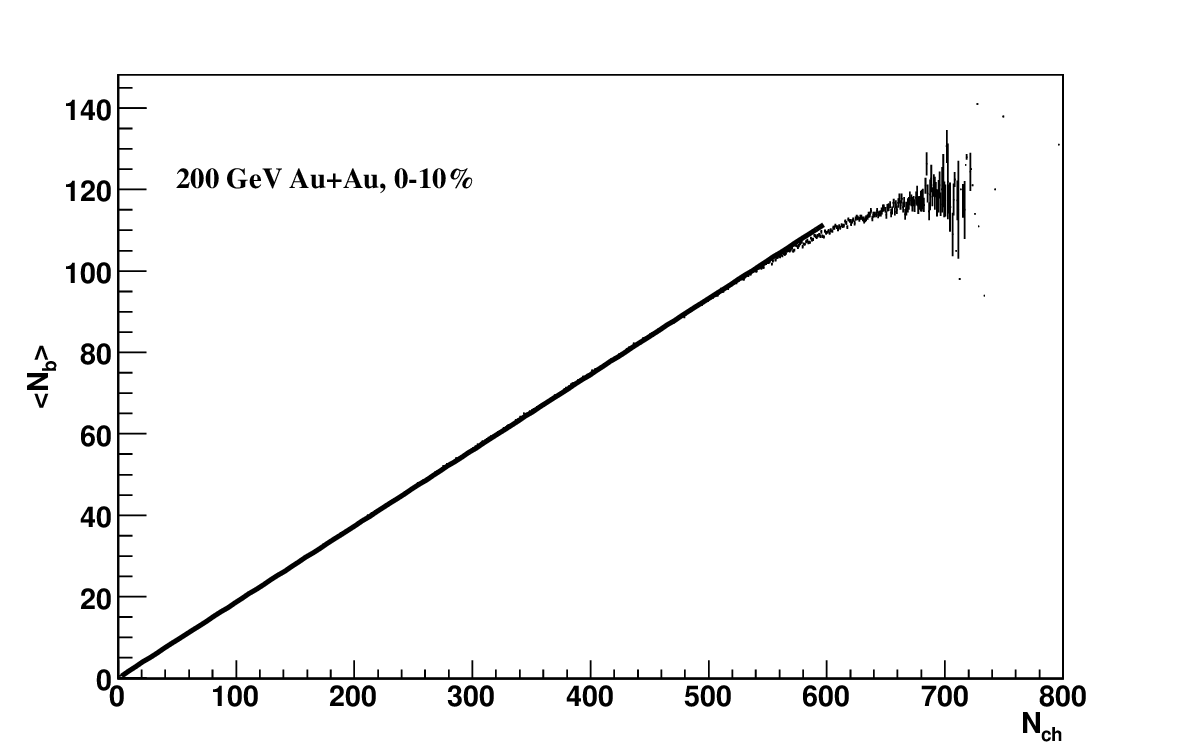} 
\label{Nb}
}
\subfigure[]{
\includegraphics[width=2.85in]{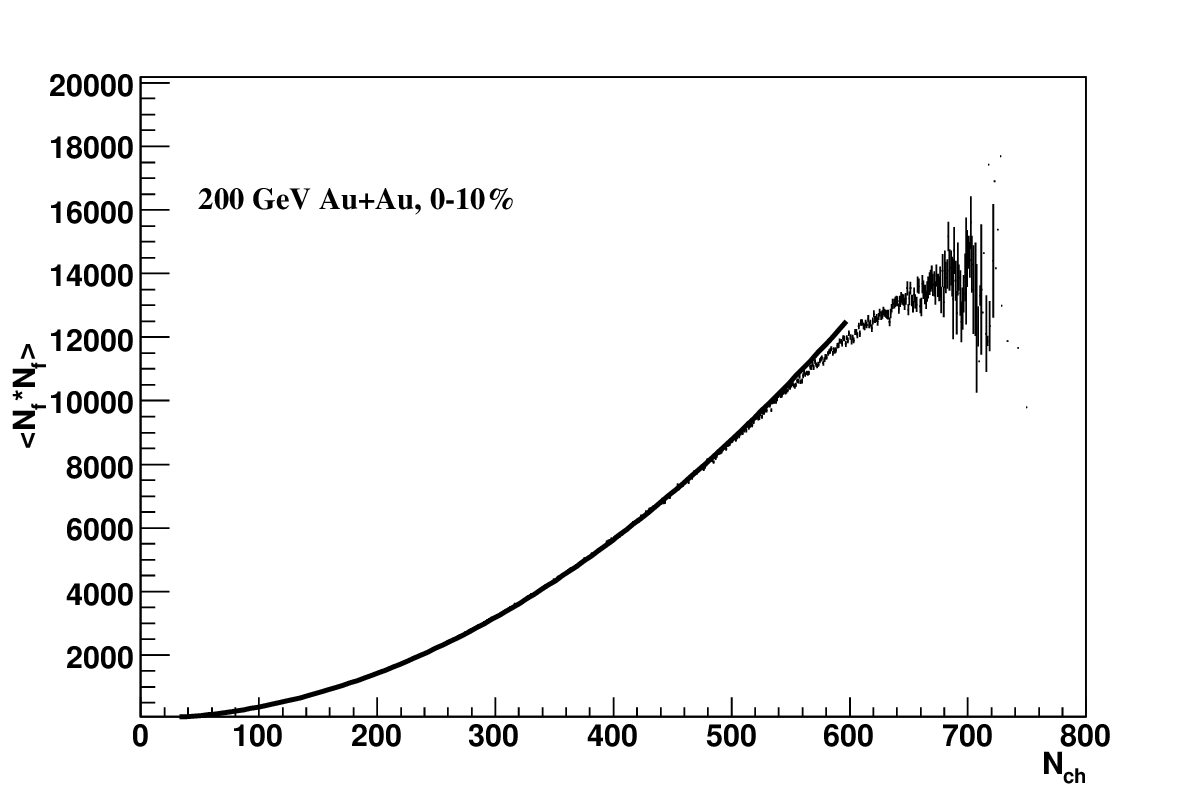} 
\label{NfNf}
}
\subfigure[]{
\includegraphics[width=2.85in]{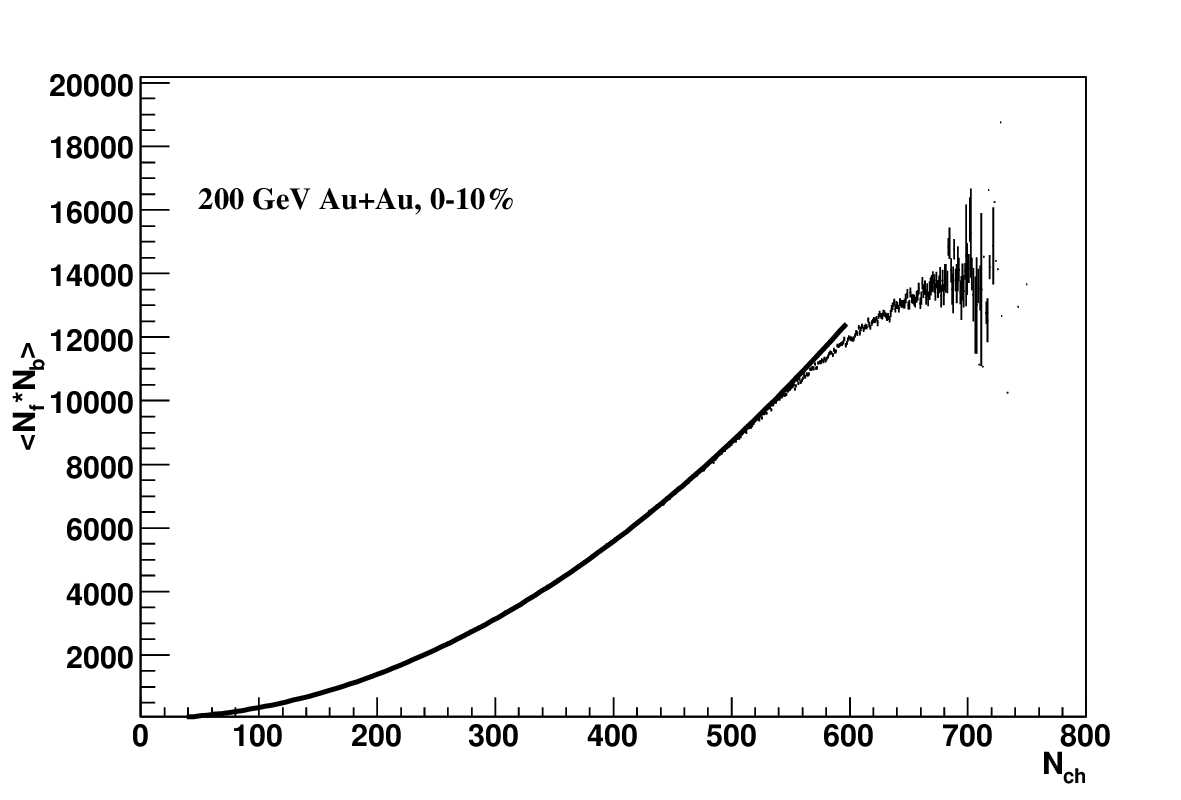} 
\label{NfNb}
}
\caption[Measured multiplicities versus total charged particle multiplicity.]{(a) Mean forward charged particle multiplicity ($<N_{f}>$) and (b) Mean backward charged particle multiplicity ($<N_{b}>$) versus total charged particle multiplicity fitted with a linear polynomial up to $N_{ch}$ = 0-600. (c) $<N_{f}*N_{f}>$) and (d) $<N_{f}*N_{b}>$) versus total charged particle multiplicity fitted with a second order polynomial up to $N_{ch}$ = 0-600.}
\label{Profiles}
\end{figure}

A strong bias exists if the measurement of $N_{ch}$ overlaps with the measurement interval in $\eta$ space. Therefore, there are three $\eta$ regions used to determine reference multiplicity: 

\begin{itemize}
\item The FB strength calculated for $\Delta\eta > 1.0$ used the charged particle multiplicity ($N_{ch}$) measured in the STAR TPC $\eta$ range from $|\eta| < 0.5$.
\item The FB strength calculated in the interval $\Delta\eta < 1.0$ used $N_{ch}$ measured in the STAR TPC from $0.5 < |\eta| < 1.0$.
\item The FB strength calculated for $\Delta\eta = 1.0$ was complicated by the fact that the measurement window overlapped both of the previously utilized $N_{ch}$ measurements. Therefore, this particular data point used the charged particle multiplicity measured in the STAR TPC from $|\eta| < 0.3 + 0.6 < |\eta| < 0.8$. 
\end{itemize}

\begin{figure}
\centering
\includegraphics[width=5in]{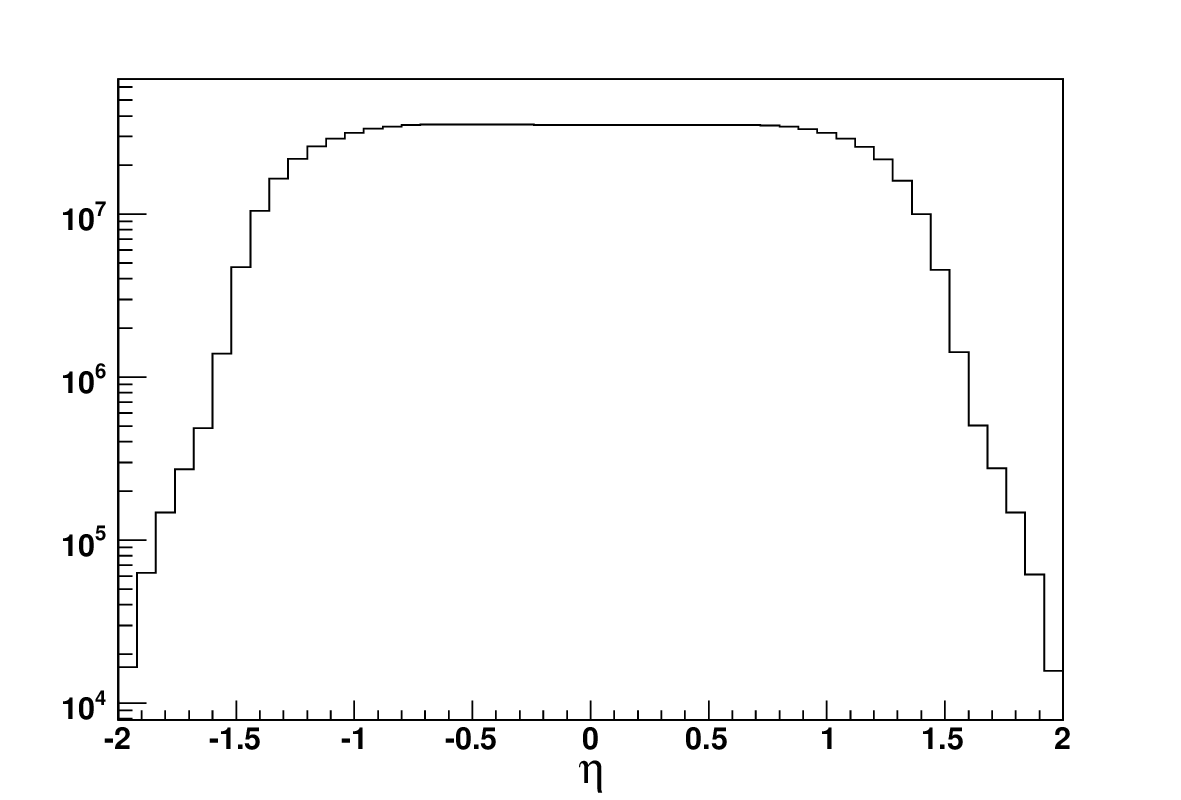}
\caption[Pseudorapidity ($\eta$) distribution of charged particles in Au+Au.]{The pseudorapidity ($\eta$) distribution of charged particles in minimum bias Au+Au data at $\sqrt{s_{NN}}$ = 200 GeV.}
\label{TPC_eta}
\end{figure}

This is possible due to the flat $\eta$ distribution of the STAR TPC within the measurement ranges being considered. The $\eta$ distribution for charged particles in the TPC for Au+Au collisions at $\sqrt{s_{NN}}$ = 200 GeV is shown in Figure \ref{TPC_eta}.

For this analysis, the forward (backward) interval is taken as positive (negative) $\eta$. In the case of symmetric collisions, these definitions are purely ancillary. 
\\

In order to eliminate the effect of statistical impact parameter (centrality) fluctuations on the measurement of the FB correlation strength, each relevant quantity (the uncorrected mean quantities $\left<N_{f}\right>_{uncorr}$, $\left<N_{b}\right>_{uncorr}$, $\left<N_{f}\right>^{2}_{uncorr}$, and $\left<N_{f}N_{b}\right>_{uncorr}$) was obtained on an event-by-event basis as a function of STAR reference multiplicity, $N_{ch}$ (Figure \ref{Profiles}). A linear fit to $\left<N_{f}\right>_{uncorr}$ and $\left<N_{b}\right>_{uncorr}$, or a second order polynomial fit to $\left<N_{f}\right>^{2}_{uncorr}$ and $\left<N_{f}N_{b}\right>_{uncorr}$, was used to extract the average of these quantities as functions of $N_{ch}$. Tracking efficiency and acceptance corrections were then applied to each event (Section \ref{EffCalc}). Due to statistical limitations, it is not possible to apply corrections for every value of $N_{ch}$. One value of the correction, calculated for each centrality bin, is applied to $\left<N_{f}\right>_{uncorr}, \left<N_{b}\right>_{uncorr}, \left<N_{f}\right>^{2}_{uncorr}$, and $\left<N_{f}N_{b}\right>_{uncorr}$ for every event in that centrality bin. Therefore, all events falling within a particular centrality have the same efficiency correction. The corrected values of $\left<N_{f}\right>$, $\left<N_{b}\right>$, $\left<N_{f}\right>^{2}$, and $\left<N_{f}N_{b}\right>$ were then used to calculate the backward-forward and forward-forward dispersions, $D_{bf}^{2}$ and $D_{ff}^{2}$, binned according to the STAR centrality definitions in Table \ref{2001AuCentrality} and normalized by the total number of events in each bin. 

\subsection{Efficiency Correction}\label{EffCalc}

Because the detectors do not operate perfectly (the tracker may miss particles, especially in a central heavy ion events, the tracks may be split or merged, etc.) and do not cover a complete volume (e.g., due to gaps between sectors), a correction for tracking efficiency and geometrical detector acceptance must be computed. This efficiency correction is defined as,

\begin{equation}
\epsilon = \frac{N_{rec}}{N_{simu}}
\end{equation}

where $N_{rec}$ is the number of reconstructed charged particles and $N_{simu}$ is the total number of simulated charged particles. When similar cuts are applied to the simulated and real tracks, the acceptance effect is implicitly included. The simulated particles are produced with the heavy ion event generator, HIJING \cite{HIJING}. These simulated particles are propagated through a virtual representation of the STAR detector constructed in the GEANT simulation framework \cite{GEANT}. The simulated particles interact with the virtual detector and its constituent materials. Some particles are lost to conversions in the material or other mechanisms, while the rest can be found by the reconstruction algorithm. The efficiency is a function of collision centrality, particle $p_{T}$, and pseudorapidity. The TPC is least efficient in high density central collisions ($\approx$ 80\%), but improves to over 90\% in peripheral events. A sample of the correction values is shown in Table \ref{CentralEffTable} for 0-10\% $\sqrt{s_{NN}}$ = 200 GeV Au+Au collisions. In central collisions at this energy the single particle tracking efficiency is $\approx$ 80\% while the two particle efficiency drops to $\approx$ 65\%. 

\begin{table}
\begin{center}
\begin{tabular}{|*{5}{c|}}
\hline
$\Delta\eta$ & $\epsilon_{N_f}$ & $\epsilon_{N_b}$ & $\epsilon_{N_f*N_b}$ & $\epsilon_{N_f*N_f}$ \\
\hline
       0.2 &      0.816 &      0.804 &      0.651 &      0.662 \\
\hline
       0.4 &      0.825 &      0.804 &      0.659 &      0.678 \\
\hline
       0.6 &      0.833 &      0.809 &      0.670 &      0.691 \\
\hline
       0.8 &      0.844 &      0.821 &      0.689 &      0.709 \\
\hline
       1.0 &      0.854 &      0.834 &      0.708 &      0.726 \\
\hline
       1.2 &      0.863 &      0.844 &      0.725 &      0.743 \\
\hline
       1.4 &      0.873 &      0.855 &      0.744 &      0.761 \\
\hline
       1.6 &      0.873 &      0.853 &      0.742 &      0.761 \\
\hline
       1.8 &      0.847 &      0.821 &      0.691 &      0.716 \\
\hline
\end{tabular}
\end{center}
\caption[Representative efficiency corrections for central Au+Au data.]{Efficiency corrections as a function of $\Delta\eta$ for Au+Au, $\sqrt{s_{NN}}$ = 200 GeV, 0-10\% centrality, $|v_{z}| < 30$ cm, and fit points $>$ 10.} 
\label{CentralEffTable}
\end{table}

\subsubsection{Forward-Backward Multiplicity Correlation Efficiency}

To accurately determine the forward-backward correlation strength ({\it b}) both single particle and two particle tracking efficiency must be determined. From the definition of the correlation strength Equation \ref{b},

\begin{eqnarray}
b = \frac{<N_{f}N_{b}>-<N_{f}><N_{b}>}{<N_{f}^{2}>-<N_{f}>^{2}} \nonumber
\end{eqnarray}

it is seen that the single particle efficiency is necessary to correct $N_{f}$ and $N_{b}$. The efficiency for finding two particles in the forward pseudorapidity hemisphere ($\epsilon_{N_f*N_f}$) and one particle in the forward and one in the backward hemispheres ($\epsilon_{N_f*N_b}$), is also required. For each of the four quantities, a correction factor is obtained as discussed in Section \ref{EffCalc} for each centrality, pseudorapidity interval, z-vertex cut, and number of fit points on the track. This correction was applied to $N_{f}$, $N_{b}$, $N_{f}*N_{f}$, and $N_{f}*N_{b}$ in every event.

\subsection{Error Studies}\label{ErrorStudy}

Statistical and systematic uncertainties have been studied. The error in the measurement of the forward-backward (FB) correlation strength is predominantly due to systematic effects.

\subsubsection{Statistical Errors}\label{StatErrors}

A large sampling of minimum bias data is available for analysis. For central Au+Au data, the statistical error is less than 1\%, while for peripheral it is less than 3\%. This and other contributions from uncertainties in the efficiency estimates are taken into account by including an additional 20\% in the overall error estimate for the FB correlation strength.

\subsubsection{Systematic Errors}\label{SysErrors}

Three main sources of systematic error were considered: the z-vertex position with respect to the center of the TPC (3), the fit range (in $N_{ch}$) of the polynomial functions to the raw data (3), and the number of TPC fit points on each track (2). This provides $3\times3\times2$ = 18 different combinations that can be used to calculate systematic deviations from the mean. The total error on the FB correlation strength {\it b} is calculated from the errors on $D_{bf}^{2}$ and $D_{ff}^{2}$ assuming they are uncorrelated, but that the error propagated to {\it b} is correlated. This is accomplished by calculating the standard deviations of $D_{bf}^{2}$ and $D_{ff}^{2}$ from the 18 separate values. When the final calculation of {\it b} is made, these errors are treated as correlated since there is overlap in the measurements of $D_{bf}^{2}$ and $D_{ff}^{2}$ on an event-by-event basis using the same $\eta$ window in the forward direction. This is shown in Equation \ref{errors}, the formula used for calculating the correlated errors on {\it b}.

\begin{equation}\label{errors}
\sigma_{b} = b\left(\left(\frac{\sigma_{D_{ff}^2}}{D_{ff}^{2}}\right)^{2}+\left(\frac{\sigma_{D_{bf}^2}}{D_{bf}^{2}}\right)^{2}-\frac{2\sigma_{D_{ff}^2}\sigma_{D_{bf}^2}}{D_{ff}^{2}D_{bf}^{2}}\right)^{1/2}
\end{equation}
\\

There is a strong interplay between the three main sources of systematic error as a function of $\Delta\eta$. This is predominantly due to the physical limitations of the STAR TPC and the tracking algorithm. This has a large effect near the center of the TPC and at the outer edges of the acceptance. At small values of $\Delta\eta$, close to the center of the TPC, one track may be split into two if it crosses the central membrane of the TPC. At large values of $\Delta\eta$ there are tracks that may fall outside the acceptance if the z-vertex position is substantially shifted from z = 0. The effect on the errors is seen as follows (in central Au+Au data):

\begin{enumerate}
\item Changes in the z-vertex position from $\pm$ 10 to $\pm$ 30 lower the value of {\it b} at large $\Delta\eta$ by $\approx$ 15\% when the number of fit points on track is 15, but only by $\approx$ 6\% for tracks with 10 fit points. There is a loss of tracks when the z-vertex is highly shifted from the center of the TPC and longer tracks (more fit points) are required. At small $\Delta\eta$ for both fit point cuts as a function of $v_{z}$, the change is less than a few percent. 
\item Changing the fit range affects the 0-10\% centrality bin more than the peripheral bins. Figure \ref{Profiles} shows that for large values of $N_{ch}$ the fit begins to deviate from the data. 
\item Adjusting the number of fit points on the track predominantly affects the FB correlation strength at small and large values of $\Delta\eta$. This is related both to the position of the z-vertex (for small $\Delta\eta$ and due to track loss at the outer edge of the measurement ($\Delta\eta$ = 1.0)).
\end{enumerate}


\subsubsection{Average Value}

The final value of the correlation strength for each $\Delta\eta$ position and centrality is the arithmetic mean of the 18 values, determined by varying the cuts as discussed in Section \ref{SysErrors}.

\chapter{Experimental Results and Discussion}

The following sections present the results for the forward-backward (FB) correlation strength {\it b}, calculated as discussed in Chapter \ref{ExpAnalysis}. All plots of $D_{bf}^{2}$, $D_{ff}^{2}$, and {\it b} are corrected for efficiency and acceptance (Section \ref{EffCalc}) and shown with the calculated statistical and systematic errors (Section \ref{ErrorStudy}). 

\section{FB Correlation Strength in Proton-Proton Interactions}
\subsection{Energy Dependence of the FB Correlation Strength in {\it pp}}

The forward-backward (FB) correlation strength in {\it pp} collisions has been analyzed. The data from {\it pp} collisions provides a baseline for the measurement of the FB correlation strength that is not contaminated by initial state effects of the colliding nuclei or final state effects from the possible production of a quark-gluon matter in A+A collisions. 
Data was obtained at three energies: $\sqrt{s_{NN}}$ = 400, 200, and 62.4 GeV.
\\

\begin{figure}
\centering
\includegraphics[width=5.5in]{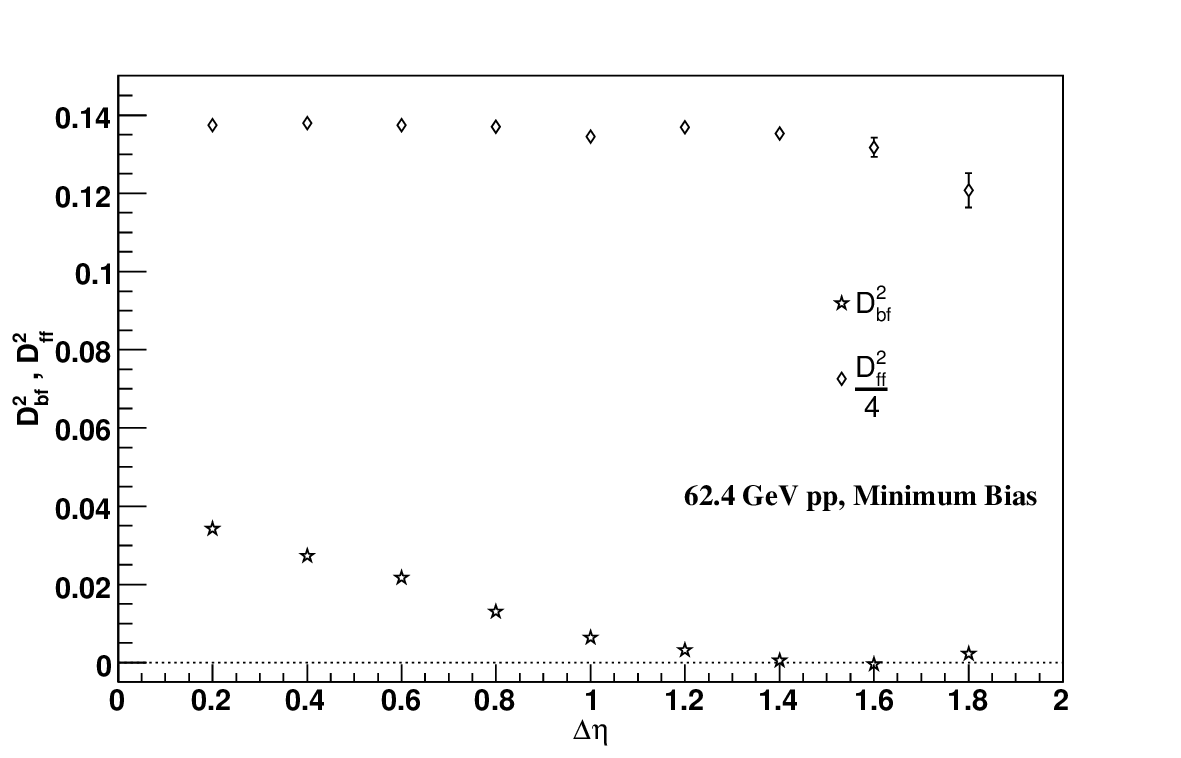}
\caption[Dispersions measured from MB {\it pp} data at $\sqrt{s_{NN}}$ = 62.4 GeV.]{Backward-forward (stars) and forward-forward (diamonds) dispersions ($D_{bf}^{2}$ and $D_{ff}^{2}$) as a function of the pseudorapidity gap $\Delta\eta$ in minimum bias {\it pp} data at $\sqrt{s_{NN}}$ = 62.4 GeV.}
\label{Dbf_pp_62_0-80}
\end{figure}

\begin{figure}
\centering
\includegraphics[width=5.5in]{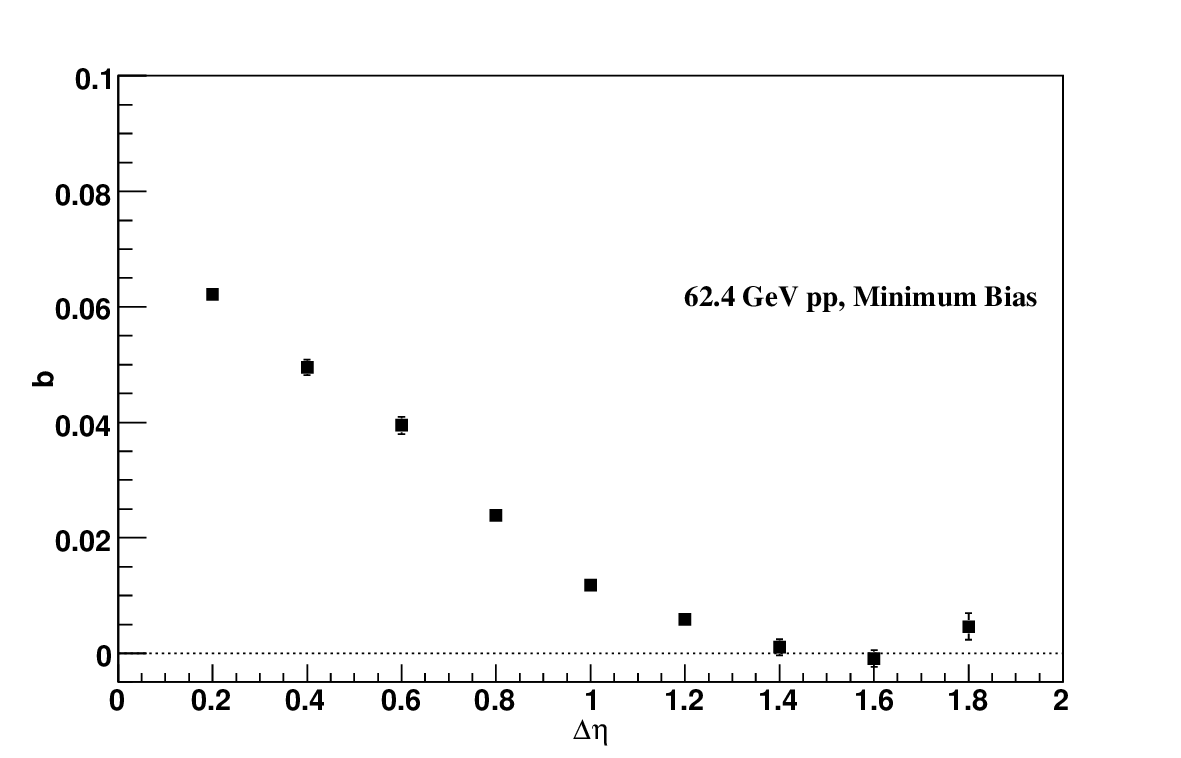}
\caption[{\it b} measured from MB {\it pp} data at $\sqrt{s_{NN}}$ = 62.4 GeV.]{Forward-backward correlation strength {\it b} as a function of the pseudorapidity gap $\Delta\eta$ in minimum bias {\it pp} data at $\sqrt{s_{NN}}$ = 62.4 GeV.}
\label{b_pp_62_MB}
\end{figure}

The forward-forward ($D_{ff}^{2}$) and forward-backward ($D_{bf}^{2}$) dispersions for $\sqrt{s_{NN}}$ = 62.4 GeV minimum bias {\it pp} data are shown in Figure \ref{Dbf_pp_62_0-80} as a function of the pseudorapidity gap, $\Delta\eta$. $D_{ff}^{2}$ is shown by the diamond symbols, while $D_{bf}^{2}$ are shown by the stars. The ratio $D_{bf}^{2}/D_{ff}^{2}$ is the FB correlation strength {\it b}. $D_{ff}^{2}$ is approximately flat as a function of $\Delta\eta$, while $D_{bf}^{2}$ exhibits a decreasing trend. This is reflected in Figure \ref{b_pp_62_MB}, the FB correlation strength {\it b} for $\sqrt{s_{NN}}$ = 62.4 GeV minimum bias {\it pp} collisions. The FB correlation strength shows only a short-range contribution ($\Delta\eta <$ 1.0). For $\Delta\eta >$ 1.0 the FB correlation strength approaches zero. 

\begin{figure}
\centering
\includegraphics[width=5.5in]{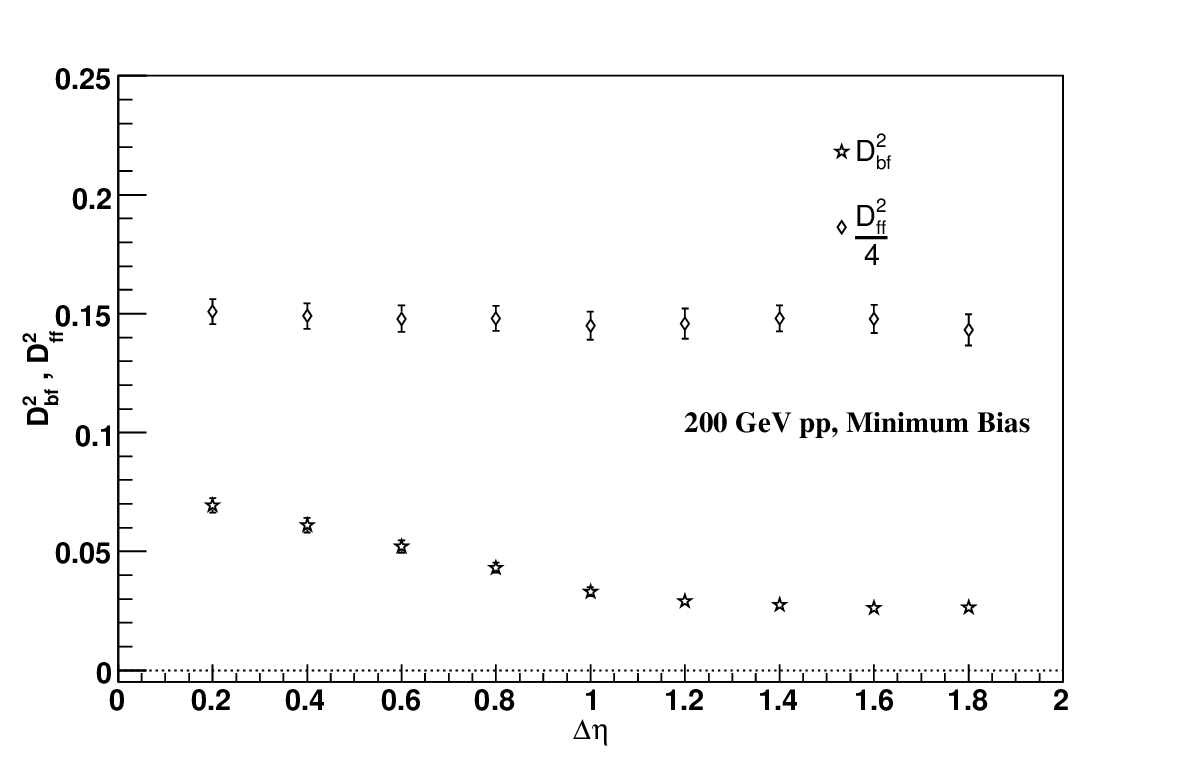}
\caption[Dispersions measured from MB {\it pp} data at $\sqrt{s_{NN}}$ = 200 GeV.]{Backward-forward (stars) and forward-forward (diamonds) dispersions ($D_{bf}^{2}$ and $D_{ff}^{2}$) as a function of the pseudorapidity gap $\Delta\eta$ in minimum bias $\sqrt{s_{NN}}$ = 200 GeV {\it pp} data.}
\label{Dbf_pp_200_0-80}
\end{figure}

\begin{figure}
\centering
\includegraphics[width=5.5in]{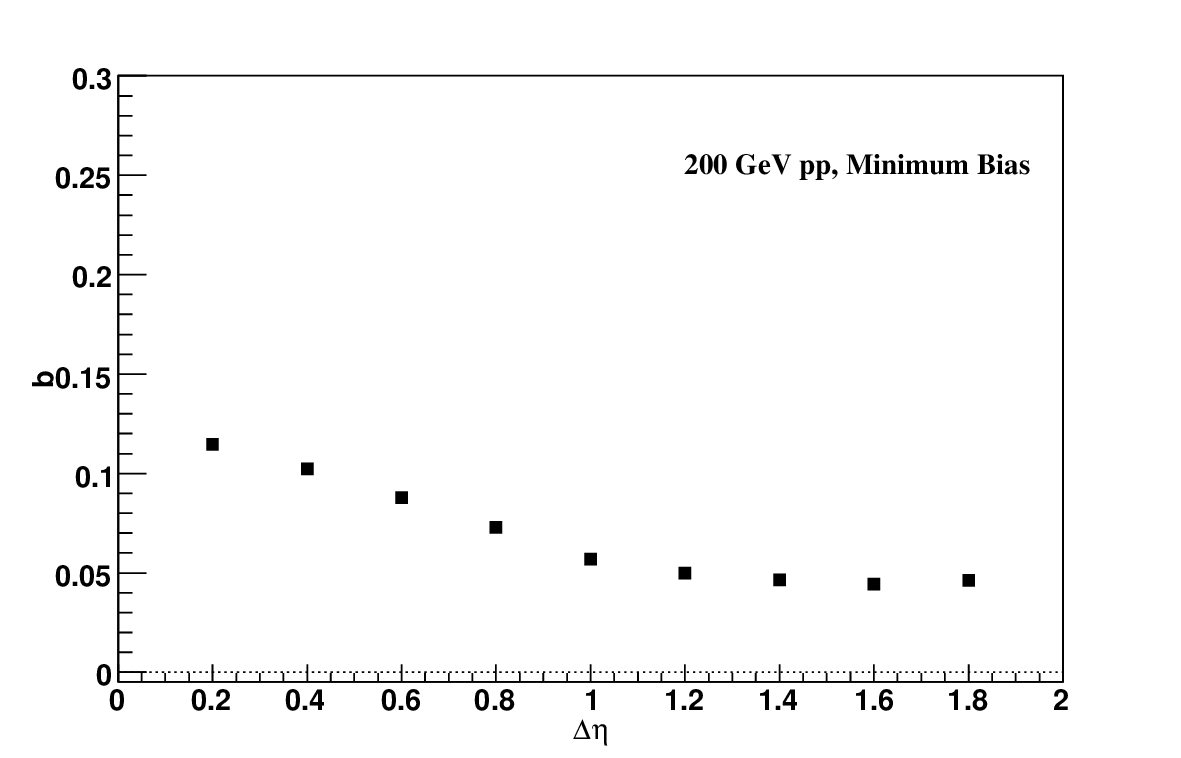}
\caption[{\it b} measured from MB {\it pp} data at $\sqrt{s_{NN}}$ = 200 GeV.]{Forward-backward correlation strength {\it b} as a function of the pseudorapidity gap $\Delta\eta$ in minimum bias $\sqrt{s_{NN}}$ = 200 GeV {\it pp} data.}
\label{b_pp_200_MB}
\end{figure}

Data from $\sqrt{s_{NN}}$ = 200 GeV {\it pp} minimum bias collisions for $D_{bf}^{2}$, $D_{ff}^{2}$, and {\it b} are shown in Figures \ref{Dbf_pp_200_0-80} and \ref{b_pp_200_MB}. 
Similar to {\it pp} at $\sqrt{s_{NN}}$ = 62.4, $D_{ff}^{2}$ is flat as a function of $\Delta\eta$. 
At energies as high as $\sqrt{s_{NN}}$ = 200 GeV, $D_{bf}^{2}$ displays a decreasing trend as a function of $\Delta\eta$. As demonstrated in Figure \ref{b_pp_200_MB}, this directs the behavior of {\it b} , which shows only a short-range component of the FB correlation strength for $\Delta\eta <$ 1.0. 

The results for minimum bias {\it pp} collisions at $\sqrt{s_{NN}}$ = 400 GeV closely resemble those at $\sqrt{s_{NN}}$ = 200 GeV. 
The FB correlation strength ({\it b}) is very similar at both energies. At $\Delta\eta$ = 0, $b$ = 0.115 $\pm$ 0.002 at $\sqrt{s_{NN}}$ = 200 GeV and 0.137 $\pm$ 0.007 at $\sqrt{s_{NN}}$ = 400 GeV. The FB correlation strength at $\Delta\eta$ = 0 is predominantly driven by short-range correlations. It is interesting to note that the value of {\it b} at large values of $\Delta\eta$ plateaus at the same, small non-zero value for both $\sqrt{s_{NN}}$ = 200 and 400 GeV. This trend is broken at the lower energy $\sqrt{s_{NN}}$ = 62.4 GeV, where {\it b} approaches zero at large values of $\Delta\eta$. The breakdown of Koba-Nielsen-Olesen (KNO) scaling in $pp$ interactions at high energy \cite{KNO1,Walker2004} provides a model independent correlation between the onset of long-range FB correlations and multiparton interactions in $pp$ collisions. The FB correlation strength at $\Delta\eta$ = 0 is predominantly driven by short-range correlations. It is interesting to note that the value of {\it b} at large values of $\Delta\eta$ plateaus at the same, small non-zero value for both $\sqrt{s_{NN}}$ = 200 and 400 GeV. This trend is broken at the lower energy $\sqrt{s_{NN}}$ = 62.4 GeV, where {\it b} approaches zero at large values of $\Delta\eta$. The breakdown of Koba-Nielsen-Olesen (KNO) scaling in $pp$ interactions at high energy \cite{KNO1,Walker2004} provides a model independent correlation between the onset of long-range FB correlations and multiparton interactions in $pp$ collisions.


\begin{figure}
\centering
\includegraphics[width=5.5in]{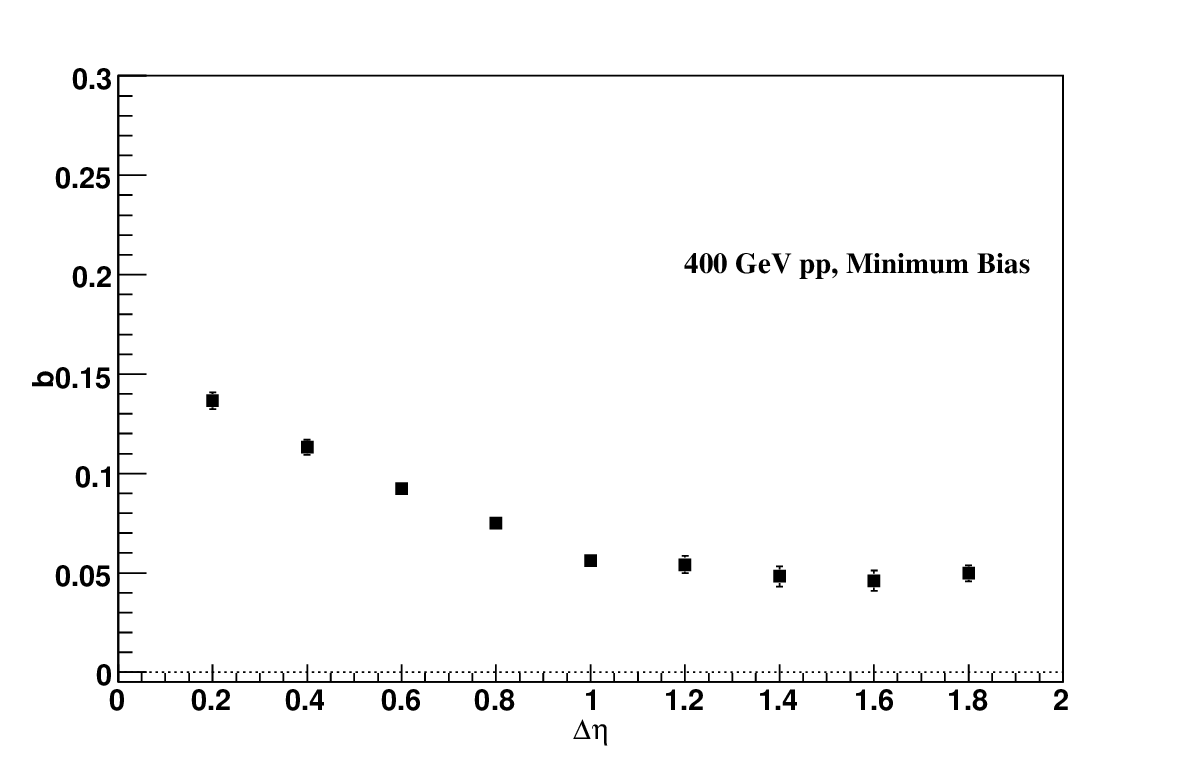}
\caption[{\it b} measured from MB {\it pp} data at $\sqrt{s_{NN}}$ = 400 GeV.]{Forward-backward correlation strength {\it b} as a function of the pseudorapidity gap $\Delta\eta$ in minimum bias {\it pp} data at $\sqrt{s_{NN}}$ = 400 GeV.}
\label{b_pp_400_MB}
\end{figure}

\section{Forward-Backward Correlation Strength in Nucleus-Nucleus Collisions}
\subsection{Au+Au}

\begin{figure}
\centering
\includegraphics[width=5.5in]{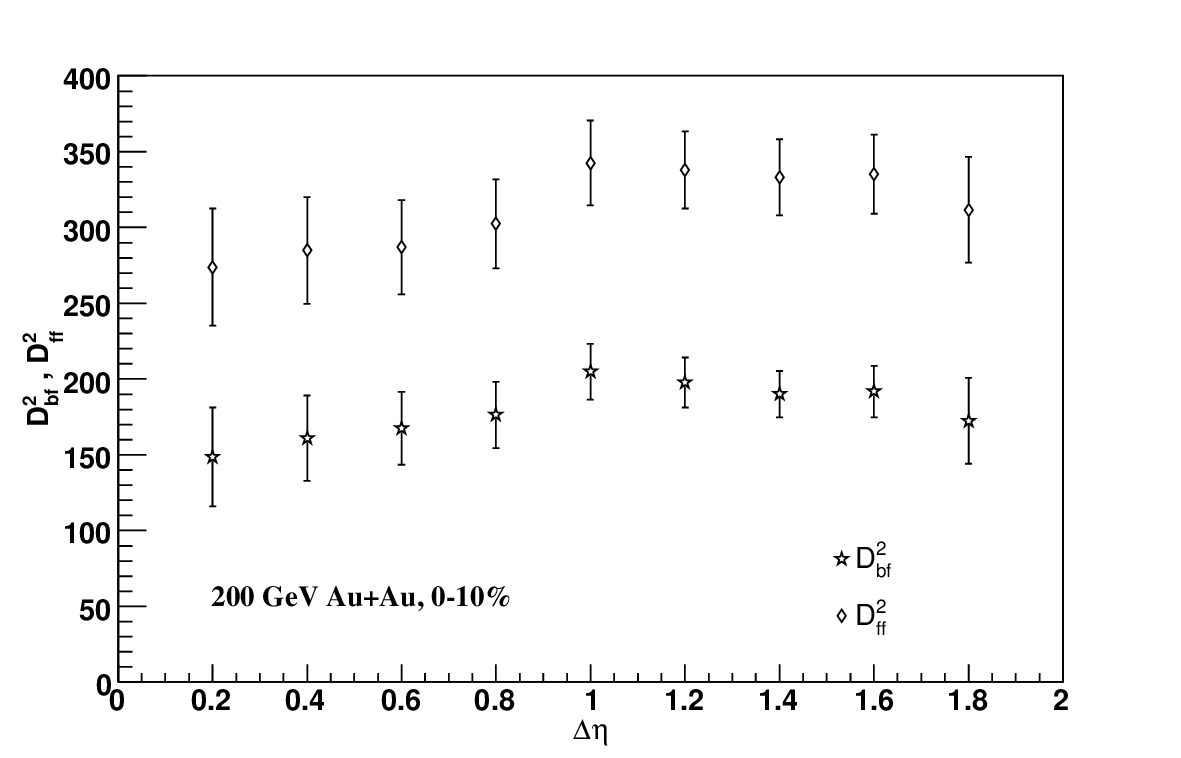}
\caption[Dispersions measured from 0-10\% Au+Au data at $\sqrt{s_{NN}}$ = 200 GeV.]{Backward-forward (stars) and forward-forward (diamonds) dispersions ($D_{bf}^{2}$ and $D_{ff}^{2}$) as a function of the pseudorapidity gap $\Delta\eta$ in central (0-10\%) $\sqrt{s_{NN}}$ = 200 GeV Au+Au data.}
\label{Dbf_AuAu_200_0-10}
\end{figure}

\begin{figure}
\centering
\includegraphics[width=5.5in]{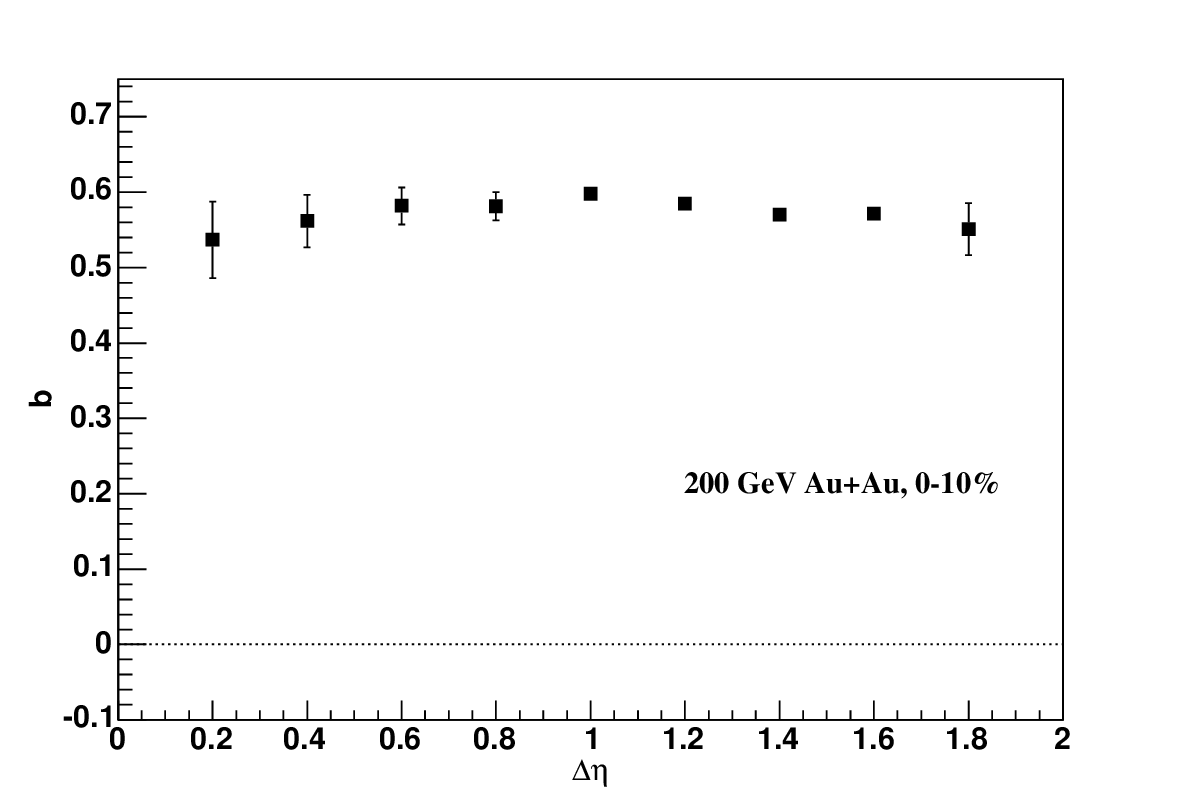}
\caption[{\it b} measured from 0-10\% Au+Au data at $\sqrt{s_{NN}}$ = 200 GeV.]{Forward-backward correlation strength {\it b} as a function of the pseudorapidity gap $\Delta\eta$ in central (0-10\%) $\sqrt{s_{NN}}$ = 200 GeV Au+Au data.}
\label{b_AuAu_200_0-10}
\end{figure}

The forward-forward ($D_{ff}^{2}$) and forward-backward ($D_{bf}^{2}$) dispersions for 0-10\% $\sqrt{s_{NN}}$ = 200 GeV Au+Au collisions are shown in Figure \ref{Dbf_AuAu_200_0-10} as a function of the pseudorapidity gap, $\Delta\eta$. $D_{ff}^{2}$ is shown by the diamond symbols, while $D_{bf}^{2}$ are shown by the stars. The ratio $D_{bf}^{2}/D_{ff}^{2}$ is the FB correlation strength {\it b}. Within the total error $D_{ff}^{2}$ and $D_{bf}^{2}$ is flat as a function of $\Delta\eta$. This is reflected in Figure \ref{b_AuAu_200_0-10}, the FB correlation strength {\it b} for 0-10\% $\sqrt{s_{NN}}$ = 200 GeV Au+Au collisions as a function of $\Delta\eta$. The FB correlation strength is also flat as a function of $\Delta\eta$ within the total error, reflecting the behavior of $D_{ff}^{2}$ and $D_{bf}^{2}$. In the ratio {\it b}, some of the individual errors on $D_{ff}^{2}$ and $D_{bf}^{2}$ are reduced.

The short-range correlations (SRC) are defined as the correlation between particles separated by less than $\Delta\eta = 1.0$. The long-range correlation (LRC) is taken as a correlation between particles separated by greater than one unit in $\Delta\eta$. Figure \ref{b_AuAu_200_0-10} shows a strong LRC as a function of $\Delta\eta$. If only short-range correlations (from sources such cluster formation, jets, resonance decay, etc.) are present, the expectation is a quickly decreasing FB correlation strength as a function of $\Delta\eta$. 
\\

\begin{figure}
\centering
\includegraphics[width=5.5in]{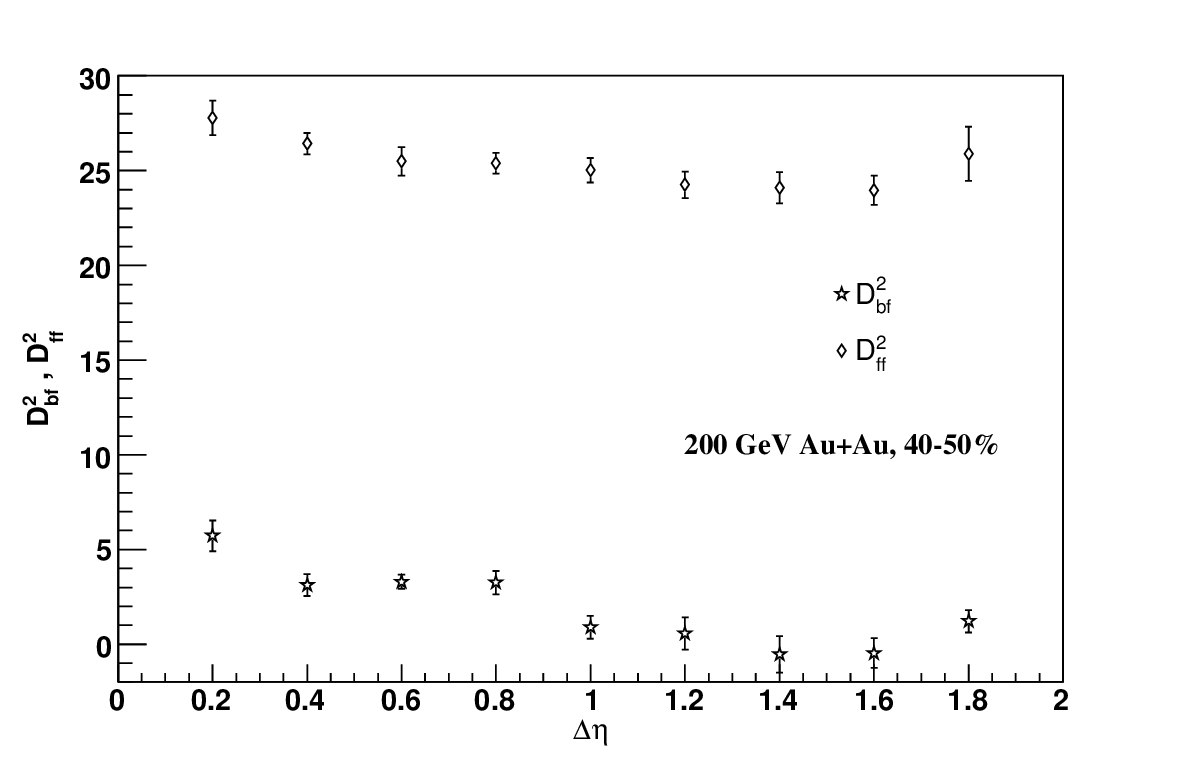}
\caption[Dispersions measured from 40-50\% Au+Au data at $\sqrt{s_{NN}}$ = 200 GeV.]{Backward-forward (stars) and forward-forward (diamonds) dispersions ($D_{bf}^{2}$ and $D_{ff}^{2}$) as a function of the pseudorapidity gap $\Delta\eta$ in mid-peripheral (40-50\%) $\sqrt{s_{NN}}$ = 200 GeV Au+Au data.}
\label{Dbf_AuAu_200_40-50}
\end{figure}

\begin{figure}
\centering
\includegraphics[width=5.5in]{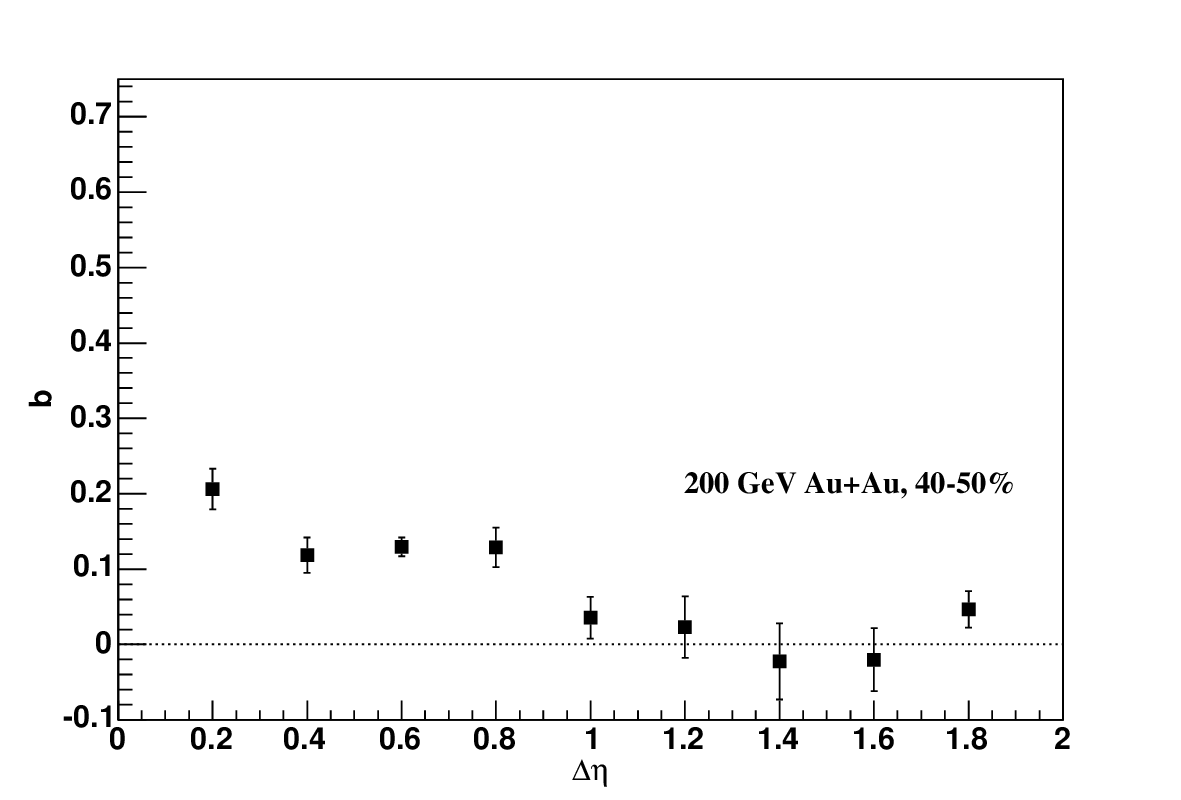}
\caption[{\it b} measured from 40-50\% Au+Au data at $\sqrt{s_{NN}}$ = 200 GeV.]{Forward-backward correlation strength {\it b} as a function of the pseudorapidity gap $\Delta\eta$ in mid-peripheral (40-50\%) $\sqrt{s_{NN}}$ = 200 GeV Au+Au data.}
\label{b_AuAu_200_40-50}
\end{figure}

Figures \ref{Dbf_AuAu_200_40-50} and \ref{b_AuAu_200_40-50} show $D_{bf}^{2}$, $D_{ff}^{2}$, and {\it b} as functions of $\Delta\eta$ for 40-50\% $\sqrt{s_{NN}}$ = 200 GeV Au+Au collisions. Unlike in the 0-10\% case, the FB correlation strength is not flat as a function of $\Delta\eta$, but decreases rapidly as expected in the case where predominantly short-range correlations are present, as in {\it pp} at an energy of $\sqrt{s_{NN}}$ = 62.4 GeV. From Figure \ref{b_AuAu_200_40-50} it appears that mid-peripheral Au+Au collisions has the expected behavior if only short-range correlations are present and does not exhibit a FB long-range correlation. 

Figure \ref{Dbf_AuAu_200_40-50} demonstrates that it is the numerator $D_{bf}^{2}$ that drives the behavior of the FB correlation strength. Figure \ref{Dbf_AuAu_200_40-50} shows that $D_{ff}^{2}$ is approximately flat as a function of $\Delta\eta$, whereas $D_{bf}^{2}$ is maximum at $\Delta\eta$ = 0 and decreases as a function of $\Delta\eta$. 

\begin{figure}
\centering
\includegraphics[width=5.5in]{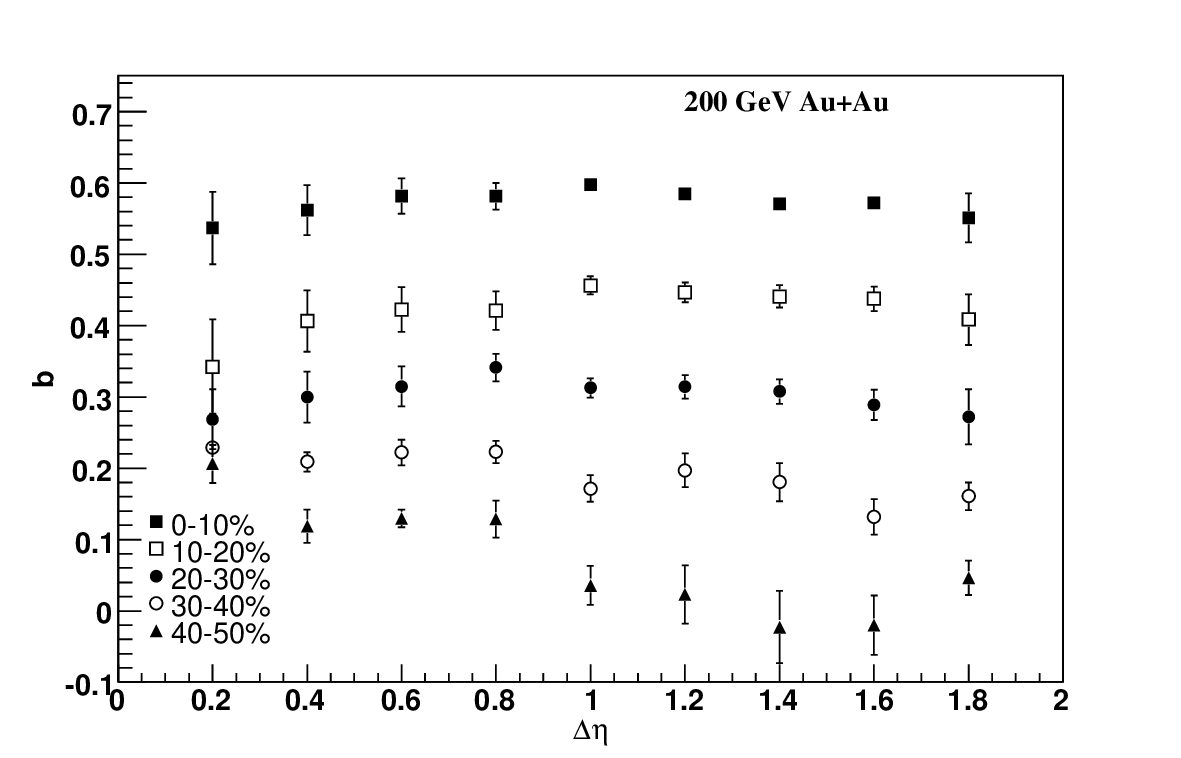}
\caption[Centrality dependence of {\it b} from Au+Au data at $\sqrt{s_{NN}}$ = 200 GeV.]{Forward-backward correlation strength {\it b} for 0-10\% (closed squares), 10-20\% (open squares), 20-30\% (closed circles), 30-40\% (open circles), and 40-50\% (closed triangles) most central $\sqrt{s_{NN}}$ = 200 GeV Au+Au data as a function of the pseudorapidity gap $\Delta\eta$.}
\label{b_AuAu_200_Cent_Dependence}
\end{figure}

\begin{figure}
\centering
\includegraphics[width=5.5in]{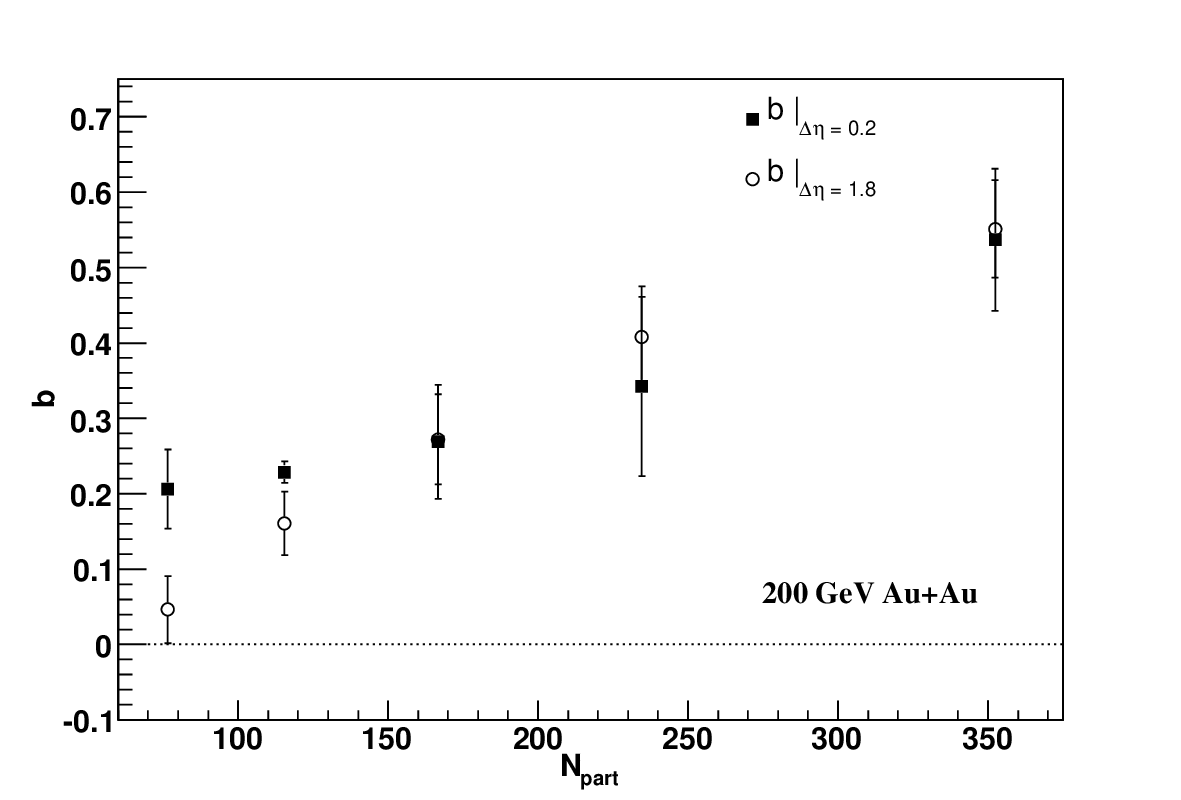}
\caption[{\it b} versus $N_{part}$ from Au+Au data at $\sqrt{s_{NN}}$ = 200 GeV.]{Centrality ($N_{part}$) dependence of the forward-backward correlation strength {\it b} evaluated at two $\Delta\eta$ values: $\Delta\eta$ = 0.2 (closed squares) and 1.8 (open circles). The long-range correlation strength ($\Delta\eta$ = 1.8) reaches the value of the short-range correlation strength ($\Delta\eta$ = 0.2) for central collisions ($N_{part} \approx$ 350).}
\label{b_AuAu_200_Npart_Dependence}
\end{figure}

The centrality dependence of the FB correlation strength {\it b} in Au+Au data at $\sqrt{s_{NN}}$ = 200 GeV is shown in Figure \ref{b_AuAu_200_Cent_Dependence}. It is seen that the FB correlation strength for central and mid-central (0-10, 10-20, and 20-30\%) Au+Au collisions at $\sqrt{s_{NN}}$ = 200 GeV exhibits a long-range component that is approximately flat across $\Delta\eta$. At about the 30-40\% centrality bin the long-range component begins to disappear and a slight decrease as a function of $\Delta\eta$ can be seen. By 40-50\% it appears that all remnants of a long-range correlation have vanished, leaving only the short-range component that decreases rapidly with $\Delta\eta$. 
Figure \ref{b_AuAu_200_Npart_Dependence} shows the FB correlation strength as a function of centrality evaluated at two values of $\Delta\eta$: $\Delta\eta$ = 0.2 and 1.8.
The overall trend shows an increase of the FB correlation strength with centrality, as shown in Figures \ref{b_AuAu_200_Cent_Dependence} and \ref{b_AuAu_200_Npart_Dependence}. 

\subsubsection{Energy Dependence of the Forward-Backward Correlation Strength in Au+Au}

\begin{figure}
\centering
\includegraphics[width=5.5in]{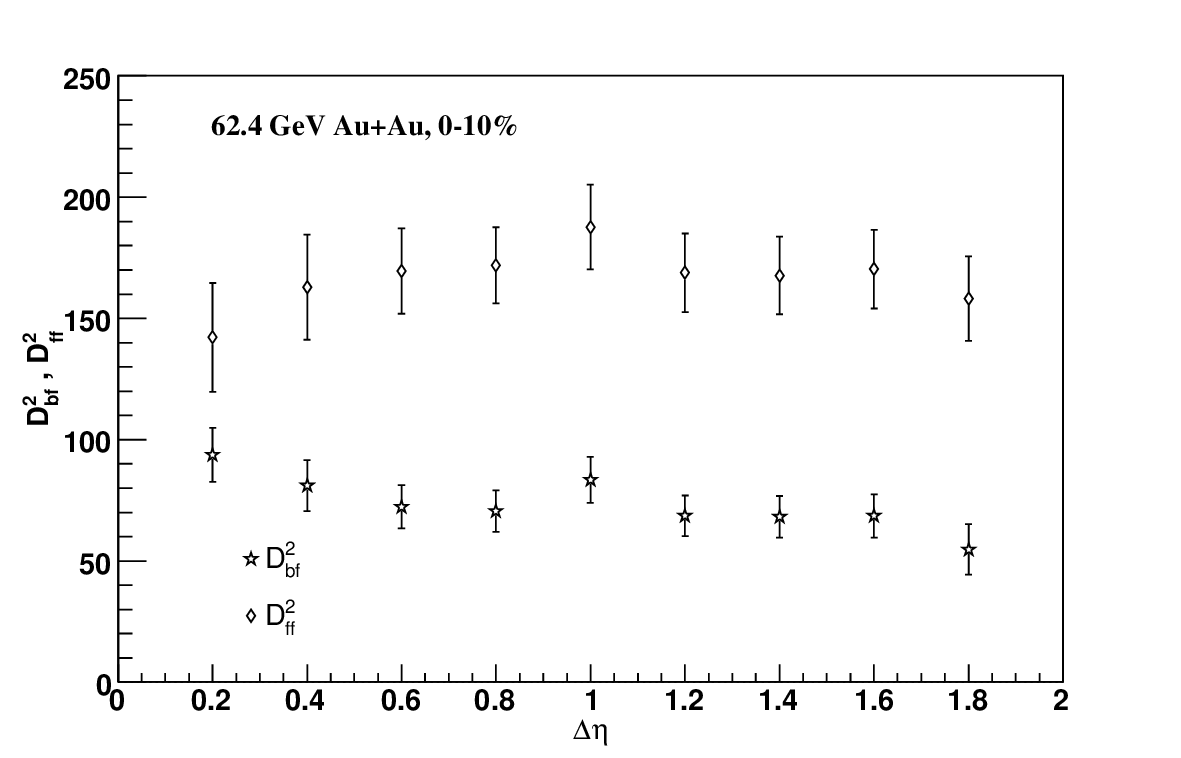}
\caption[Dispersions measured from 0-10\% Au+Au data at $\sqrt{s_{NN}}$ = 62.4 GeV.]{Backward-forward (stars) and forward-forward (diamonds) dispersions ($D_{bf}^{2}$ and $D_{ff}^{2}$) as a function of the pseudorapidity gap $\Delta\eta$ in central (0-10\%) $\sqrt{s_{NN}}$ = 62.4 GeV Au+Au data.}
\label{Dbf_AuAu_62_0-10}
\end{figure}

\begin{figure}
\centering
\includegraphics[width=5.5in]{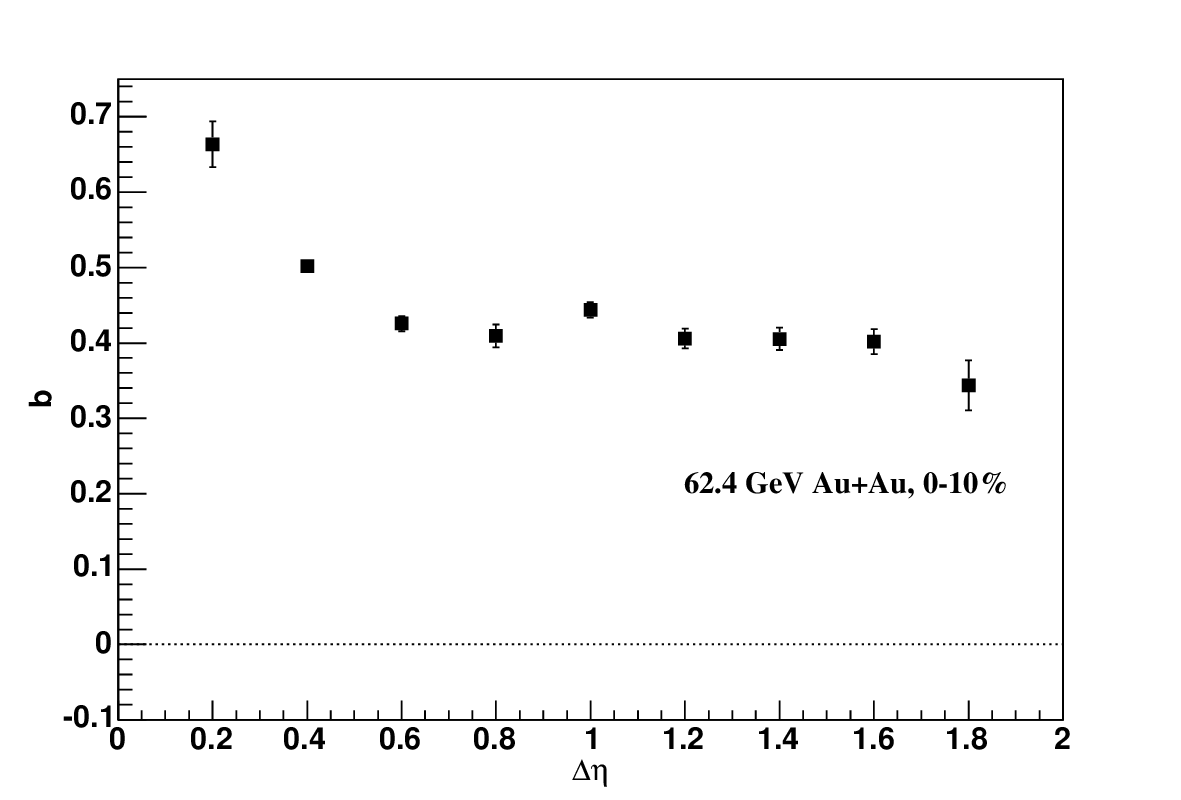}
\caption[{\it b} measured from 0-10\% Au+Au data at $\sqrt{s_{NN}}$ = 62.4 GeV.]{Forward-backward correlation strength {\it b} as a function of the pseudorapidity gap $\Delta\eta$ in central (0-10\%) $\sqrt{s_{NN}}$ = 62.4 GeV Au+Au data.}
\label{b_AuAu_62_0-10}
\end{figure}

Additional study of the FB correlation strength was carried out in Au+Au collisions at a lower energy of $\sqrt{s_{NN}}$ = 62.4 GeV. As Figures \ref{Dbf_AuAu_62_0-10} and \ref{b_AuAu_62_0-10} show, the presence of a long-range correlation in central Au+Au collisions persists at this lower energy. For small values of $\Delta\eta$ Figure \ref{Dbf_AuAu_62_0-10} shows $D_{bf}^{2}$ trending down as a function of small $\Delta\eta$ ($<$ 1.0), while the opposite is true of $D_{ff}^{2}$. This results in the short-range FB correlation strength in Figure \ref{b_AuAu_62_0-10} at $\Delta\eta$ = 0 having a value close to that shown in Au+Au at an energy of $\sqrt{s_{NN}}$ = 200 GeV (Figure \ref{b_AuAu_200_0-10}). This is also a result of the tracking limitations as discussed in Section \ref{SysErrors}.
The long-range FB correlation strength is lower by approximately 30\% at $\sqrt{s_{NN}}$ = 62.4 GeV (Figure \ref{b_AuAu_62_0-10}) compared to $\sqrt{s_{NN}}$ = 200 GeV (Figure \ref{b_AuAu_200_0-10}). Though the ratio of the energies is approximately a factor of three, the ratio of the multiplicities differ by about a factor of 1.5. The normalization of {\it b} by $D_{ff}^{2}$ contributes to this, since $D_{ff}^{2}$ goes approximately with the multiplicity. Therefore, it is the quantity $D_{bf}^{2}$ that carries the dynamical information. This approximate scaling with multiplicity does not hold when comparing different colliding systems (e.g., Au+Au and Cu+Cu). 

\begin{figure}
\centering
\includegraphics[width=5.5in]{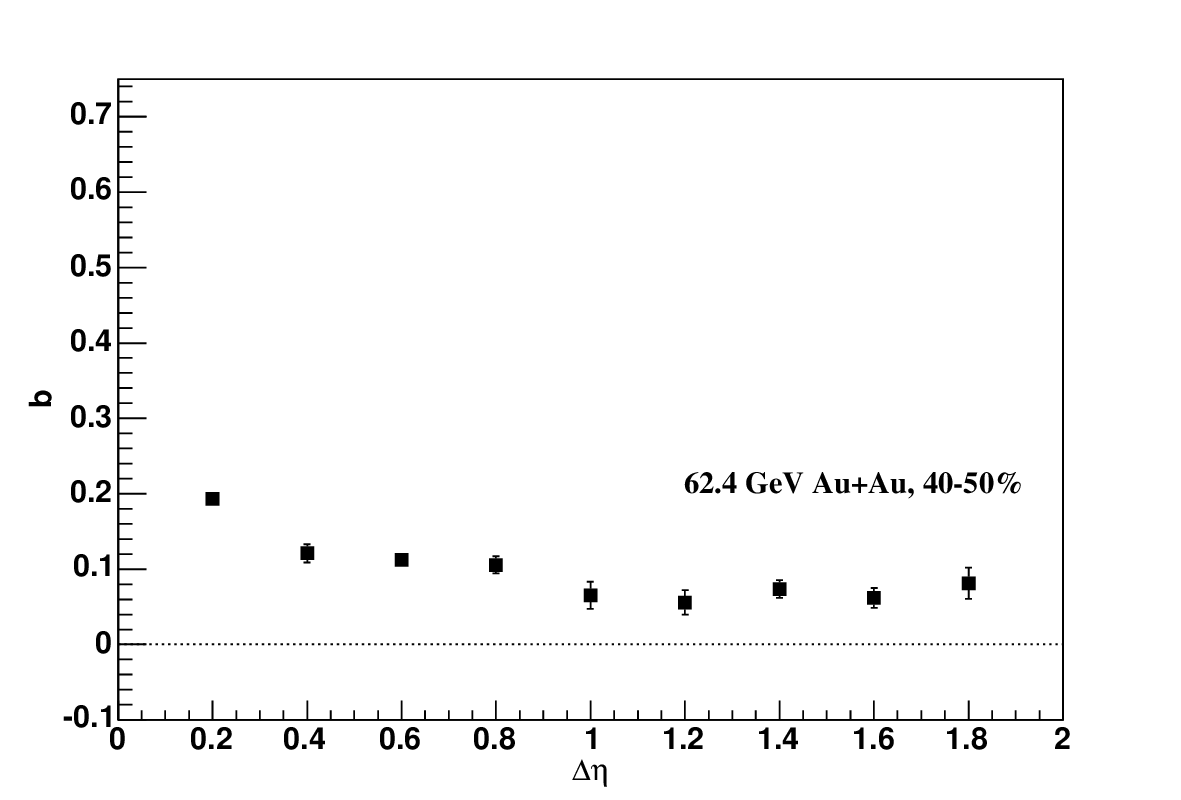}
\caption[{\it b} measured from 40-50\% Au+Au data at $\sqrt{s_{NN}}$ = 62.4 GeV.]{Forward-backward correlation strength {\it b} as a function of the pseudorapidity gap $\Delta\eta$ in mid-peripheral (40-50\%) $\sqrt{s_{NN}}$ = 62.4 GeV Au+Au data.}
\label{b_AuAu_62_40-50}
\end{figure}

For comparison purposes, the 40-50\% centrality bin for $\sqrt{s_{NN}}$ = 62.4 GeV Au+Au is shown in Figure \ref{b_AuAu_62_40-50}.
As seen in the $\sqrt{s_{NN}}$ = 200 GeV Au+Au result, the mid-peripheral data does not exhibit a long-range component.


\begin{figure}
\centering
\includegraphics[width=5.5in]{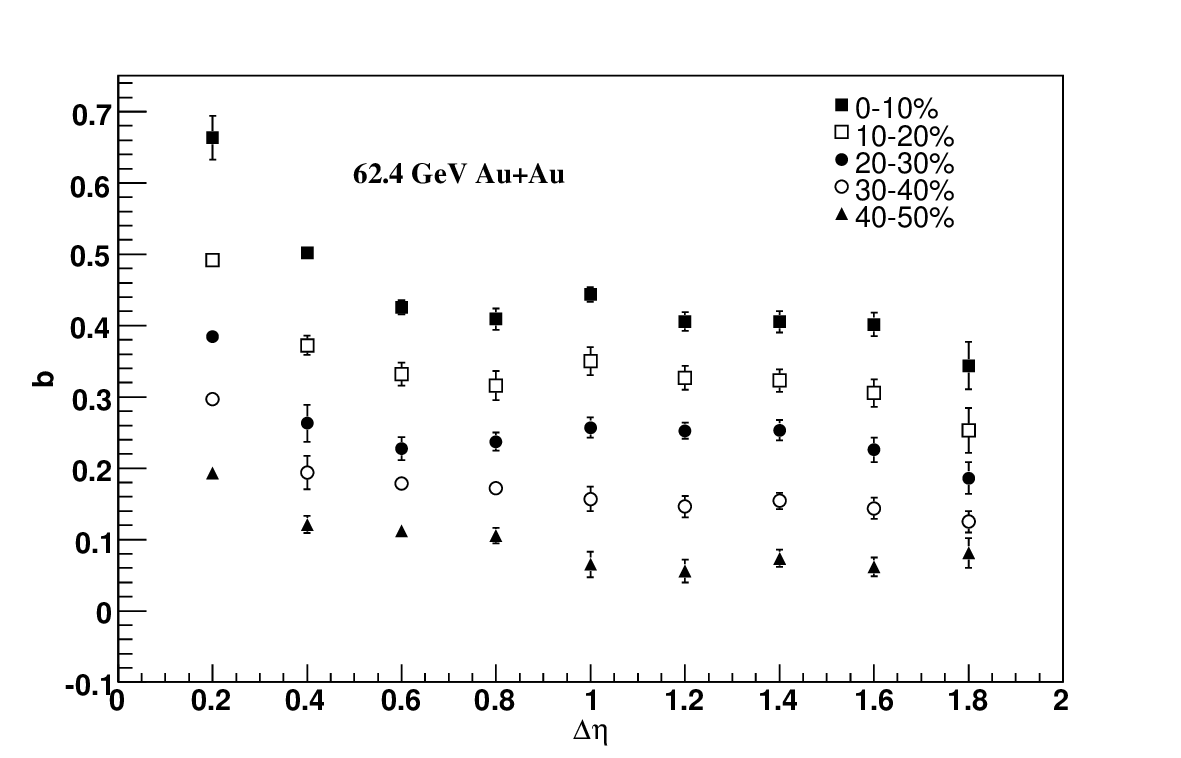}
\caption[Centrality dependence of {\it b} from Au+Au data at $\sqrt{s_{NN}}$ = 62.4 GeV.]{Forward-backward correlation strength {\it b} for 0-10\% (closed squares), 10-20\% (open squares), 20-30\% (closed circles), 30-40\% (open circles), and 40-50\% (closed triangles) most central $\sqrt{s_{NN}}$ = 62.4 GeV Au+Au data as a function of the pseudorapidity gap $\Delta\eta$.}
\label{b_AuAu_62_Cent_Dependence}
\end{figure}

The centrality dependence of the FB correlation strength at $\sqrt{s_{NN}}$ = 62.4 GeV (Figure \ref{b_AuAu_62_Cent_Dependence}) demonstrates the progressive decrease in the long-range component as a function of centrality. Additionally, the FB correlation strength in central collisions exhibits a dependence on the collision energy. The long-range FB correlation strength is smaller at lower energies.

\subsection{Cu+Cu}

To supplement the studies of Au+Au collisions at RHIC, collisions of the lighter nuclei Cu+Cu were examined at three energies: $\sqrt{s_{NN}}$ = 200, 62.4, and 22.4 GeV. The STAR centrality definitions for $\sqrt{s_{NN}}$ = 200 Cu+Cu data are shown in Table \ref{2005Cu200Centrality}.

\begin{table}
\begin{center}
\begin{tabular}{|*{4}{c|}}
\hline
Centrality &   $N_{ch}$ &      $N_{part}$ &       $N_{bin}$ \\
\hline
    0-10\% & $N_{ch}\geq140$           &  98.4 +1.0 -1.0 &  185.7 +5.9 -5.2 \\
\hline
   10-20\% &  $103\leq N_{ch} \leq139$ &  74.8 +2.5 -2.2 &  126.7 +6.7 -7.1 \\
\hline
   20-30\% &  $74\leq N_{ch} \leq102$ &  54.4 +2.8 -2.5 &  81.5 +6.0 -6.1 \\
\hline
   30-40\% &  $53\leq N_{ch} \leq73$ & 38.5 +2.5 -3.1 &  51.0 +4.8 -5.4 \\
\hline
   40-50\% &  $37\leq N_{ch} \leq52$ &  26.3 +2.4 -3.3 &  30.6 +3.9 -4.6 \\
\hline
   50-60\% &   $25\leq N_{ch} \leq36$ &  17.6 +2.6 -3.1 &  18.2 +3.5 -3.4 \\
\hline
\end{tabular}
\caption[Centrality definition for Cu+Cu at $\sqrt{s_{NN}}$ = 200 GeV.]{Centrality definition in terms of the number of charged particles for Cu+Cu at $\sqrt{s_{NN}}$ = 200 GeV (P06ib production). The number of nucleons participating in the collision ($N_{part}$) and the number of binary nucleon-nucleon collisions ($N_{bin}$) estimated from Monte Carlo Glauber simulations \protect\cite{STARGlauber} are also shown.}
\label{2005Cu200Centrality}
\end{center}
\end{table}

\begin{figure}
\centering
\includegraphics[width=5.5in]{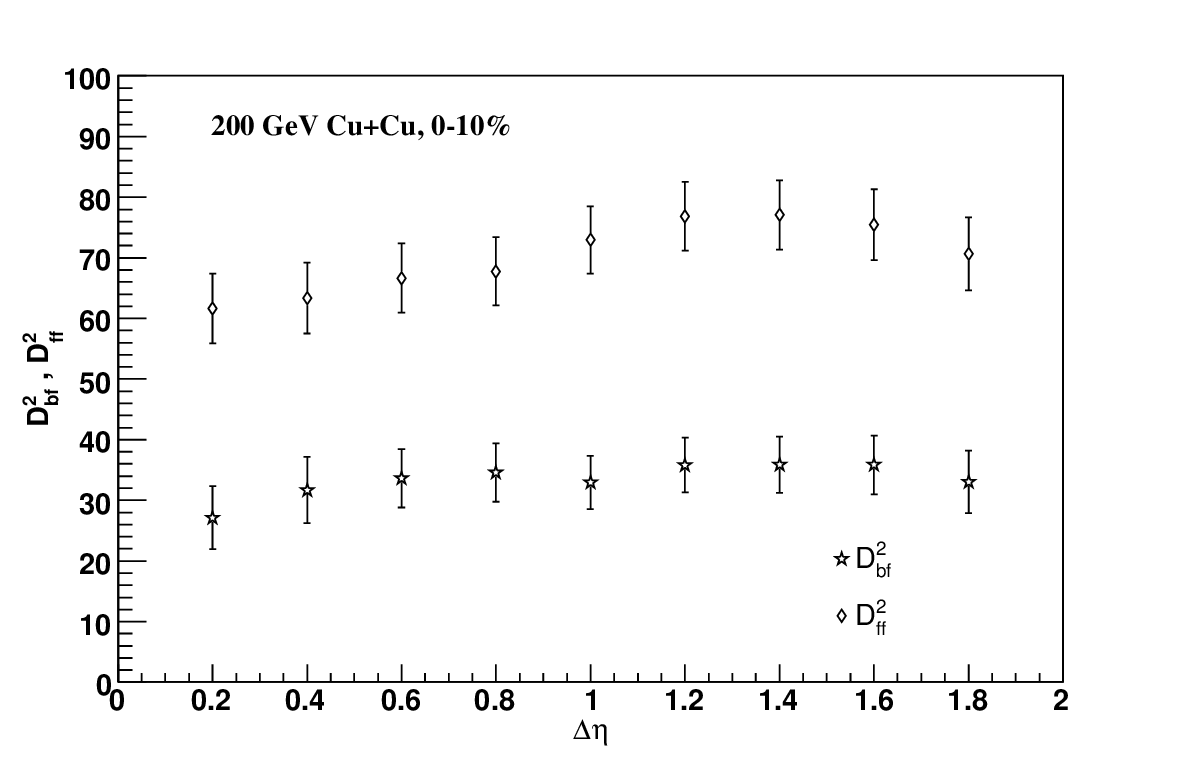}
\caption[Dispersions measured from 0-10\% Cu+Cu data at $\sqrt{s_{NN}}$ = 200 GeV.]{Backward-forward (stars) and forward-forward (diamonds) dispersions ($D_{bf}^{2}$ and $D_{ff}^{2}$) as a function of the pseudorapidity gap $\Delta\eta$ in central (0-10\%) Cu+Cu data at $\sqrt{s_{NN}}$ = 200 GeV.}
\label{Dbf_CuCu_200_0-10}
\end{figure}

The study of the FB correlation strength {\it b} for Cu+Cu was accomplished in the same manner as that for Au+Au. Figure \ref{Dbf_CuCu_200_0-10} shows $D_{bf}^{2}$ and $D_{ff}^{2}$ for 0-10\% most central $\sqrt{s_{NN}}$ = 200 GeV Cu+Cu collisions. Both quantities are flat as a function of $\Delta\eta$, similar to central Au+Au collisions at the same energy (Figure \ref{Dbf_AuAu_200_0-10}). Though the qualitative trend is the same, the values of $D_{bf}^{2}$ and $D_{ff}^{2}$ are both much smaller in central Cu+Cu than in Au+Au.

\begin{figure}
\centering
\includegraphics[width=5.5in]{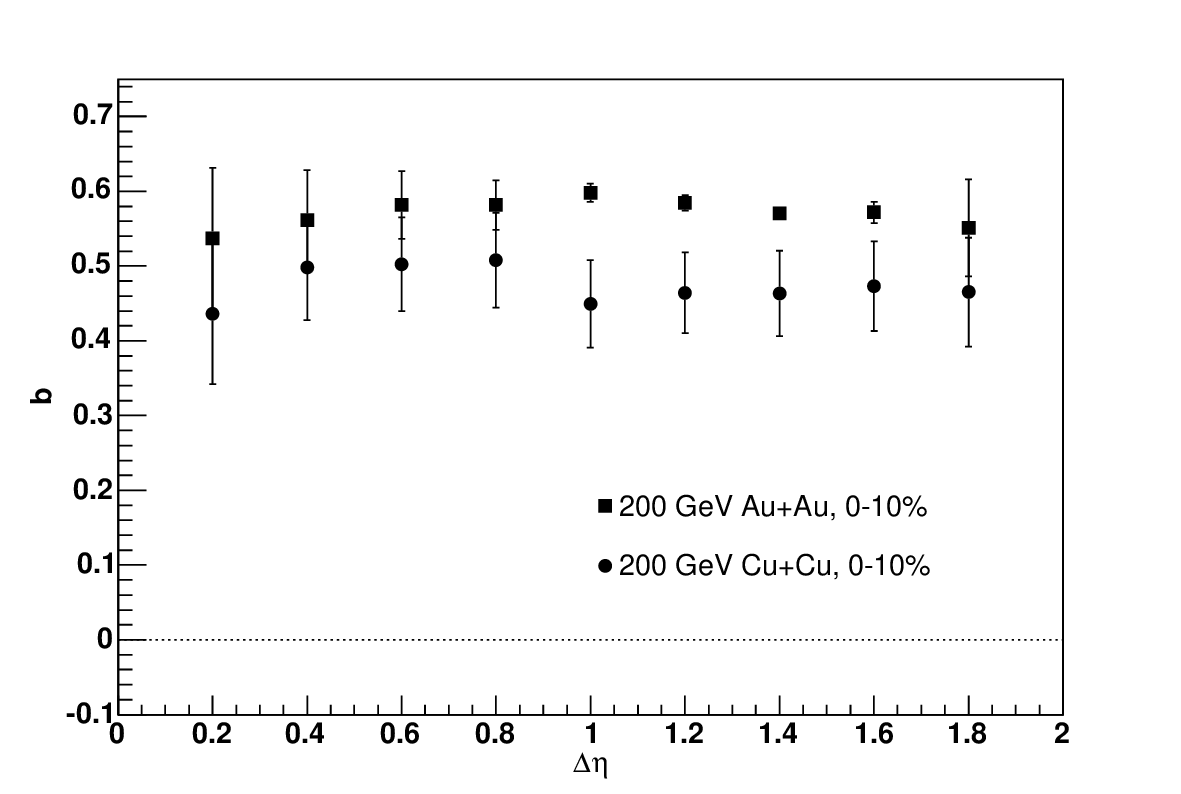}
\caption[{\it b} measured from 0-10\% Cu+Cu data at $\sqrt{s_{NN}}$ = 200 GeV.]{Forward-backward correlation strength {\it b} as a function of the pseudorapidity gap $\Delta\eta$ in central (0-10\%) Cu+Cu data (circles) at $\sqrt{s_{NN}}$ = 200 GeV. For comparison, the same quantity is plotted for central (0-10\%) Au+Au data at the same energy (squares).}
\label{b_CuCu_200_0-10}
\end{figure}

\begin{figure}
\centering
\includegraphics[width=5.5in]{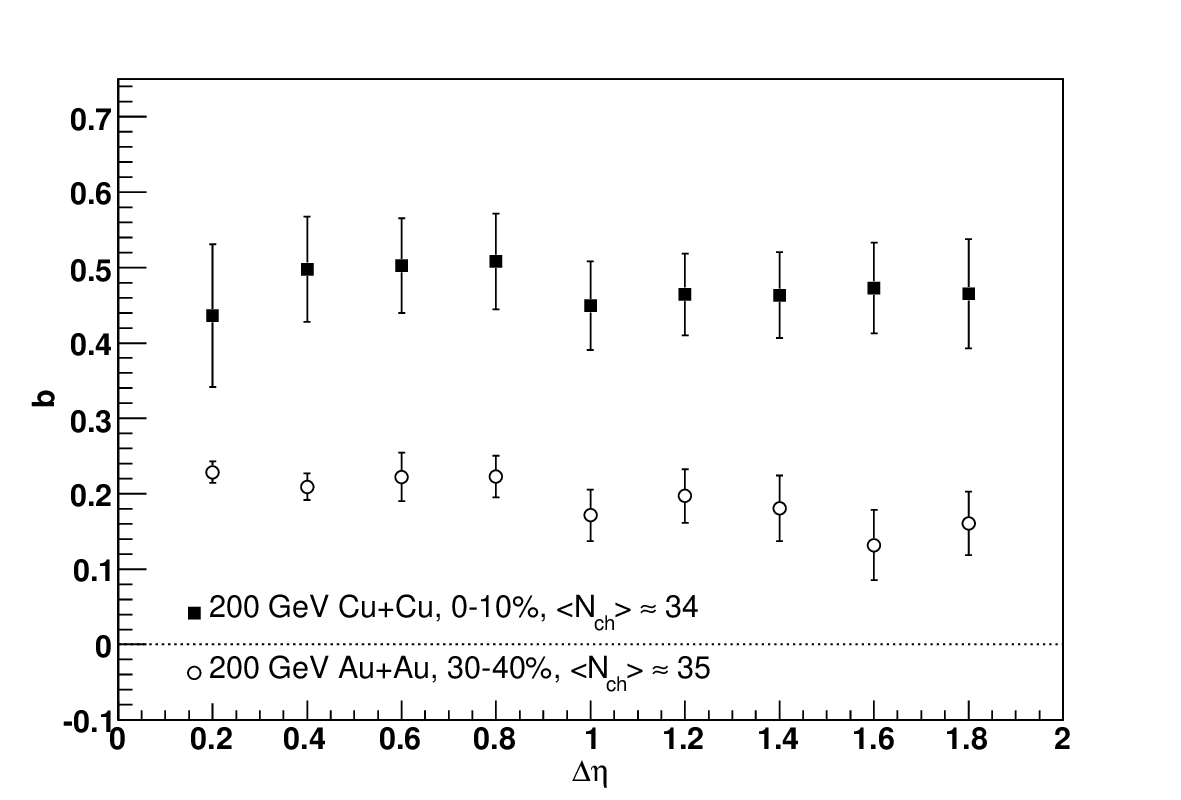}
\caption[{\it b} measured for 0-10\% Cu+Cu compared to the value in 30-40\% Au+Au.]{Comparison of the forward-backward correlation strength {\it b} as a function of the pseudorapidity gap $\Delta\eta$ in central (0-10\%) Cu+Cu (squares) and mid-peripheral (30-40\%) Au+Au (Au+Au) data at an energy of $\sqrt{s_{NN}}$ = 200 GeV. The two centralities have approximately the same average multiplicity, but exhibit a different evolution as a function of $\Delta\eta$.}
\label{b_AuAu_30-40_CuCu_0-10_200}
\end{figure}

Figure \ref{b_CuCu_200_0-10} is the FB correlation strength {\it b} for the 0-10\% most central $\sqrt{s_{NN}}$ = 200 GeV Cu+Cu collisions. As with $D_{bf}^{2}$ and $D_{ff}^{2}$, the qualitative trend of {\it b} is the same for Cu+Cu and Au+Au at the same energy and centrality. Both exhibit a strong, long-range component to the FB correlation strength. The plateau of the FB correlation strength is $\approx$ 15\% lower in Cu+Cu than Au+Au at the same energy (Figure \ref{b_AuAu_200_0-10}), while the multiplicity is $\approx$ 70\% lower in Cu+Cu ($<N_{ch}>_{Au+Au} \approx$ 116 and $<N_{ch}>_{Cu+Cu} \approx$ 34). The average multiplicity in 0-10\% Cu+Cu data at $\sqrt{s_{NN}}$ = 200 GeV is most closely matched by the 30-40\% centrality bin of Au+Au. Comparison to the 30-40\% Au+Au centrality from Figure \ref{b_AuAu_200_Cent_Dependence} shows a large difference between the FB correlation strength of the two systems at the same average multiplicity. The comparison of the FB correlation strength in 0-10\% Cu+Cu and 30-40\% Au+Au is shown in Figure \ref{b_AuAu_30-40_CuCu_0-10_200}.

\begin{figure}
\centering
\includegraphics[width=5.5in]{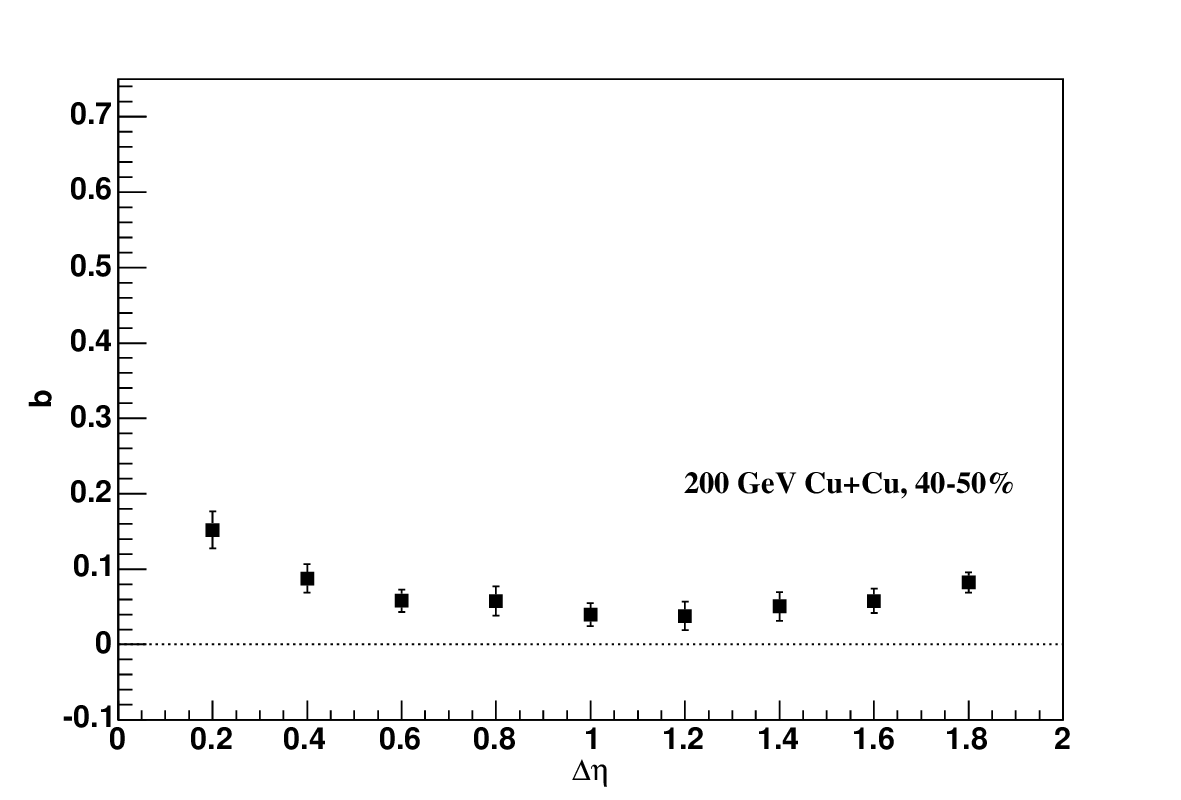}
\caption[{\it b} measured from 40-50\% Cu+Cu data at $\sqrt{s_{NN}}$ = 200 GeV.]{Forward-backward correlation strength {\it b} as a function of the pseudorapidity gap $\Delta\eta$ in mid-peripheral (40-50\%) Cu+Cu data at $\sqrt{s_{NN}}$ = 200 GeV.}
\label{b_CuCu_200_40-50}
\end{figure}

The FB correlation strength in mid-peripheral $\sqrt{s_{NN}}$ = 200 GeV Cu+Cu is also very similar to Au+Au at the same energy. Figure \ref{b_CuCu_200_40-50} shows the Cu+Cu results for the 40-50\% centrality bin. 
The FB correlation strength in mid-peripheral Cu+Cu and Au+Au (Figure \ref{b_AuAu_200_40-50}) both show the presence of only short-range correlations. Both systems show similar qualitative and quantitative evolution as a function of $\Delta\eta$. There is also a close similarity to the FB correlation strength in {\it pp} at the same energy.

\begin{figure}
\centering
\includegraphics[width=5.5in]{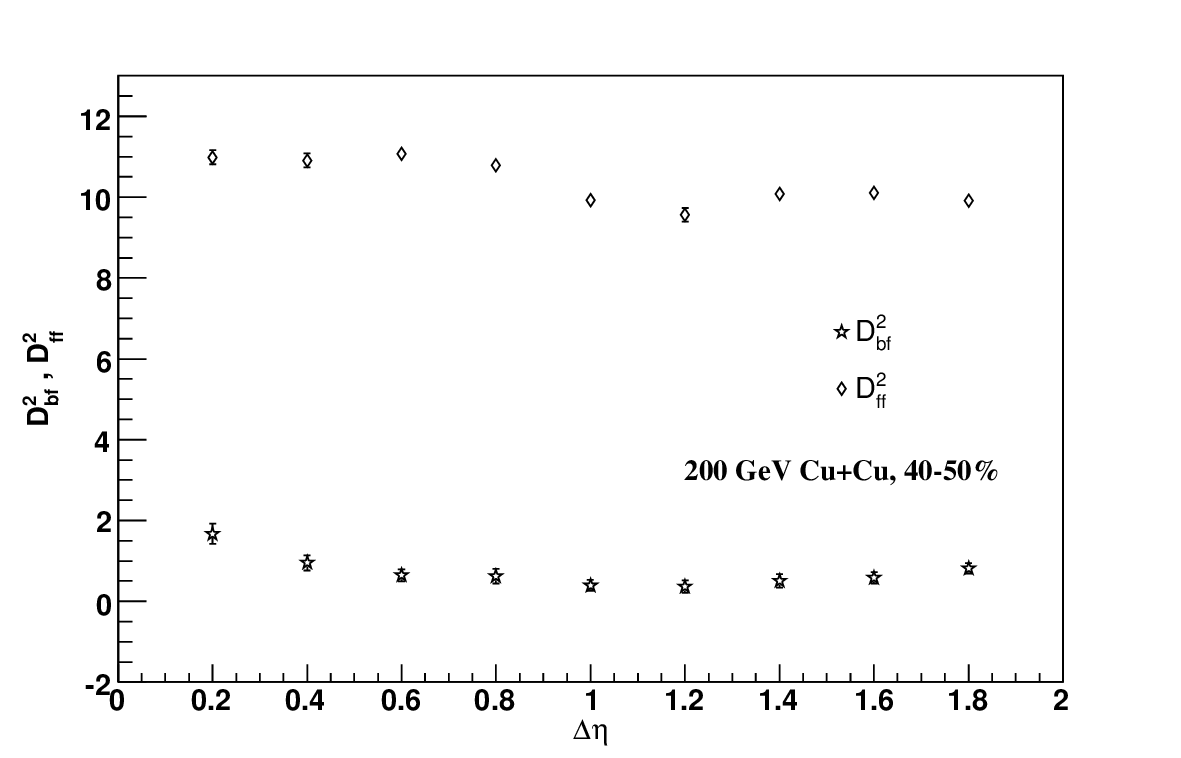}
\caption[Dispersions measured from 40-50\% Cu+Cu data at $\sqrt{s_{NN}}$ = 200 GeV.]{Backward-forward (stars) and forward-forward (diamonds) dispersions ($D_{bf}^{2}$ and $D_{ff}^{2}$) as a function of the pseudorapidity gap $\Delta\eta$ in mid-peripheral (40-50\%) Cu+Cu data at $\sqrt{s_{NN}}$ = 200 GeV.}
\label{Dbf_CuCu_200_40-50}
\end{figure}

\begin{figure}
\centering
\includegraphics[width=5.5in]{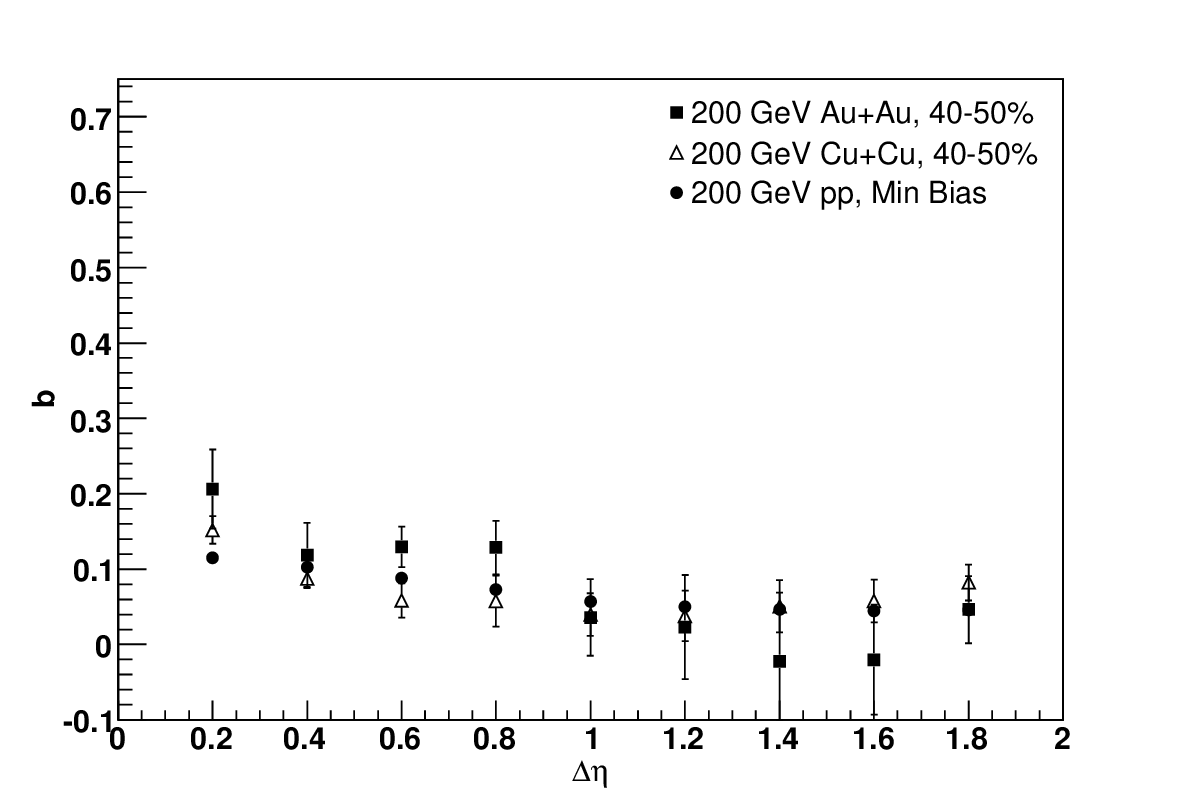}
\caption[{\it b} as measured for 40-50\% Au+Au, Cu+Cu, and MB {\it pp}.]{Forward-backward correlation strength {\it b} as a function of the pseudorapidity gap $\Delta\eta$ in mid-peripheral (40-50\%) Au+Au (squares), Cu+Cu (triangles), and minimum bias {\it pp} (circles) data at $\sqrt{s_{NN}}$ = 200 GeV. There is a general lack of long-range correlations in peripheral heavy ion data that is consistent with basic proton-proton interactions.}
\label{b_AuAu_CuCu_pp200}
\end{figure}

\begin{figure}
\centering
\includegraphics[width=5.5in]{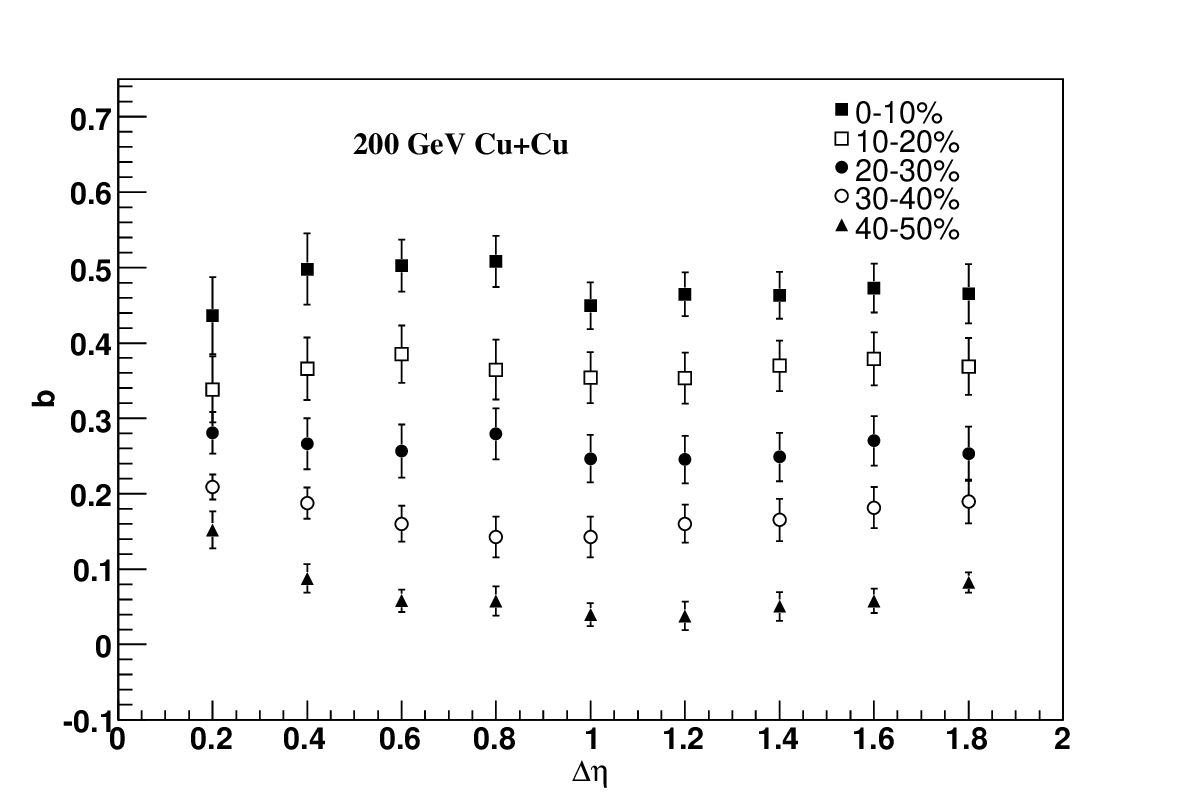}
\caption[Centrality dependence of {\it b} from Cu+Cu data at $\sqrt{s_{NN}}$ = 200 GeV.]{Forward-backward correlation strength {\it b} for 0-10\% (closed squares), 10-20\% (open squares), 20-30\% (closed circles), 30-40\% (open circles), and 40-50\% (closed triangles) most central Cu+Cu data at $\sqrt{s_{NN}}$ = 200 GeV as a function of the pseudorapidity gap $\Delta\eta$.}
\label{b_CuCu_200_Cent_Dependence}
\end{figure}

The centrality dependence of the FB correlation strength from $\sqrt{s_{NN}}$ = 200 GeV, 0-10\% to 40-50\% Cu+Cu collisions is shown in Figure \ref{b_CuCu_200_Cent_Dependence}. Though the total average multiplicity differs dramatically from Au+Au to Cu+Cu, the qualitative trend of the FB strength as a function of centrality is similar. The FB correlation strength for 40-50\% Au+Au, Cu+Cu, and minimum bias {\it pp} data at an energy of $\sqrt{s_{NN}}$ = 200 GeV are shown together in Figure \ref{b_AuAu_CuCu_pp200}. The peripheral heavy ion data exhibits a short-range correlation as a function of $\Delta\eta$, with no long-range component. This is in good agreement with {\it pp}, which is dominated by short-range correlations.

\subsubsection{Energy Dependence of the Forward-Backward Correlation Strength in Cu+Cu}

\begin{figure}
\centering
\includegraphics[width=5.5in]{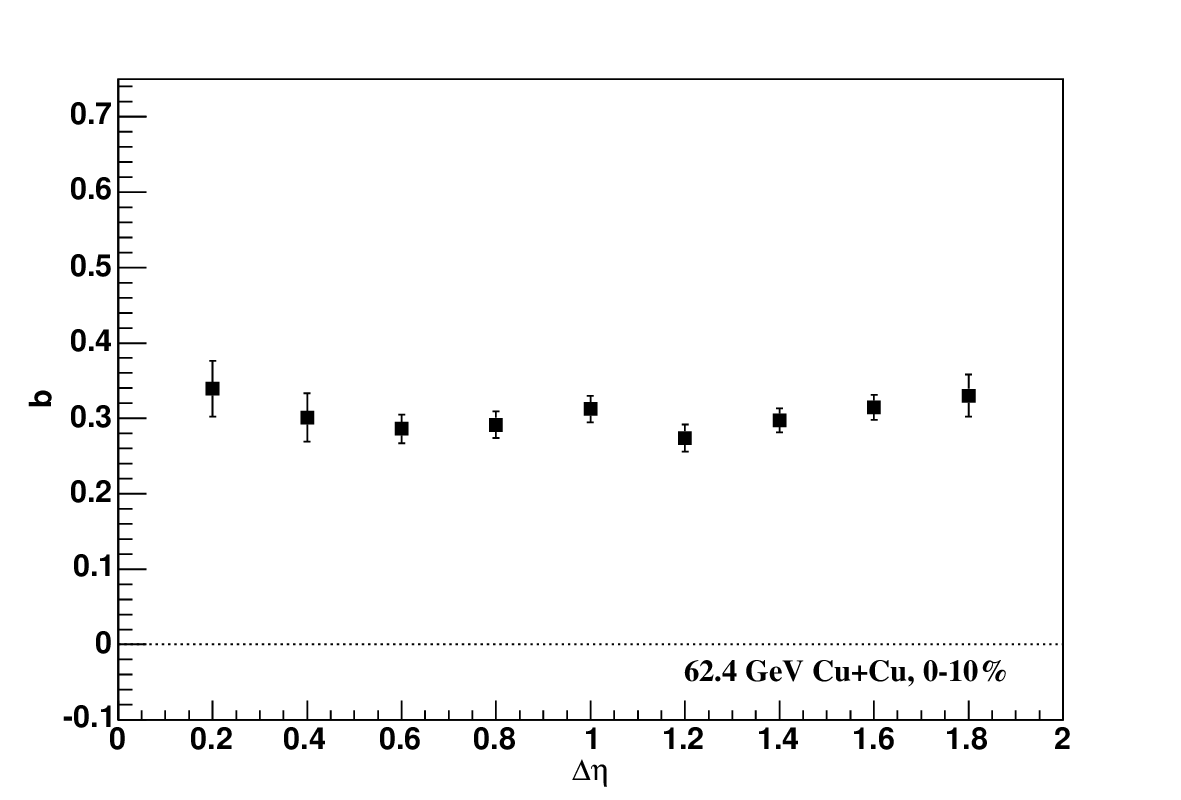}
\caption[{\it b} measured from 0-10\% Cu+Cu data at $\sqrt{s_{NN}}$ = 62.4 GeV.]{Forward-backward correlation strength {\it b} as a function of the pseudorapidity gap $\Delta\eta$ in central (0-10\%) Cu+Cu data at $\sqrt{s_{NN}}$ = 62.4 GeV.}
\label{b_CuCu_62_0-10}
\end{figure}

The FB strength was also studied for Cu+Cu collisions at $\sqrt{s_{NN}}$ = 62.4 GeV. The mean multiplicity in central (0-10\%) Cu+Cu at $\sqrt{s_{NN}}$ = 62.4 GeV is approximately equivalent to that from 10-20\% Cu+Cu at $\sqrt{s_{NN}}$ = 200 GeV and 40-50\% $\sqrt{s_{NN}}$ = 200 GeV Au+Au. As Figure \ref {b_CuCu_62_0-10} shows for central $\sqrt{s_{NN}}$ = 62.4 GeV Cu+Cu, the FB correlation strength {\it b} is about 20\% lower than 10-20\% $\sqrt{s_{NN}}$ = 200 GeV Cu+Cu (Figure \ref{b_CuCu_200_Cent_Dependence}) and over four times larger than that from 40-50\% $\sqrt{s_{NN}}$ = 200 GeV Au+Au (Figure \ref{b_AuAu_200_40-50}). 
Quantitatively, the long-range component exhibits a similarity to both 20-30\% Cu+Cu and Au+Au at $\sqrt{s_{NN}}$ = 200 GeV. 


\begin{figure}
\centering
\includegraphics[width=5.5in]{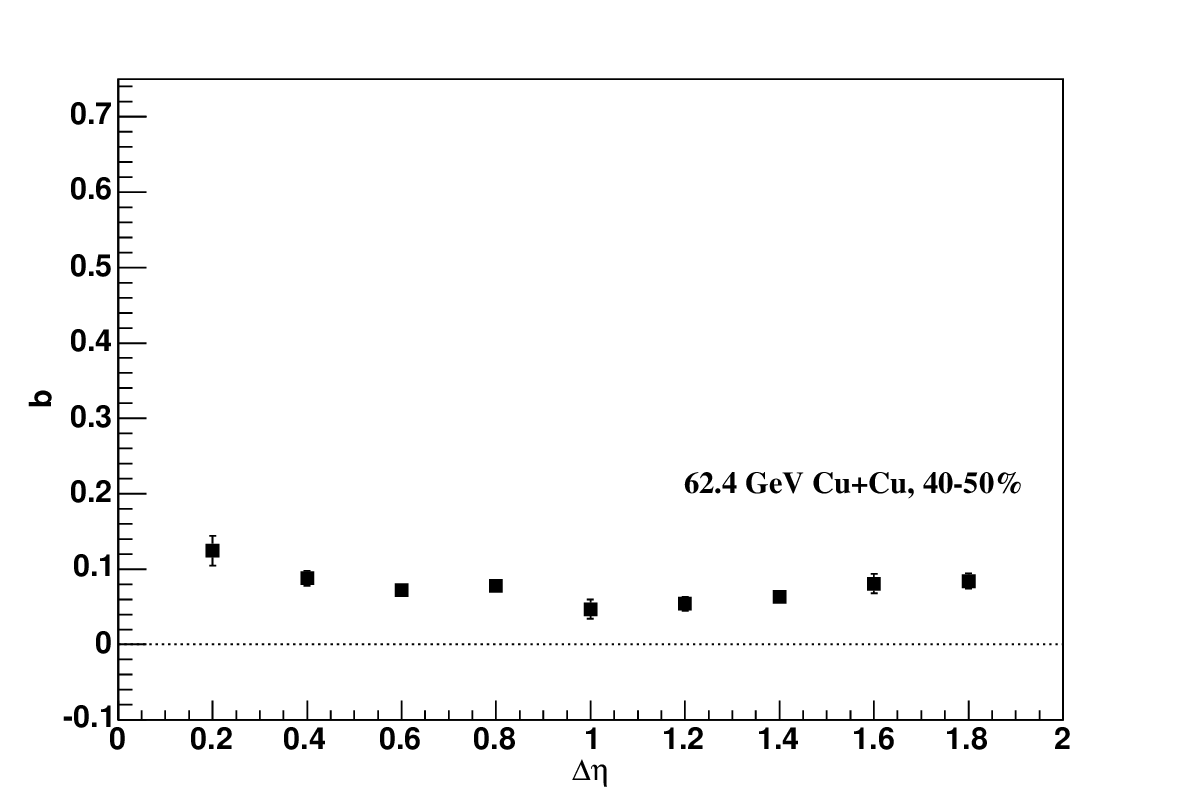}
\caption[{\it b} measured from 40-50\% Cu+Cu data at $\sqrt{s_{NN}}$ = 62.4 GeV.]{Forward-backward correlation strength {\it b} as a function of the pseudorapidity gap $\Delta\eta$ in mid-peripheral (40-50\%) Cu+Cu data at $\sqrt{s_{NN}}$ = 62.4 GeV.}
\label{b_CuCu_62_40-50}
\end{figure}

\begin{figure}
\centering
\includegraphics[width=5.5in]{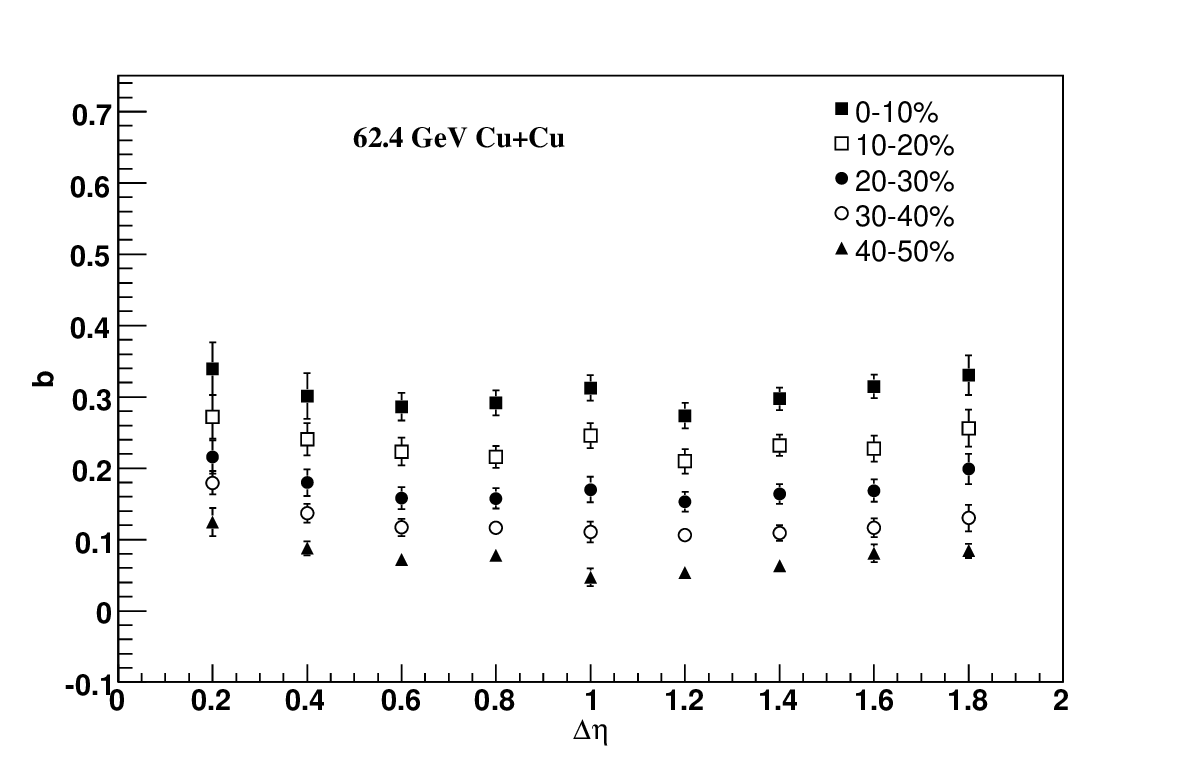}
\caption[Centrality dependence of {\it b} from Cu+Cu data at $\sqrt{s_{NN}}$ = 62.4 GeV.]{Forward-backward correlation strength {\it b} for 0-10\% (closed squares), 10-20\% (open squares), 20-30\% (closed circles), 30-40\% (open circles), and 40-50\% (closed triangles) most central Cu+Cu data at $\sqrt{s_{NN}}$ = 62.4 GeV as a function of the pseudorapidity gap $\Delta\eta$.}
\label{b_CuCu_62_Cent_Dependence}
\end{figure}

The results for mid-peripheral (40-50\% centrality) $\sqrt{s_{NN}}$ = 62.4 GeV Cu+Cu are shown in Figure \ref{b_CuCu_62_40-50}. 
As seen in the same centrality for both Cu+Cu and Au+Au at $\sqrt{s_{NN}}$ = 200 and 62.4 GeV and $\sqrt{s_{NN}}$ = 200 GeV MB {\it pp}, the FB correlation strength appears to possess only a short-range component that decreases with increasing $\Delta\eta$. The behavior as a function of centrality is shown in Figure \ref{b_CuCu_62_Cent_Dependence}.


\begin{figure}
\centering
\includegraphics[width=5.5in]{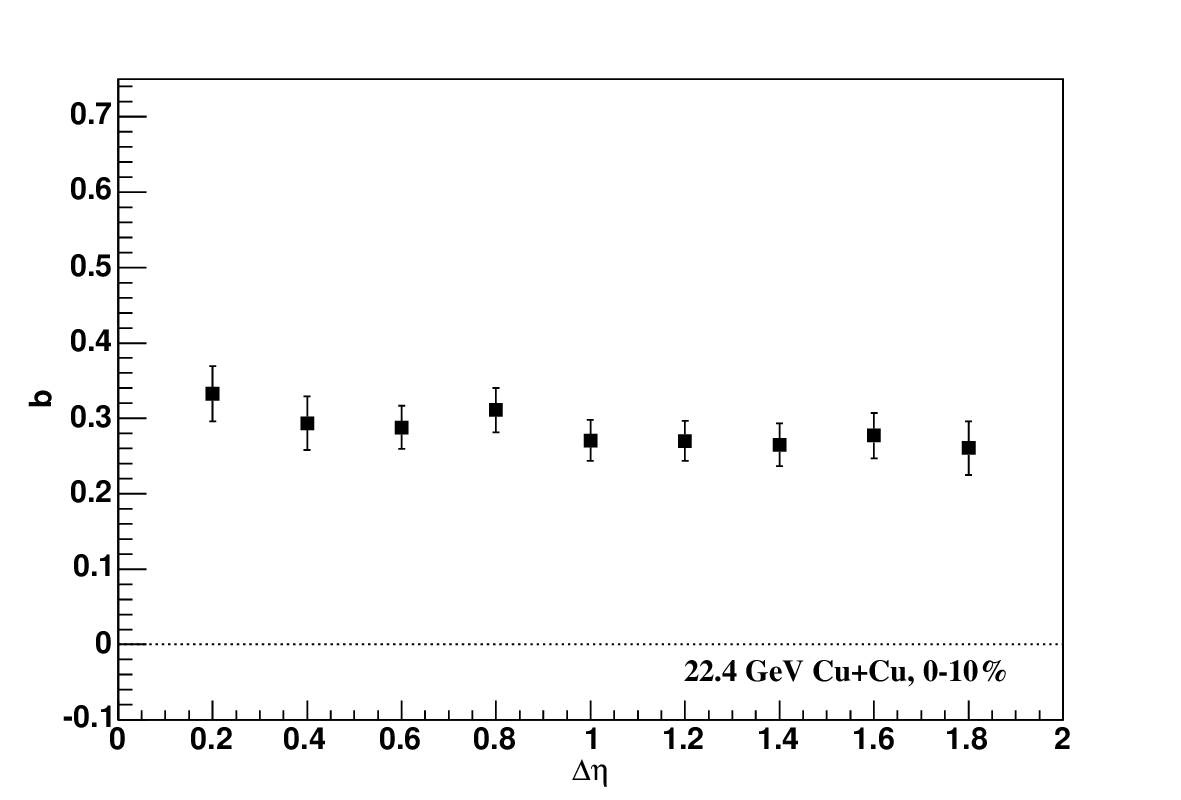}
\caption[{\it b} measured from 0-10\% Cu+Cu data at $\sqrt{s_{NN}}$ = 22.4 GeV.]{Forward-backward correlation strength {\it b} as a function of the pseudorapidity gap $\Delta\eta$ in central (0-10\%) Cu+Cu data at $\sqrt{s_{NN}}$ = 22.4 GeV.}
\label{b_CuCu_22_0-10}
\end{figure}

Additionally ,the FB correlation strength in Cu+Cu collisions was studied at the lowest available energy, $\sqrt{s_{NN}}$ = 22.4 GeV. Figure \ref{b_CuCu_22_0-10} shows the result for the most central (0-10\%) data. 
The FB correlation strength {\it b} is both qualitatively and quantitatively similar in central $\sqrt{s_{NN}}$ = 22.4  and 62.4 GeV Cu+Cu collisions (Figure \ref{b_CuCu_62_0-10}). Though the mean multiplicities are $\approx$ 30\% different, the similarity may indicate that the particle production mechanism for Cu+Cu at $\sqrt{s_{NN}}$ = 22.4 GeV is the same as that at $\sqrt{s_{NN}}$ = 62.4 GeV.


\begin{figure}
\centering
\includegraphics[width=5.5in]{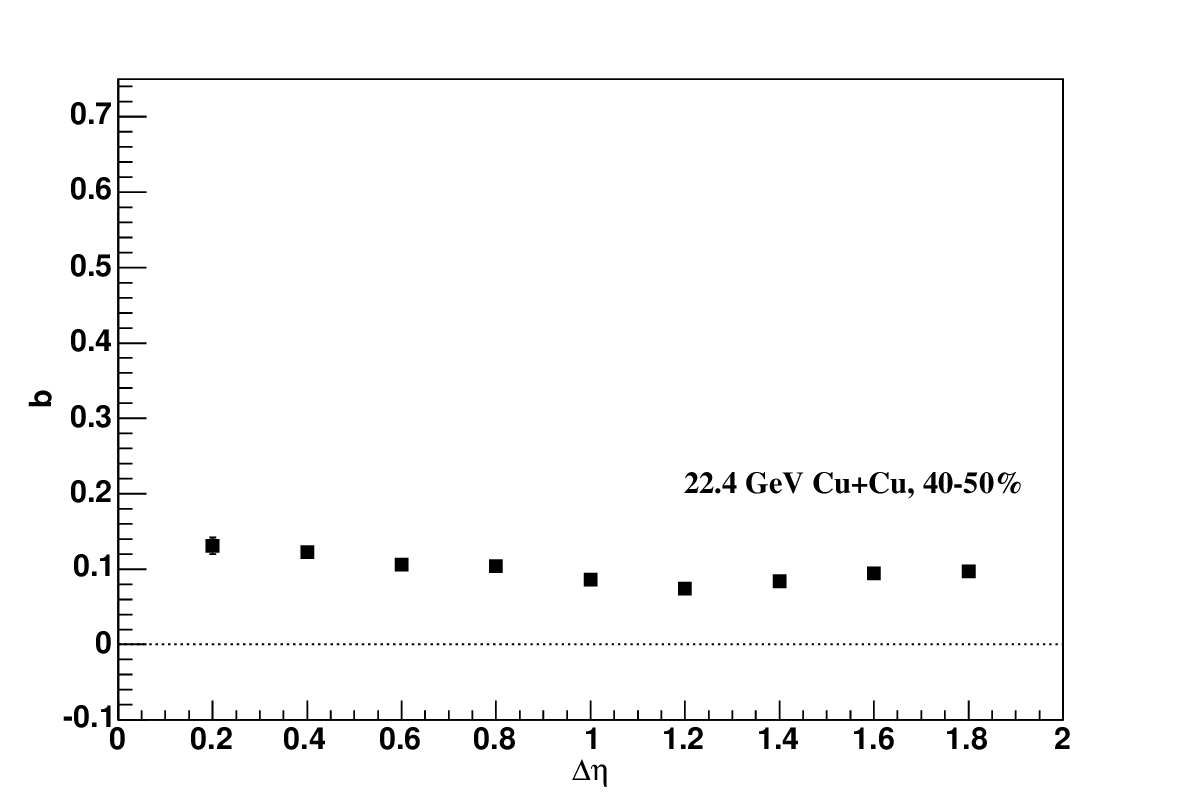}
\caption[{\it b} measured from 40-50\% Cu+Cu data at $\sqrt{s_{NN}}$ = 22.4 GeV.]{Forward-backward correlation strength {\it b} as a function of the pseudorapidity gap $\Delta\eta$ in mid-peripheral (40-50\%) Cu+Cu data at $\sqrt{s_{NN}}$ = 22.4 GeV.}
\label{b_CuCu_22_40-50}
\end{figure}

\begin{figure}
\centering
\includegraphics[width=5.5in]{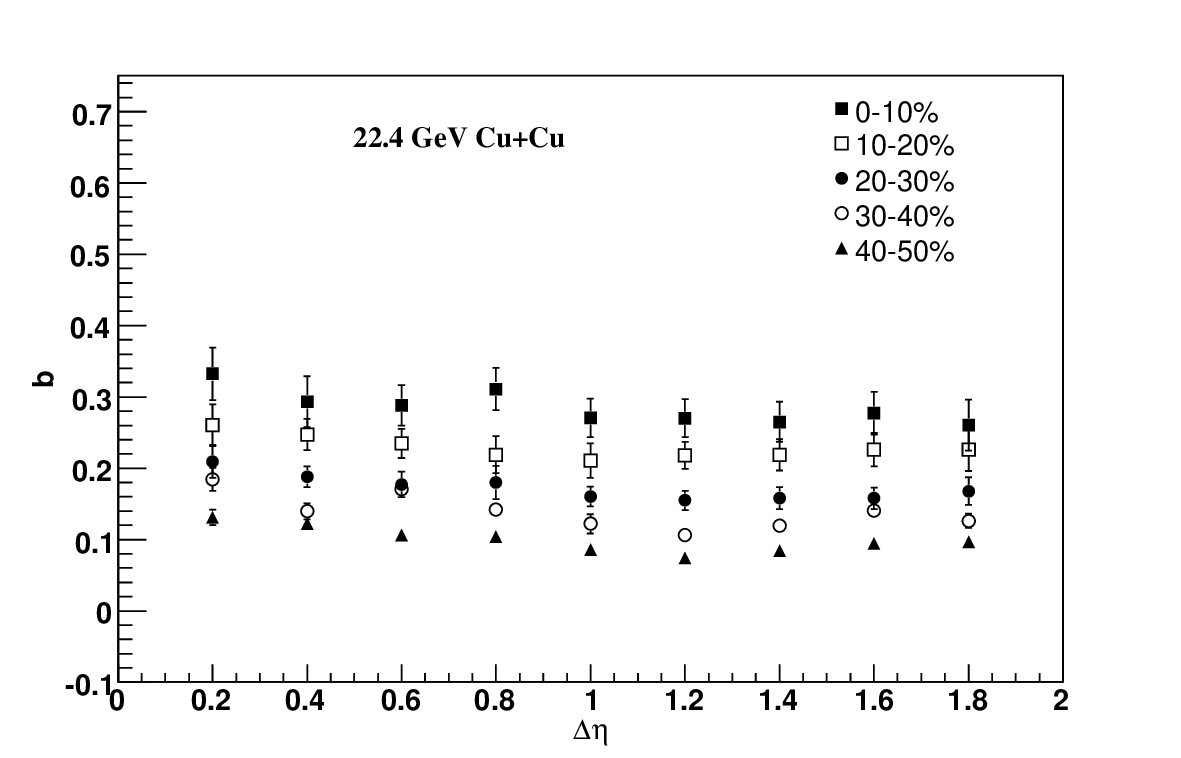}
\caption[Centrality dependence of {\it b} from Cu+Cu data at $\sqrt{s_{NN}}$ = 22.4 GeV.]{Forward-backward correlation strength {\it b} for 0-10\% (closed squares), 10-20\% (open squares), 20-30\% (closed circles), 30-40\% (open circles), and 40-50\% (closed triangles) most central Cu+Cu data at $\sqrt{s_{NN}}$ = 22.4 GeV as a function of the pseudorapidity gap $\Delta\eta$.}
\label{b_CuCu_22_Cent_Dependence}
\end{figure}

Mid-peripheral (40-50\%) $\sqrt{s_{NN}}$ = 22.4 Cu+Cu results (Figure \ref{b_CuCu_22_40-50}) 
is also in close qualitative and quantitative agreement with the corresponding measurement at $\sqrt{s_{NN}}$ = 62.4 GeV (Figure \ref{b_CuCu_62_40-50}). This also holds for the remaining centralities at $\sqrt{s_{NN}}$ = 22.4 (Figure \ref{b_CuCu_22_Cent_Dependence}) and $\sqrt{s_{NN}}$ = 62.4 (Figure \ref{b_CuCu_62_Cent_Dependence}). These are the first indications of the energy independence of the FB correlation strength in Cu+Cu below a collision energy of $\sqrt{s_{NN}}$ = 62.4.

\begin{figure}
\centering
\includegraphics[width=5.5in]{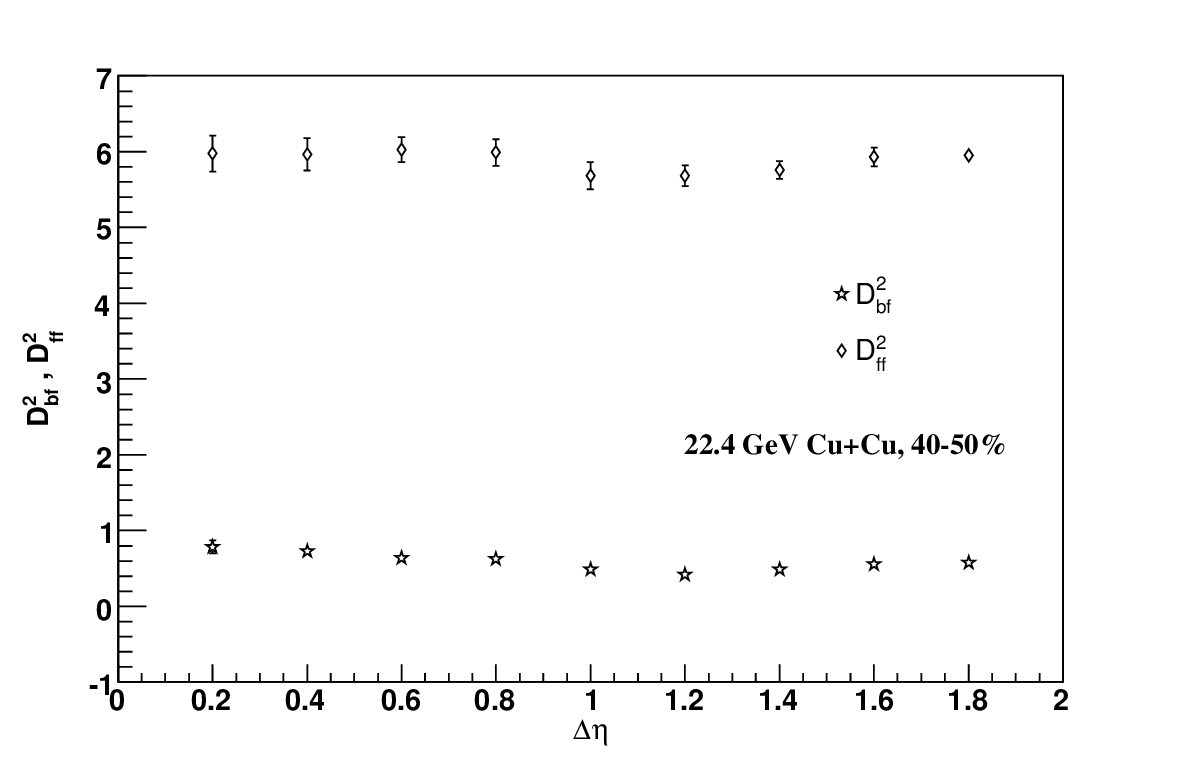}
\caption[Dispersions measured from 40-50\% Cu+Cu data at $\sqrt{s_{NN}}$ = 22.4 GeV.]{Backward-forward (stars) and forward-forward (diamonds) dispersions ($D_{bf}^{2}$ and $D_{ff}^{2}$) as a function of the pseudorapidity gap $\Delta\eta$ in mid-peripheral (40-50\%) Cu+Cu data at $\sqrt{s_{NN}}$ = 22.4 GeV.}
\label{Dbf_CuCu_22_40-50}
\end{figure}

\section{Discussion}

The study of long-range FB correlations in Au+Au, Cu+Cu, and {\it pp} reveals details of particle production for the bulk of particles in these collisions. Whether quark-gluon matter is produced or not, the vast majority of particles have transverse momentum, $p_{T} <$ 1.0 GeV. In this analysis, all charged particles with $p_{T} >$ 0.15 GeV are considered. This probes the behavior of the bulk material formed in these experiments. 
\\

The Dual Parton Model (DPM) ascribes the long-range correlation to fluctuations in the number of elementary, inelastic collisions, directed by unitarity (Section \ref{LRCinDPM}) \cite{LRCbasis}. In the DPM, the single particle inclusive spectrum in elementary {\it pp} collisions is expressed as \cite{Capella2003},

\begin{subequations}
\begin{equation}\label{SingleParticleDPM} 
\frac{dN^{pp}}{dy}(y) = \displaystyle\sum_{n}\frac{1}{\sigma_{n}}\displaystyle\sum_{n}\sigma_{n}\left(N_{n}^{qq-q_{\nu}}(y)+N_{n}^{q_{\nu}-qq}(y)=(2n-2)N_{n}^{q_{s}-\overline{q}_{s}}(y)\right)
\end{equation}
\begin{equation}\label{SingleParticleDPM2}
\widetilde{=}N_{k}^{qq-q_{\nu}}(y)+N_{k}^{q_{\nu}-qq}(y)+(2k-2)N_{k}^{q_{s}-\overline{q}_{s}}(y)
\end{equation}
\end{subequations}

where {\it k} is the average number of inelastic collisions (Equation \ref{meaninelas}). Each term contributes two strings per inelastic collision. Two strings are of the type diquark-quark, which come from the valence quarks of the proton. Any additional strings manifest between $q-\overline{q}$ from the quark sea. Recalling Equation \ref{SRCLRC} for the FB correlation in {\it pp},

\begin{eqnarray}\nonumber
<N_{f}N_{b}>-<N_{f}><N_{b}>  \:= \: <n>\left(<N_{0f}N_{0b}>-<N_{0f}><N_{0b}>\right) \nonumber\\
+ \left[\left(<n^{2}>-<n>^{2}\right)\right]<N_{0f}><N_{0b}>
\end{eqnarray}

with the first term in Equation \ref{SRCLRC} the FB correlation of particles produced in a single elementary inelastic collision, while the second term describes the fluctuation in the number of elementary inelastic collisions, identified with the long-range correlation.

The generalized version of Equation \ref{SingleParticleDPM} for nucleus-nucleus collisions at fixed impact parameter (b) is written as \cite{Capella2003}, 

\begin{equation}\label{SingleParticleDPM_AA}
\frac{dN^{AA}}{dy}(b) = n_{A}(b)\left[N_{\mu(b)}^{qq-q_{\nu}}(y)+N_{\mu(b)}^{q_{\nu}qq-}(y)+(2k-2)N_{\mu(b)}^{q_{s}-\overline{q}_{s}}\right]+\left(n(b)-n_{A}(b)\right)2kN_{\mu(b)}^{q_{s}-\overline{q}_{s}}(y)
\end{equation}

In Equation \ref{SingleParticleDPM_AA}, $n_{A}$ is the average number of participating (wounded) nucleons in the collision, $n$ is the average number of binary nucleon-nucleon collisions, and $\mu(b)$ is the average number of inelastic collisions by a particular nucleon at fixed impact parameter.	The first term in Equation \ref{SingleParticleDPM_AA} describes the nucleon-nucleon interactions outlined in Equation \ref{SingleParticleDPM2}, scaled by the average number of participating nucleons, $n_{A}$. In this case, the average number of inelastic collisions is $k \cdot n$, making the total number of strings $2k \cdot n$. The second term is required at high energies to account for the increasing influence of $q-\overline{q}$ sea strings \cite{Capella1999}.

It was shown that the FB correlation strength for {\it pp} and mid-peripheral (40-50\%) heavy ion data is dominated by short-range correlations. There is good qualitative and quantitative agreement for all energies and colliding systems when considering these mid-peripheral results. The lack of long-range FB correlations for peripheral heavy ion events indicates that there is a lack of multiple parton scattering. There is a strong suggestion that the correlation between particles in mid- and peripheral heavy ion collisions is very similar to that in {\it pp} collisions, where multiple parton interactions are expected to be small. 

It is possible that a small, residual long-range correlation is present in high energy {\it pp} collisions, perhaps related to the breakdown of KNO scaling above energies of $\sqrt{s_{NN}}$ = 100 GeV \cite{KNO1, KNO2}. It has been suggested that this breakdown in KNO scaling is related to the increasing contribution of multiple parton interactions in $pp$ and $p\overline{p}$ at higher energies \cite{Walker2004}. This provides a model independent correlation between long-range FB correlations and multiparton interactions in $pp$ collisions.
\\

In contrast, central and semi-central $\sqrt{s_{NN}}$ = 200 GeV Au+Au and Cu+Cu collisions exhibit a large FB correlation strength for $\Delta\eta > 1.0$. 
There is a substantial difference between the long-range FB correlation strength in central nucleus-nucleus collisions compared to the {\it pp}-like behavior of mid-peripheral and peripheral data. The experimental measurement of substantial long-range correlations is indicative of multiple partonic interactions in central and semi-central nucleus-nucleus collisions. 
\\

\begin{figure}
\centering
\includegraphics[width=5.5in]{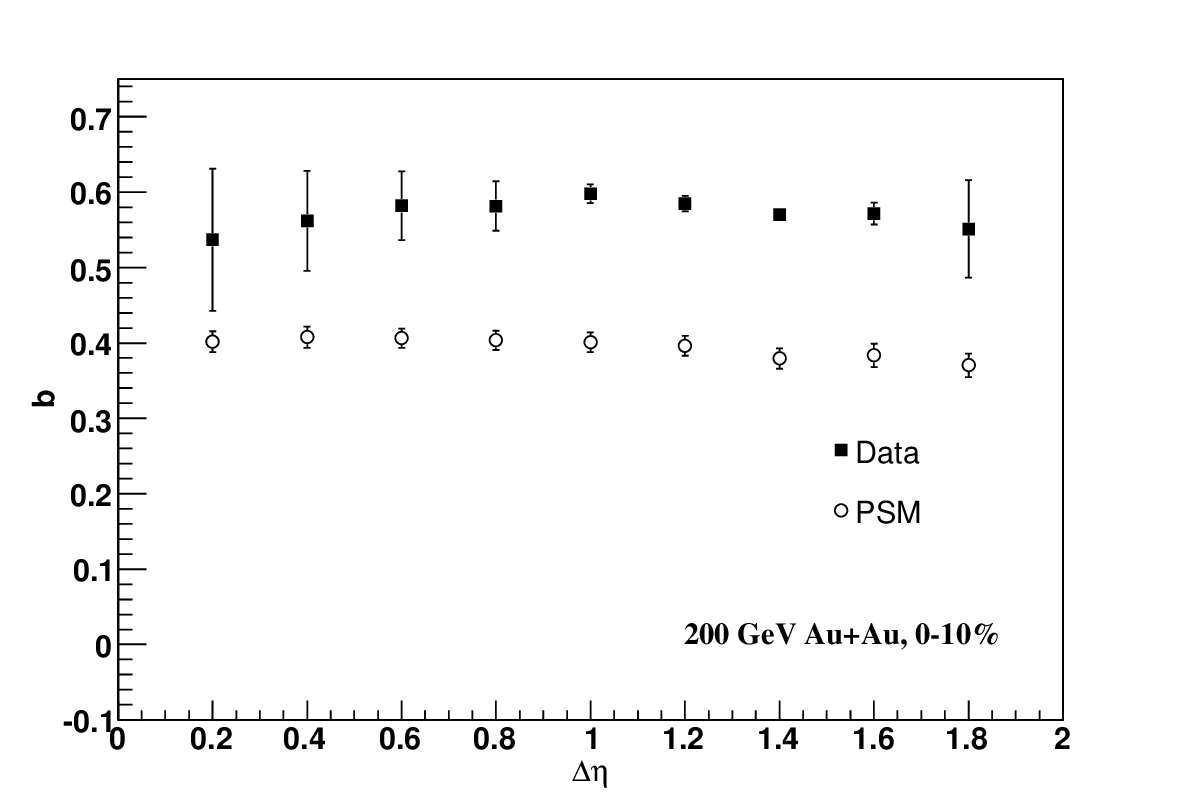}
\caption[{\it b} measured in data and the PSM for 0-10\% Au+Au data.]{Comparison of the measured forward-backward correlation strength {\it b} as a function of the pseudorapidity gap $\Delta\eta$ in central (0-10\%) Au+Au data at $\sqrt{s_{NN}}$ = 200 GeV (closed circles) to the prediction of the Parton String Model (PSM) (open circles). The PSM reproduces the qualitative trend of the long-range correlation, but under predicts the magnitude. For this simulation, the string fusion option is not enabled.}
\label{b_AuAu_200_0-10_PSM}
\end{figure}

Figure \ref{b_AuAu_200_0-10_PSM} demonstrates the comparison to the Parton String Model (PSM, Section \ref{PSM}) for central 0-10\% Au+Au at an energy of $\sqrt{s_{NN}}$ = 200 GeV. The PSM reproduces the qualitative trend of the FB correlation strength as a function of $\Delta\eta$. This provides a model dependent interpretation of the long-range FB correlation as being due to multiple partonic interactions.

A possible qualitative explanation for the centrality dependence of the FB correlation strength as shown in Figure \ref{b_AuAu_200_Cent_Dependence} relates to the idea of the formation of a central, partonic core surrounded by a hadronic corona. This has been discussed in terms of hydrodynamic properties and finite formation time of the strongly interacting quark gluon plasma (sQGP) \cite{HiranoGyulassy2006, Pantuev2006}, as well as a model to specifically quantify individual contributions from the core and corona \cite{Werner2007}. The primary difficulty is to find an experimental signal that can distinguish the respective contributions of the core and corona. The measurements of various observables integrate over all particles produced in the collision, whether they are produced from a single source (i.e., partonic system) or typical {\it pp}-like (i.e., hadronic) interactions.  If the manifestation of a large, long-range correlation is due to partonic effects (as described in the DPM), the partonic core formed in heavy ion collisions would be the primary source of this correlation. Since the hadronic corona would produce particles similar to {\it pp} interactions (with no formation of partonic matter), the contribution of the corona would consist of only short-range correlations. It is expected from geometrical arguments that this partonic core would have a larger volume in central heavy ion collisions, compared to more peripheral interactions. Therefore, the increasing influence of a partonic core would lead to a larger long-range correlation. The evolution with centrality also agrees with predictions from the DPM \cite{LRC_TTWWND, LRC_TTIWCF, LRC_TTCPOD, LRC_BKSQNP, LRC_BKSIWCF}. 
%
%

%
%


%
%

\chapter{Summary}
\section{Summary}

The energy and system size dependence of long-range multiplicity correlations has been studied using the forward-backward (FB) correlation strength {\it b}. The FB correlation strength was calculated in both nucleus-nucleus and hadron-hadron collisions for three systems: Au+Au, Cu+Cu, and {\it pp}. The energies considered were: $\sqrt{s_{NN}}$ = 200 and 62.4 GeV for Au+Au, $\sqrt{s_{NN}}$ = 200, 62.4, and 22.4 GeV for Cu+Cu, and $\sqrt{s_{NN}}$ = 400, 200 and 62.4 GeV for {\it pp}. In the most central nucleus-nucleus collisions at $\sqrt{s_{NN}}$ = 200 GeV, the magnitude of the FB correlation strength is approximately flat across a wide range in $\Delta\eta$. This includes a strong, long-range correlation for $\Delta\eta >$ 1.0. The long-range multiplicity correlations in central nucleus-nucleus collisions appear to depend less on the size of the colliding system (Au+Au or Cu+Cu) than on the energy of the incident nuclei. The data demonstrates that the long-range FB correlation strength in central heavy ion collisions is predominantly driven by the collision energy. This is the first indication of dense matter, which also exhibits partonic characteristics, in central heavy ion collisions.
\\

The long-range correlation decreases with decreasing centrality. By $\approx$ 40-50\% most central Au+Au (and Cu+Cu) collisions, the FB correlation strength is consistent with only short-range correlations, as also observed in {\it pp} interactions at the same energy. The evolution as a function of $\Delta\eta$ in semi-peripheral data shows a similar behavior across all energies and colliding systems. 
The energy dependence of the FB correlation strength in {\it pp} shows that for $\sqrt{s_{NN}}$ = 200 and 400 GeV, the FB correlation strength plateaus at a similar (small) value for $\Delta\eta > 1.0$, while $\sqrt{s_{NN}}$ = 62.4 GeV data goes smoothly toward $b = 0$. 
\\

These long-range correlations can be ascribed to multiple elementary inelastic collisions, which are predicted in the Dual Parton Model and the Color Glass Condensate/Glasma phenomenology. Additionally, the association of long-range correlations with the breakdown of KNO scaling provides a model independent interpretation in terms of multiple partonic collisions. The centrality dependence of the FB correlation strength demonstrates the increasing dominance of a partonic core in the resulting particle production.  This indicates that substantial amounts of dense 
partonic matter are formed in central Au+Au (and possibly Cu+Cu) collisions at $\sqrt{s_{NN}}$ = 200 GeV.

\section{Future Study}

Additional studies of the FB correlation strength remains to be done. Some of the items requiring further study are:

\begin{itemize}
\item Quantifying the number of particles produced from the core versus the corona.
\item The dependence of the FB correlation strength on transverse momentum ($p_{T}$).
\item The particle species dependence of the FB correlation strength, in terms of the difference between mesons and baryons.
\end{itemize} 


%
%
%

\bibliography{}
\bibliographystyle{unsrt}

\appendix
%
%

\chapter{Appendix}\label{AppendixA}

The kinematic variables used in relativistic heavy ion (and other high-energy) physics are defined in such a way that they have simple properties under Lorentz transformation. All notation makes use of natural units, $c = \hbar  = 1$. The rapidity variable, {\it y}, is defined as follows,

\begin{equation}\label{rapidity} 
y = \frac{1}{2}\left(\frac{p_{0} + p_{z}}{p_{0} - p_{z}}\right)
\end{equation}

This dimensionless quantity is defined in terms of $p_{0}$, the energy of the particle, and $p_{z}$, the longitudinal momentum of the particle. The longitudinal direction is the direction defined by the direction of the beam. An additive constant relates the rapidity of particles in different frames of reference. The relationship between $p_{0}$, $p_{z}$, and {\it y} is,

\begin{subequations}
\begin{equation}\label{yp0} p_{0} = m_{T}\cosh{y} \end{equation}
\begin{equation}\label{ypz} p_{z} = m_{T}\sinh{y} \end{equation}
\end{subequations}

where the transverse mass, $m_{T}^{2} = m^{2} + p_{T}^{2}$. The rest mass of the particle is {\it m}, and $p_{T}$ is the transverse momentum, the component of the particle momentum orthogonal to the beam direction.

To measure rapidity, an experiment must measure both energy and longitudinal momentum. A more tractable experimental measure is often the angle of the particle with respect to the beam axis. In that case, the pseudorapidity variable is useful,

\begin{equation}\label{eta}
\eta = -\ln\left[\tan\left(\frac{\theta}{2}\right)\right]
\end{equation}

where $\theta$ is the angle between particle momentum and the beam axis. Pseudorapidity and rapidity  become approximately equal for large particle momenta.
\\

In the collision of projectile A with target B, $A+B \rightarrow A+B$, the collision is referred to as elastic. If the target dissociates, $A+B \rightarrow A+X$, the reaction is called inelastic. The momentum transfer is defined as,

\begin{equation}
q = \Delta k = k - k'
\end{equation}

where k and k' are the initial and final 4-momenta of the projectile, respectively. In inclusive deep inelastic scattering experiments with electrons and protons, the scattering cross section was found to be independent of $q^{2}$ and scaled with the ratio \cite{BjorkenScaling},

\begin{equation}
x = \frac{Q^2}{2\vec{q}\cdot\vec{p}} = \frac{Q^2}{2M\nu}
\end{equation}

This was unlike the elastic cross-section, which is highly dependent on $q^{2}$. The Bjorken x is the fraction of the momentum carried by a parton in the infinite momentum frame and $\nu$ is the energy transfer $E - E'$.

\chapter{Appendix}\label{AppendixB}

\section{FTPC}

\subsection{Overview}

\begin{figure}
\centering
\includegraphics[width=5in]{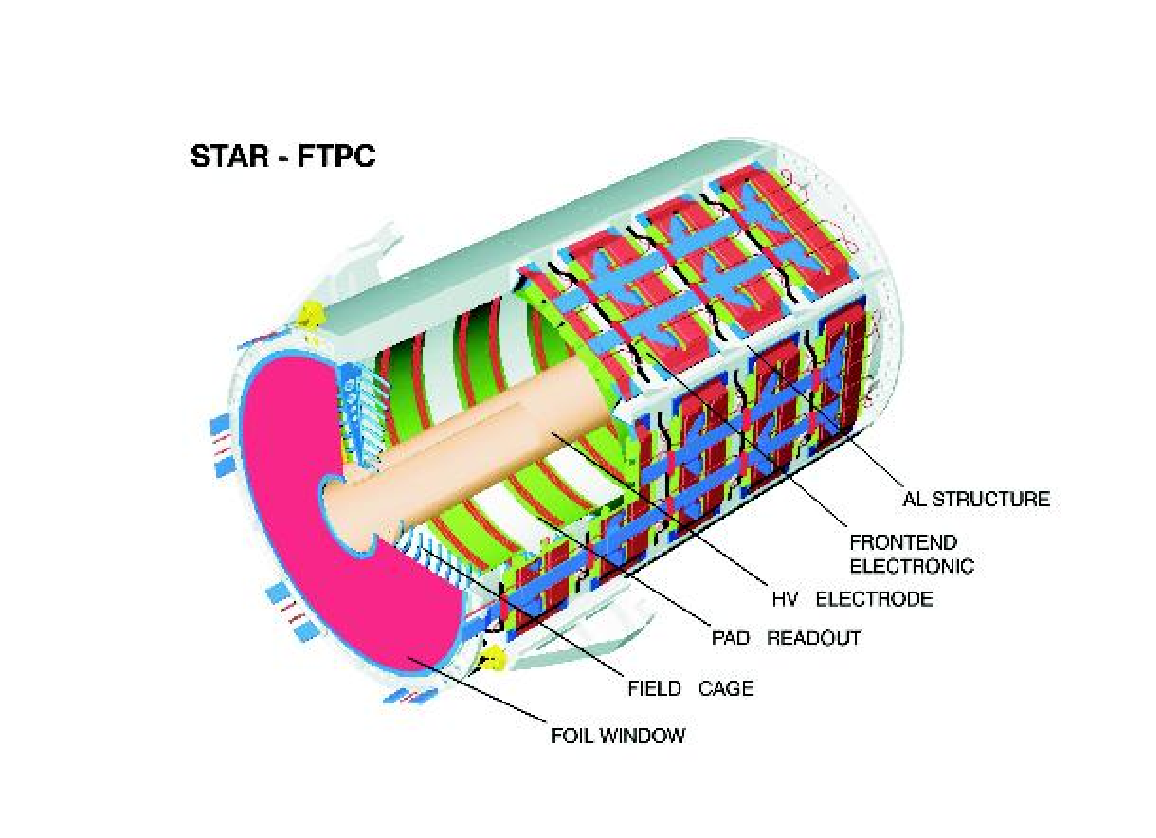} 
\caption{Schematic of a Forward Time Projection Chamber (FTPC).}
\label{FTPC}
\end{figure}

The Forward Time Projection Chamber (FTPC) subsystem at the STAR experiment consists of two cylindrical drift chambers 75 centimeters in diameter and 120 centimeters in length, mounted at $\pm$ 2.34 meters from the center of the Time Projection Chamber (TPC). These detectors provide full azimuthal (2$\pi$) charged particle tracking and momentum determination over a pseudorapidity ($\eta$) range of $2.5 < |\eta| < 4.0$. Due to the particle production rates in heavy ion collisions such as Au+Au, as well as space constraints within the TPC, the FTPC makes use of a radial drift field to maximize two-track resolution. This is unlike the main TPC, which utilizes a longitudinal electric field to drift electrons to the end caps. The structure of the FTPC is shown in Figure \ref{FTPC} \cite{STAR}. The readout chambers are curved structures located in five rings around the outer surface. These rings consist of two padrows each and are divided into six azimuthally equivalent readout chambers. The field cage consists of the inner, metallized plastic cylinder which functions as the HV-electrode, the outer cylinder wall at ground potential, and a concentric ringed, aluminum field cage that closes the field at either end. 

Because the FTPC utilizes a radial field, the drift electrons do not travel parallel to the STAR magnetic field as they do in the main TPC. Instead, they drift orthogonal to the magnetic field and experience a Lorentz force proportional to $\mathbf{\overrightarrow{E}x\overrightarrow{B}}$. This effect must be taken into account when reconstructing hit positions. At the inner radius, where the cluster density and the drift distance are maximal, the $\mathbf{\overrightarrow{E}x\overrightarrow{B}}$ effect leads to a broadening of the cluster distribution, improving the two-track resolution capability. The drift electrons are produced by ionization due to the passage of charged particles through the FTPC gas volume. The gas used is a 50\%-50\% mixture of Ar-CO$_{2}$, selected for its drift characteristics and chemical stability. The FTPC readout chambers are curved in order to maintain as close to a radial field as possible. Each FTPC has 9600 pads with 256 timebins/pad. Due to space constraints, the number of padrows in the FTPC is 10, which is also the maximum number of hits on track. Therefore, the FTPC momentum resolution is limited to $\approx$ 20\% across the range 0.1-2.0 GeV/c. Energy loss calculations, dE/dx, are also not possible with the FTPC due to the limited number of fit points. Therefore, the FTPC does not have particle ID capability.
\\

\subsection{FTPC Calibration}

The information presented in this section consists of several steps that are necessary to maintain the FTPC subsystem. In the past several years, the various software and hardware tasks to calibrate the FTPC for data taking purposes have been learned and implemented. This has required extensive time at Brookhaven National Laboratory in support of FTPC operations. All these tasks needed to be distilled down to the point where they could be carried out by one student. Previously, the support for the FTPC consisted of a larger group from another STAR institution. A recently completed, extensive write-up describing FTPC operational parameters and calibration procedures will help ease the training of additional people in FTPC procedures. This work is considered service work for the STAR collaboration.

\subsection{FTPC Software and Operational Tasks}

The FTPC software is an extensive set of C++ programs integrated into the STAR computing environment. The major programs can be broken down into four general categories: 

\begin{enumerate}
\item Cluster finding;
\item Particle tracking;
\item Laser calibration;
\item Diagnostic/operations.
\end{enumerate}

The FTPC must be calibrated for every run. Additionally, if the collider energy or particle species is changed, several parameters may need to be adjusted. There are several required procedures to both calibrate the detector and maintain detector effectiveness during and after data taking, including:

\begin{enumerate}
\item Laser analysis to verify gas composition;
\item Inner cathode correction of the internal detector geometry;
\item Transverse (x, y) FTPC vertex offset with respect to the TPC, which takes into account rotational corrections about FTPC mounting points;
\item Longitudinal (z) FTPC vertex offset with respect to the TPC, which indicates changes in electronics timing (t$_{0}$);
\item Analysis procedure for determining effective inner radius for cluster reconstruction (cluster radial step position), which provides a check on measured temperatures and t$_{0}$;
\item Gain tables to mask dead/noisy front-end electronics (FEEs);
\item Embedding to verify the tracking efficiency of the FTPCs;
\item Working knowledge of the FTPC control software, including interactive graphical user interfaces (GUIs) that are used to directly monitor and control the real-time operation of the detector;
\item Continual monitoring of quality assurance (QA) plots to diagnose potential problems during several months of data taking;
\item Troubleshooting of problems that arise both during the run and in fully reconstructed data.
\end{enumerate}

\subsection{FTPC Laser System}

\begin{figure}
\centering
\includegraphics[width=5in]{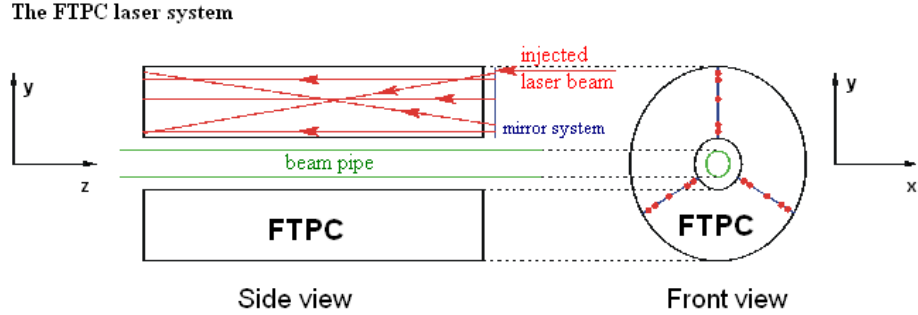} 
\caption{Laser configuration in one FTPC.}
\label{Laser}
\end{figure}

\begin{figure}
\centering
\subfigure[Clockwise from upper left: Laser ADC vs. radial position in the FTPC; \# hits on track; Phi position; Laser radial position vs. z-position.]{
\includegraphics[width=5in]{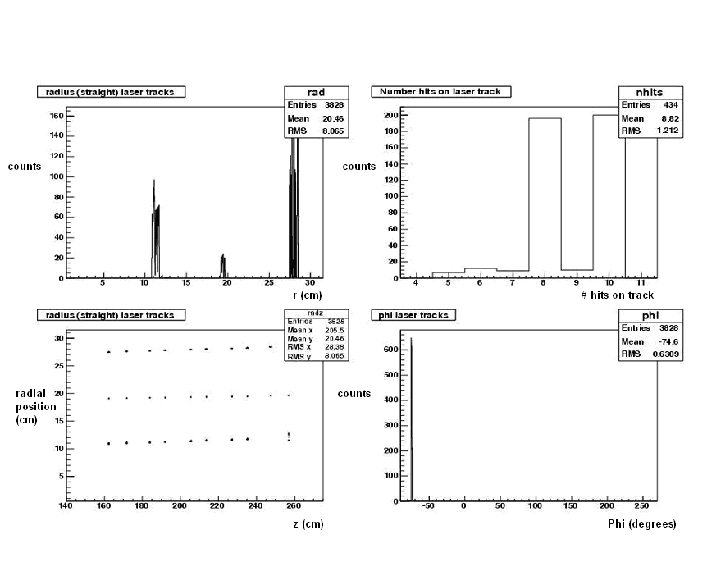} 
\label{Laser1}
}
\subfigure[Clockwise from upper left: x, y, r, and phi laser track residuals]{
\includegraphics[width=5in]{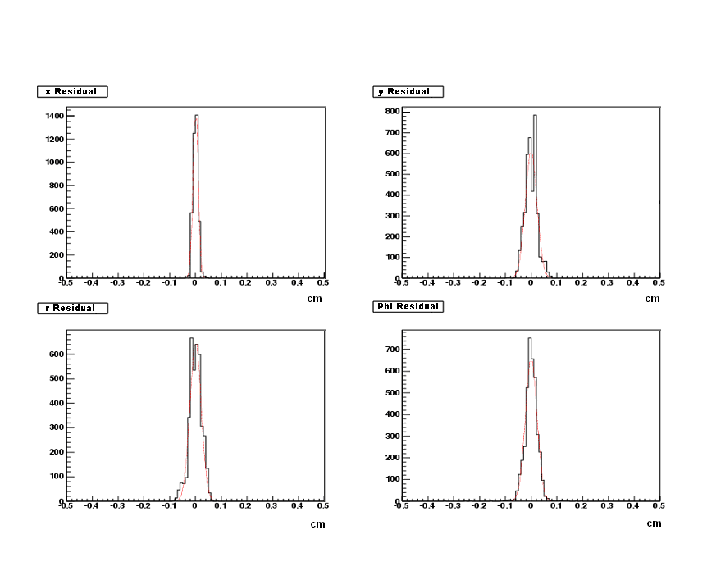} 
\label{Laser2}
}
\caption{Results from a laser run showing all three radial beams in one sector.}
\label{FTPC Laser}
\end{figure}

The gas composition of the FTPC determines the drift characteristics of the detector. Therefore, as part of the FTPC calibration process, it must be determined whether the gas composition has changed over the course of the data run. The FTPC laser system is integral for checking the gas composition and $\mathbf{\overrightarrow{E}x\overrightarrow{B}}$ corrections. The FTPC is sensitive enough to see changes in the gas composition on the order of 1\%. There are a maximum of 15 laser tracks per FTPC, 5 each in three laser sectors. In each sector are 3 straight tracks that run parallel to the beam pipe at a particular radius, and 2 inclined (diagonal) tracks. The measured radial positions of the 3 straight tracks are 11.91 cm, 19.55 cm, and 28.56 cm from the beam line. Laser runs are taken every few days during the run, or as needed. The events are processed through a modified version of the FTPC reconstruction chain that is optimized for laser runs. Once reconstructed, the data is processed once again to plot the positions and residuals for the laser tracks, after corrections have been made for magnetic field, t$_{0}$, and gas composition. The laser configuration and an example of the data from a representative laser run are shown in Figures \ref{Laser}, \ref{Laser1} and \ref{Laser2}.

The straight beams are used to check the gas composition and t$_{0}$. Because the inner beam has the longest drift distance, it will be more affected by changes to the gas composition. Conversely, the outer beam will not be as affected by the gas composition, but will be influenced more by changes to t$_{0}$. Ideally, the positions of the straight tracks will agree (or be close to) the theoretical positions with no change in either t$_{0}$ or gas composition. If that is not the case, iterating through various values of t$_{0}$ and gas composition is required. The residuals of the inclined tracks are used to check the $\mathbf{\overrightarrow{E}x\overrightarrow{B}}$ corrections.
\\

The high radiation operational environment in heavy ion collisions takes a physical toll on detectors. Occasionally, a large amount of stray particle flux from degrading beam conditions or the heavy ion collisions themselves can create currents in electronics beyond the level deemed safe. The affected anodes ``trip'', or discharge all their energy rapidly to hopefully prevent damage to current sensitive electronics. This is not always successful, and damage to Front End Electronics (FEEs), or even entire Readout Boards (RDOs), is possible. Dead and noisy electronics reduce FTPC tracking efficiency. Dead areas can be corrected with efficiency results from embedding, but noisy electronics are also useless when reconstructing data. The solution is to mask out these noisy electronics after data taking, but before data reconstruction. This is accomplished using a gain table. The gain table multiplies each pad by a calculated gain factor. If a pad exceeds a user defined noise cut (ADC count), the gain factor is set to 0 and the noisy pad is now a ``dead'' region.  
\\

There are additional steps that must be taken to calibrate the FTPCs to produce physics data. The FTPCs use tracks to independently reconstruct the primary vertex. These FTPC vertices will not be identical to the TPC vertex due to several factors. This includes changes in t$_{0}$ (which affects the z-component of the vertex position), the long lever arm of FTPC tracks to the primary vertex (small errors propagated over large distances), and a slight physical shift (or rotation) about the FTPC mounting points. This slight shift is exacerbated by the long lever arm when projecting tracks to the primary vertex, resulting in an offset of several millimeters in the transverse (x, y) plane. This offset needs to be corrected to ensure that the FTPC and TPC vertices match as closely as possible for proper data reconstruction. In 2004 it was discovered that the reconstructed FTPC vertices were different for various STAR magnetic field settings. There appeared to be an almost independent, orthogonal shift in the vertex position for each FTPC (one moved mostly in the x-direction, the other in the y-direction). When the magnetic field polarity was returned to its initial setting, the reconstructed FTPC vertices almost, but not quite, returned to their initial values. This could possibly be due to a small movement of the entire STAR detector and magnet coils due to a change in the force vector after a magnet polarity flip.
\\

\begin{figure}
\centering
\includegraphics[width=5in]{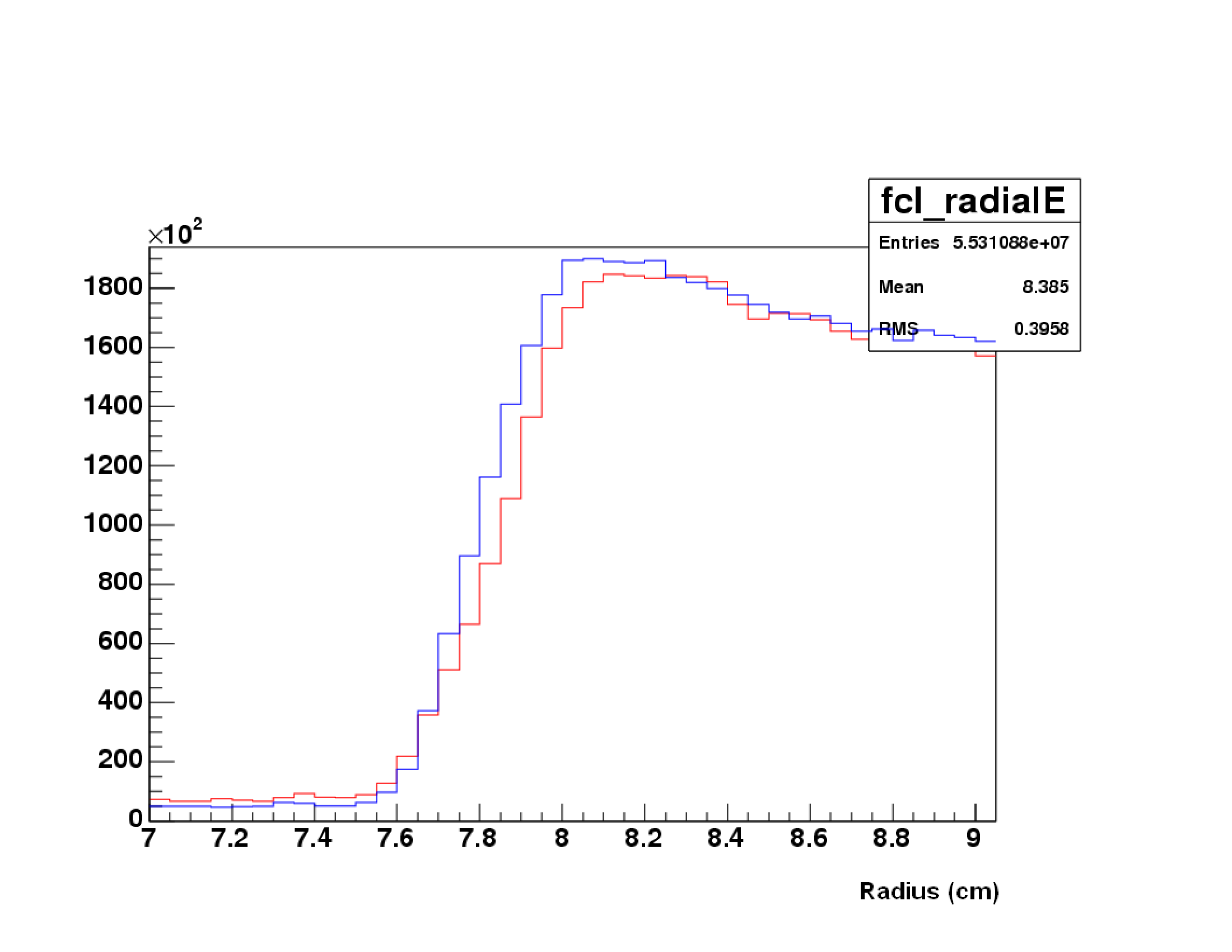} 
\caption[FTPC cluster radial position.]{The FTPC cluster radial position (radial step) at $\approx$ 7.8 cm, the start of the inner volume of the FTPC.}
\label{radialstep}
\end{figure}

The usable, inner volume of the FTPC begins at approximately 7.80 cm from the beam line. Therefore, good, reconstructed clusters should not be found at distances smaller than 7.80 cm. There are a small amount of bad clusters, either from electronics noise, beam background, or other sources, that are reconstructed below 7.80 cm. At $\approx$ 7.80 cm, there is a rapid rise in the cluster count, seen in Figure \ref{radialstep}.
\\

\begin{figure}
\centering
\includegraphics[width=5.5in]{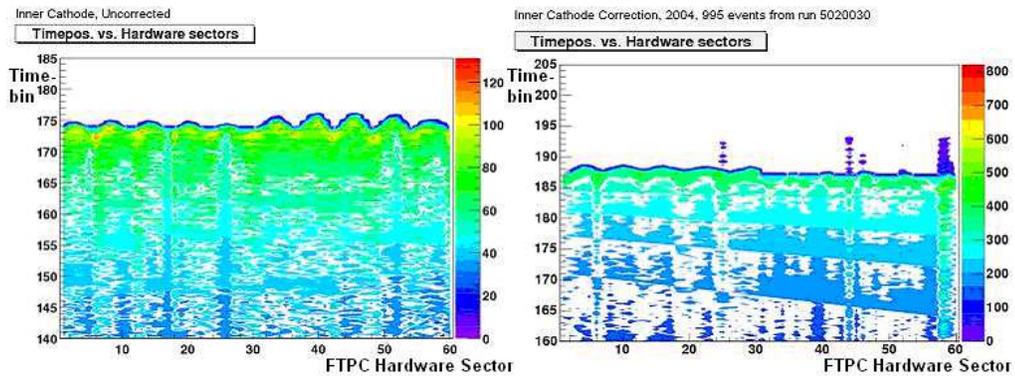} 
\caption[Time position of clusters at the outer radius of the FTPC.]{The time position of clusters at the outer radius of the FTPC as a function of hardware sector. The affect of the inner cathode offset is clearly seen as an oscillatory structure in both FTPCs (0-30, FTPC W, 30-60, FTPC E) (Left panel). After corrections are applied, the magnitude of the oscillations is reduced (Right panel).}
\label{cathode}
\end{figure}

As previously mentioned, the FTPC utilizes a cylindrical cathode located at the inner radius to produce the drift field. This would ideally be located in the direct center of the FTPC and provide perfect cylindrical symmetry for the produced electric field. However, due to slight machining errors, the cathode is not perfectly centered in the FTPC. This shift is on the order of 0.25 mm, but corresponds to a factor of 10 increase at the outer radius, or 2 mm. Additionally, the detector is sensitive enough that the effect of gravity on the inner cathode, causing a slight warping in the vertical direction, is also noticeable. The inner cathode offset has two, sympathetic, effects. It increases (decreases) the electric field in a particular hemisphere, while simultaneously decreasing (increasing) the drift distance, thereby affecting the drift time. This effect was seen as the oscillatory structure in the left panel of Figure \ref{cathode}. Fortunately, it is possible to correct for this effect.

\end{document}